\documentclass[namedreferences]{solarphysics}

\usepackage[optionalrh,solaromanenum]{spr-sola-addons} % For Solar Physics 
\usepackage{graphicx}                    % For eps figures, newer & more powerfull
\usepackage{amssymb}                     % useful mathematical symbols
\usepackage{color}                       % For color text: \color command
\usepackage{breakurl}                    % For breaking URLs easily trough lines
\usepackage{comment}
\usepackage{setspace}
\usepackage[utf8]{inputenc}
\usepackage[usenames,dvipsnames]{xcolor}
\renewcommand{\deg}{$^\circ$}
\newcommand{\mdeg}{^\circ}

\renewcommand{\l}{\lambda_{\rm N}}%\lambda_
\newcommand{\mrsun}{{\rm R_\odot}}
\newcommand{\med}{{\rm Md}}
\newcommand{\avgTe}{\left<\Tm\right>}
\newcommand{\MK}{{\rm MK}}
\newcommand{\cm}{{\rm cm}}
\newcommand{\cminvc}{\cm^{-3}}
\newcommand{\cminvs}{\cm^{-2}}
\newcommand{\erg}{{\rm erg}}
\newcommand{\s}{{\rm s}}

\newcommand{\LDEM}{{\rm LDEM}}
\newcommand{\FBE}{{\rm FBE}}
\newcommand{\TRF}{{\rm TRF}}
\newcommand{\lN}{\lambda_N}
\newcommand{\NCB}{N_{\rm CB}}

\newcommand{\dr}{\triangledown_r}
\newcommand{\er}{\mathbf{e}_r}
\newcommand{\Te}{T_{\rm e}}
\newcommand{\Tm}{T_{\rm m}}
\newcommand{\aTm}{\left<\Tm\right>}
\newcommand{\Tmi}{T_{{\rm m},i}}
\newcommand{\Ne}{N_{\rm e}}
\newcommand{\Nm}{N_{\rm m}}
\newcommand{\Nsqmi}{N^2_{{\rm m},i}}
\newcommand{\Nsqm}{N^2_{{\rm m}}}
\newcommand{\rhoTr}{\rho(T,r)}

\newcommand{\sqravgN}{\sqrt{\Nsqm}}
\newcommand{\Tc}{T_{\rm c}}

\newcommand{\etal}{{\it et al.}}
\newcommand{\Pl}{\texttt{+}}
\newcommand{\Mi}{\texttt{-}}

\begin{document}

\begin{article}

\begin{opening}

\title{Thermodynamic Structure of the Solar Corona: Tomographic Reconstructions and MHD Modeling}

\author[addressref={aff1},corref,email={dlloveras@iafe.uba.ar}]{\inits{D.G.}\fnm{Diego G.}~\lnm{Lloveras}}%\orcid{0000-0003-1402-0398}}

\author[addressref={aff1,aff2},email={albert@iafe.uba.ar}]{\inits{A.M.}\fnm{Alberto M.}~\lnm{V\'asquez}}%\orcid{0000-0003-3401-6409}} 

\author[addressref={aff1,aff3},email={federico@iafe.uba.ar}]{\inits{F.A.}\fnm{Federico A.}~\lnm{Nuevo}}%\orcid{0000-0003-2355-5853}} 

\author[addressref={aff1,aff2},email={cmaccormack@iafe.uba.ar}]{\inits{C.}\fnm{Cecilia}~\lnm{Mac Cormack}}%\orcid{0000-0003-1173-503X}}

\author[addressref={aff4},email={nishthas@umich.edu}]{\inits{N.}\fnm{Nishtha}~\lnm{Sachdeva}}%\orcid{0000-0001-9114-6133}}

\author[addressref={aff4},email={chipm@umich.edu}]{\inits{W.}\fnm{Ward}~\lnm{Manchester IV}}%\orcid{0000-0003-0472-9408}}

\author[addressref={aff4},email={bartvand@umich.edu}]{\inits{B.}\fnm{Bartholomeus}~\lnm{Van der Holst}}%\orcid{0000-0001-5260-3944}}

\author[addressref={aff4},email={rfrazin@umich.edu}]{\inits{R.A.}\fnm{Richard A.}~\lnm{Frazin}}%\orcid{0000-0002-0281-7677}}

\runningauthor{D.G.~Lloveras~\etal}
\runningtitle{Comparative Coronal Study of CR-2082 and CR-2208}

\address[id=aff1]{Instituto de Astronom\'{\i}a y F\'{\i}sica del Espacio (IAFE), CONICET-UBA, CC 67 - Suc 28,  (C1428ZAA) Ciudad Aut\'onoma de Buenos Aires, Argentina}

\address[id=aff2]{Universidad Nacional de Tres de Febrero (UNTREF). Departamento de Ciencia y Tecnolog\'{\i}a, Sáenz Peña, Argentina.}

\address[id=aff3]{Ciclo Básico Común (CBC), Universidad de Buenos Aires (UBA), Buenos Aires, Argentina}

\address[id=aff4]{Department of Climate and Space Sciences and Engineering (CLaSP), University of Michigan, 2455 Hayward Street, Ann Arbor, MI 48109-2143, USA}

\begin{abstract}
{We carry out a study of the global three-dimensional (3D) structure of the electron density and temperature of the quiescent inner solar corona ($r<1.25\,\mrsun$) by means of tomographic reconstructions and magnetohydrodynamic simulations. We use differential emission measure tomography (DEMT) and the Alfv\'en Wave Solar Model (AWSoM), in their latest versions. Two target rotations were selected from the solar minimum between solar cycles (SCs) 23 and 24 and the declining phase of SC 24. We report in quantitative detail on the 3D thermodynamic structure of the core and outer layers of the streamer belt, and of the high latitude coronal holes (CH), as revealed by the DEMT analysis. We report on the presence of two types of structures within the streamer belt, loops with temperature decreasing/increasing with height (dubbed down/up loops), as reported first in previous DEMT studies. We also estimate the heating energy flux required at the coronal base to keep these structures stable, found to be or order $10^5\,\erg\,\cminvs\,\s^{-1}$, consistently with previous DEMT and spectroscopic studies. We discuss how these findings are consistent with coronal dissipation of Alfvén waves. We compare the 3D results of DEMT and AWSoM in distinct magnetic structures. We show that the agreement between the products of both techniques is the best so far, with an overall agreement $\lesssim 20\%$, depending on the target rotation and the specific coronal region. In its current implementation the ASWsoM model can not reproduce down loops though. Also, in the source region of the fast and slow components of the solar wind, the electron density of the AWSoM model increases with latitude, opposite to the trend observed in DEMT reconstructions.}
\end{abstract}


\end{opening}

\section{Introduction}\label{intro} 

Being the {region} where the solar atmosphere plasma is {heated to million degree temperatures, the solar wind accelerated}, and where impulsive events such as solar flares and coronal mass ejections are energized, observing and modeling of the solar corona are of great relevance to improving our understanding of the Sun-Earth environment. To advance our knowledge of the physics of the solar corona, as well as {to enhance} and validate three-dimensional (3D) models, information derived from observational data plays a key role. Solar rotational tomography (SRT) is currently the sole observational technique able to provide a quantitative empirical description of the 3D distribution of some fundamental plasma parameters of the solar {corona at a global} scale.

To study the {3D structure of the quiet-Sun global corona,} SRT has proven to be a powerful tool. {In SRT, solar rotation is taken advantage of, so that instruments gather time series of images covering all viewing angles of the solar corona. This allows posing an inversion problem to solve for the unknown 3D distribution of specific quantities of the solar corona. Based on extreme ultraviolet (EUV) images, taken in several channels sensitive to different temperatures, differential emission measure tomography (DEMT) allows reconstruction of the 3D distribution of the differential emission measure (DEM). The final product of DEMT is in the form of 3D maps of electron density and temperature, covering the range of heliocentric heights $\lesssim 1.25\,\mrsun$.} The technique was first developed by \citet{frazin_2009}, and first applied to the observational study of coronal structures by \citet{vasquez_2009}. {A recent review on DEMT was published by \citet{vasquez_2016}. The technique is summarized in Section \ref{demt}.} 

{Non-tomographic studies of localized regions of the quiet-Sun corona have been carried out by means of DEM analysis. \citet{mackovjak_2014} used regularized inversion techniques to study characteristic temperatures in the quiet-Sun. \citet{lopez_2019} used a parametric method to study EUV dimmings after coronal mass ejections (CMEs) to estimate the coronal mass evacuated by the events. At a global scale, DEM analysis has been used by \citet{morgan_2017} to characterize the evolution of the temperature of the quiet-Sun corona during most of solar cycle (SC) 24.}

{The combination of DEMT with global magnetic models provides insight} into the 3D thermodynamical structure of the global corona. {DEMT was first combined with a potential field source surface (PFSS) model by \citet{huang_2012} and \citet{nuevo_2013}. More recently,} \citet{lloveras_2017} combined DEMT with PFSS {models to study the thermodynamics of the global solar corona in specific magnetic structures for two target rotations selected from the last two solar minimum epochs. Also combining DEMT with a PFSS model,} \citet{maccormack_2017} developed a new DEMT product that estimates the energy input flux {required at the coronal base to maintain stable coronal loops}. In this article, {DEMT is first combined with the magnetic field of a magnetohydrodynamic (MHD) model.}

The Alfv\'{e}n Wave Solar atmosphere Model (AWSoM) within the Space Wea\-ther Modeling Framework (SWMF) is a three-dimensional (3D) physics-based, data-driven MHD model extending from the upper chromosphere, to the upper corona and to {1 AU and beyond} \citep{vander_2010, Van2014}. {The only data input of the model is a magnetogram of the global corona, used as boundary condition for the simulation. As new improvements are implemented, the model is continuously being validated with observations.} DEMT results were used by \citet{jin_2012} and \citet{oran_2015} to validate AWSoM results finding an agreement within 50\% in density and electron temperature in the low corona. More recently, \citet{sachdeva_2019} compared the results of the latest version of AWSoM model with DEMT products in a global fashion. 

{In this work, the AWSoM model is used with two purposes. Firstly, to provide an MHD model of the coronal magnetic field to be used to study the DEMT results along magnetic field lines. Secondly, to provide thermodynamic results to be compared with those reconstructed by DEMT.}

{Combining the DEMT and AWSoM models, we carry out a detailed {quantitative} analysis of two target rotations. We selected Carrington rotation} (CR)-2082 (2009, 05 April through 03 May) during the SC 23/24 minimum, and CR-2208 (2018, 02 September through 29 September) during {the end of the declining} phase of SC 24. {In the case of CR-2082, the DEMT analysis is based on data} taken by the \textit{Extreme UltraViolet Imager Behind} {(EUVI-B: \citealt{wuelser_2004})} on board the \textit{Solar TErrestrial RElations Observatory} {(STEREO), while the AWSoM model uses the synoptic magnetogram provided by the Global Oscillation Network Group (GONG: \citealt{gong}). In the case of CR-2208, the DEMT analysis is based {on} data taken} by the \textit{Atmospheric Imaging Assembly instrument} {(AIA: \citealt{lemen_2012})} on board the \textit{Solar Dynamics Observatory} {(SDO), while the AWSoM model uses the magnetogram provide by the Air Force Data Assimilation Photospheric flux Transport (ADAPT)-GONG model.}

Section \ref{demt} and \ref{awsom} {summarize the DEMT technique and the AWSoM model, respectively.} Section \ref{trace} {details} the method used to trace {the DEMT} results along the field lines of the magnetic model. Section \ref{energia} details {the method that allows determination of the energy input flux at the coronal base.} In Section \ref{demt_res} the quantitative {detailed DEMT analysis of both target rotations is shown,} and in Section \ref{awsom_res} the AWSoM and DEMT results are compared. {Finally, Section} \ref{discu} summarizes and discusses the main conclusions of this analysis, and anticipates further planned work. 

\section{Methodology}\label{meto}   

\subsection{{DEMT Reconstructions}}\label{demt}

{As detailed in Section \ref{intro}, rotations CR-2082 and CR-2208 were selected to carry out DEMT reconstructions, based on data taken by the STEREO/EUVI-B and SDO/AIA instruments, respectively.} The EUVI and AIA data were prepared using the latest processing tools and calibration corrections provided by their teams through the SolarSoft package. {In the case of EUVI data, stray-light contamination is removed by deconvolution of the point spread function (PSF), carefully determined for each detector by \citet{shearer_2012}. In the case of AIA data, we have not yet implemented such a procedure as we were not aware of reliable determinations of their PSF, and we also understand that stray light contamination is expected to be less important for this instrument. A recent study by \citet{saqri_2020} indicates that the effect is noticeable in DEM analysis of coronal holes (CHs). That is also the case for DEM analysis of EUVI images. Nonetheless, as shown by \citet{lloveras_2017}, due to the temporal and spatial binning of the images used in tomography the effect of stray-light removal in DEMT products turns out to be mild, being $\lesssim 10\%$ for density products and negligible for temperature products. In the future, we will explore the effect of stray-light removal in AIA images on DEMT tomography, which we expect to be smaller than for EUVI images.} For this work, we introduced two improvements in the implementation of the DEMT technique, as described next.

{While in all previous DEMT studies full-disk data was used to perform tomography, in this work we {opt} to only use off-limb data. In this way, the smallest scale and brighter coronal features seen on disk (most typically in the 171\AA\ band) are not included. This has two implications. Firstly, the fast dynamics that typically characterizes those structures is absent from the data. Secondly, only half synodic rotation worth of data is needed to constrain the inversion problem for the whole coronal volume. As a result, {artifacts induced by coronal dynamics} are reduced compared to previous DEMT reconstructions.}

{The solution of the tomographic problem involves {inversion of} a very large sparse matrix. Such inversion problems are characterised by spurious {high-fre\-quen\-cy} artifacts in the solution, which can be mitigated through {regularization} techniques \citep{frazin_2000}. In the case of DEMT, all previous efforts used the 2D scheme implemented by \citet{frazin_2009}, using a finite difference matrix operator to approximate angular derivatives in both latitude and longitude. Also new to the present work, is the implementation of an expanded 3D regularization scheme, which adds to the previous scheme a finite difference matrix operator to approximate radial derivatives. In this way, the tomographic inversion problem is performed penalizing nonphysical high-frequency artifacts in all three spatial directions. As a result, tomographic reconstructions behave more smoothly close to the radial boundaries of the computational grid {when compared to previous reconstructions.}}

{In DEMT, the inner corona in the range of heliocentric heights $\lesssim 1.25\,\mrsun$ is discretized in a spherical computational grid. The size of the tomographic grid cell (or voxel) is typically set to $0.01\,\mrsun$ in the radial direction and 2\deg\ in both the latitudinal and longitudinal directions. The cadence of the data time-series is set to 6\,hr. The main product of the technique is the local DEM (LDEM) at each voxel, a measure of the temperature distribution of the plasma contained in it. We summarize next the main aspects of DEMT required for the analysis of this work.}

In a first step, the time series of EUV images is used to solve a solar SRT problem, for each EUV band independently. As a result, the 3D distribution of the so called \emph{filter band emissivity} (FBE) is determined for each band separately. The FBE, {an emissivity-type quantity}, is defined as the wavelength integral of the coronal EUV spectral emissivity and the telescope's passband function of each EUV channel. Line-of-sight (LOS) integration of the FBE provides synthetic images that can be quantitatively compared to the real data in the time series. To find the FBE, the tomographic problem is posed as a global optimization problem in which the quadratic norm of the difference between all {pairs of synthetic and real images is minimized.}

Due to unresolved coronal dynamics, tomographic reconstructions exhibit negative values of the reconstructed FBE, or zero when the solution is constrained to positive values \citep{frazin_2000,frazin_2009}. These non-reconstructed voxels are indicated {in black color in the latitude-longitude (Carrington) maps of DEMT} results in Section \ref{resu}.

{In a second step, the FBE values obtained for all bands in each voxel {of the tomographic grid are used to constrain the determination of the local-DEM (LDEM) which, as described in Section \ref{intro}, describes the temperature distribution of the plasma within the individual voxel.} Specifically, at each tomographic voxel $i$, the FBE of the band $k$ is related to the LDEM of the voxel according to}

\begin{equation}\label{fbeldem}
\FBE_i^{(k)}  = \int\,\textrm{d}T\,\LDEM_i(T)\,{\TRF}^{(k)}(T), \ \ \ {k=1,...,K} \,,
\end{equation}

\noindent
{where $K$ is the number EUV bands, and $\TRF^{(k)}$ is the \emph{temperature response function} of the $k$-th detector. In this work, the TRFs are computed based on the (known) channel’s passband times the coronal emissivity at that temperature (normalized by the squared electron density). The emissivity model used here is provided by the latest version of the CHIANTI atomic database and plasma emission model \citep{delzanna_2015,landi_2013}.}

{In this work,} data from three EUV bands {was used:} 171, 193 and {211 \AA\ in} {the} case of AIA, and 171, 195 and {284 \AA\ bands} in {the} case of EUVI. When using data from {three bands}, a Gaussian model for the LDEM is able to accurately predict {the FBEs \citep{nuevo_2015}}. In each tomographic voxel, the problem is then reduced to finding the values of the three free parameters of the Gaussian (centroid, standard deviation, and area) that best reproduce the three tomographically reconstructed values of FBE in that voxel.

{As the LDEM describes the temperature distribution of the plasma in a specific voxel, it does not deal with different large scale structures, as it may be the case for the DEM describing the plasma along a full LOS. As a result, LDEMs are usually successfully modeled with simpler profiles (such as Gaussian) than those returned by DEM studies constrained by LOS-integrated intensities. Parametric techniques are also used for DEM analysis of narrowband images, such as in the works by \citet{aschwanden_2011,plowman_2013,delzanna_2013}. Other methods applied to DEM analysis of narrow band images include Monte Carlo Markov Chain (MCMC) techniques \citep{schmeltz_2016}, regularized inversion techniques \citep{hannah_2012}, and iterative solvers \citep{pickering_2019,morgan_2019} that use the known TRFs of all filters as a functional base. The latter work in particular, introduced a fast iterative solver named SITES, which can easily be adapted for its use in DEMT. We will explore this in the future and compare results with those provided by our parametric technique.} 

Once the LDEM is determined at each voxel, {the LDEM-averaged squared electron density $\Nm^2$ and electron temperature $\Tm$} in the voxel can be computed by taking its zeroth and first moments over temperature. More specifically, at the $i$-th voxel,
{
\begin{eqnarray}\label{momento1}
 \Nsqmi = \langle \Ne^2 \rangle_i &=& \int\,\textrm{d}T\,\LDEM_i(T),\label{Ne} \\ 
\label{momento2}
 \Tmi  = \langle \Te \rangle_i &=& \frac{1}{\langle \Ne^2 \rangle_i}\,\int\,\textrm{d}T\,T\,\LDEM_i(T).\label{Tm} 
\end{eqnarray}
}

Next, we define a measure of the accuracy of the LDEM model to predict the tomographic FBEs in each voxel, as

\begin{equation}\label{R}
R_i \, \equiv \ (1/K) \, \sum_{k=1}^K \left| \, 1 - \FBE^{(k)}_{i,\rm syn}\,/\,\FBE^{(k)}_{i,\rm tom}\, \right| \,,
\end{equation}

\noindent 
{being the} {average absolute relative difference} between the tomographic and the synthetic FBEs. {The final product of DEMT is in the form of 3D maps of the LDEM-averaged quantities $\sqravgN$ and $\Tm$, as well as of the measure $R$.} {For a full description of the {DEMT technique} we refer the reader to \citet{frazin_2009}.}

\subsection{{AWSoM Simulations}}\label{awsom} 

{AWSoM is a three-temperature}, MHD model of the solar corona and inner heliosphere which provides the 3D distribution of density and temperatures as well as the 3D magnetic structure and velocity of the solar wind. In this work, we use AWSoM model simulated results below 1.25 $\mrsun$ to correspond to the DEMT analysis region.

{Heating of the solar corona is addressed} by including the non-linear interaction of forward and counter-propagating (reflected) Alfv\'{e}n waves which results in a turbulent cascade. This dissipated turbulent energy is distributed over anisotropic (parallel and perpendicular) proton temperatures and isotropic electron temperature using theories of linear wave damping and stochastic heating.

{The model accounts} for both collisional and collision-less electron heat conduction and does not use ad-hoc heating functions. The extended MHD equations including radiative cooling, heat conduction and wave turbulence within AWSoM \citep{Van2014} are solved using the Block-Adaptive-Tree-Solarwind-Roe-Upwind Scheme (BATS-R-US, \citealp{Pow1999, Tot2012}) numerical scheme. 

{In a previous} version of the model, the cascade time of the major wave was used to determine the wave damping rate \citep{Cha2011, Van2014}. {In its present version, the {energy partitioning is improved}} by using the Alfv\'{e}n wave number associated with the damping rate as determined by the critical balance condition, which uses the cascade time of the minor wave \citep{Lit2007}. This leads to more electron heating and less solar wind acceleration \citep{Van2019b}.

The inner boundary of AWSoM is located at the base of the transition region (at $\approx$1.0\,$\mrsun$). In reality, the thin {transition region (TR)} has steep gradients in temperature and density as a result of the balance between coronal heating, heat conduction and the radiative losses. To resolve these gradients in a global model would require excessive numerical resources. As described in \citet{Lio2009} and \citet{Sok2013}{, the TR is artificially {broadened} to be resolved with a} finest grid resolution of 0.001\,$\mrsun$.  To ensure that the base of the TR is not affected by chromospheric evaporation we overestimate the density at the inner boundary, $N_{e}=N_{i}=N_{\odot} =2\times10^{17}$\,m$^{-3}$ corresponding to the isotropic temperature values, $T_{e}=T_{i}=T_{i\parallel}=T_{\odot}=50,000$\,K{, where the subscripts represents electrons and ions}. The upper chromosphere is required to extend radially for the density to fall rapidly to correct (lower) values \citep{Lio2009}.  At this level, the radiative losses are sufficiently low so that the temperature can increase monotonically with height and form the transition region. Since the broadening of the transition region pushes the corona outwards, {the AWSoM model achieves coronal conditions at height $\approx 1.05\,\mrsun$, below which results can not then be compared to coronal tomographic reconstructions.}

To drive {the AWSoM model, estimates} of the photospheric magnetic field of the Sun {are the main} input. Synoptic magnetograms are used to specify the initial and the boundary conditions of the {magnetic field.} We use the PFSS model to extrapolate the 3D magnetic field (from the 2D photospheric magnetic field maps) using spherical harmonics. The source surface is taken to be at 2.5 $\mrsun$. GONG provides synoptic full-disk surface maps of the radial magnetic field component of the Sun. However, since the polar regions are not well observed from the ecliptic, GONG estimates the polar fields by fitting a polynomial to neighboring observed latitudes, which might lead to inaccuracies. An improvement over these maps is provided by the Air Force Data Assimilation Photospheric Flux Transport (ADAPT) model \citep{Wor2000}, which creates synchronic-synoptic maps by incorporating supergranulation, meridional circulation, and differential rotation. These maps provide a physics-based description of the unobserved polar magnetic fields \citep{Arg2010, Hen2012}. In this work, we use the GONG synoptic map as input for CR-2082 (ADAPT-GONG maps are unavailable for CR-2082) and the ADAPT-GONG global magnetic field map for CR-2208. {Based on results from previous efforts, for CR-2082} the magnetic field from the GONG map is scaled up by a factor of 1.85 {for weak fields ($B_r< 5\,\rm{G}$), while no modification is applied in the case of} the ADAPT-GONG map for CR-2208. The AWSoM steady-state simulation set-up and input parameters for both rotations are described below. 

To account for the energy partitioning between electrons and protons, based on \citet{Cha2011}, we set the stochastic heating exponent {equal to} 0.21 and the amplitude equal to {0.18, for both} rotations. The boundary condition for Alfv\'{e}n wave energy is given by the {poynting flux ($S_{A}$)} of the outgoing {wave, $({S_{A}/}{B})_{\odot} = 1.1\times 10^{6}$\,W\,m$^{-2}$\,T$^{-1}$ and $1.0\times 10^{6}$\,W\,m$^{-2}$\,T$^{-1}$ for CR-2082 and CR-2208, respectively, with $B_{\odot}$ being the field} strength at the inner boundary. The correlation length of the Alfv\'{e}n waves is set {to, $L_{\perp} \sqrt{B} = 1.5 \times 10^{5}$m\,$\sqrt{T}$}{, where $L_{\perp}$ is transverse to the magnetic field direction.}

{The computational domain of the solar corona} extends from 1 to {$24\,\mrsun$}. {The spherical grid has an adaptive grid that has fine resolution near the Sun, and increases outward} with the z-axis aligned with the rotation axis in Heliographic Rotation coordinates. The Adaptive Mesh refinement (AMR) resolves the angular cell size to $1.4\mdeg$ between $1 - 1.7\,\mrsun$ and to $2.8\mdeg$ outside this radius range. The solar corona component uses about 3 million cells on $6 \times 8 \times 8$ grid blocks (for CR-2208) and $6 \times 4 \times 4$ grid blocks for CR-2082. Local time stepping is used to speed up the steady state convergence. {The AWSoM simulation of the solar corona for solar minimum conditions represented by CR-2082 and CR-2208 are compared to the DEMT in the results section.}

\subsection{{Tracing Results Along Magnetic Field Lines}}\label{trace} 

{To determine} the electron density and temperature along individual magnetic field lines, first both the thermodynamic results and the magnetic field obtained with the AWSoM model were interpolated into the DEMT grid. Then, the geometry of the field lines is determined by numerical integration of the first order differential equations  ${\rm d}r/B_r = {r\,{\rm d}\theta}/B_\theta = r\,\sin(\theta)\,{\rm d}\phi/B_\phi$, both inwards and outwards, from the specified 3D coordinates of a starting point. In order to evenly sample the whole volume spanned by the DEMT reconstructions, one starting point is selected at the center of each tomographic cell at 6 uniformly spaced heights, ranging from $1.025$ to $1.225\,\mrsun$, and every 2$^\circ$ in both latitude and longitude, {for a total of $97,200$ starting points.}

{For analysis purposes, {the traced magnetic field lines are classified as open or closed according to their full geometry. Each closed field} line is further classified as ``small" or ``large", according to its coronal length $L$ being {respectively smaller or larger than the median value of the whole population, which is ${\rm Md}(L)\approx 0.5\,\mrsun$ for both rotations.} Finally, each closed magnetic field line is separated in its two ``legs", defined as the two segments that go from each of its two footpoints (\textit{i.e.} their location at $r=1\,\mrsun$) to its apex (\textit{i.e.}, the location of maximum height).}

At this {stage, DEMT and AWSoM products} {are} traced along open and closed magnetic field lines. Once the field line geometry is computed in high spatial resolution, only one sample point per tomographic cell is kept (the median one). As a result, for each field line one data point per tomographic cell {is obtained. This approach} was first used by \citet{huang_2012} to {study temperature} structures in the solar minimum {corona and by} \citet{nuevo_2013} to expand that analysis to rotations with different level of activity.

For each open field line and {for each closed field leg}, an exponential fit {was applied} to the electron density {data points} and a linear fit applied to the electron temperature {data points}. {For DEMT the data points used are} $\sqravgN(r)$ and $\Tm(r)$, and in case of {the AWSoM models} the data points used {are} $\Ne(r)$ and $\Te(r)$. The exponential and linear fit equations are described by

\begin{eqnarray}\label{Nfit}
\sqravgN &=& N_0\,\exp{\left[\,-\,\left(h/\l\right)\,/\,\left(r/\mrsun\right)\,\right]} \,,\\
\label{Tfit}
\Tm &=& T_0\,+\,a\,h\,,
\end{eqnarray}

\noindent
where $h \equiv r - 1\,\mrsun$ is the coronal height measured from the photosphere. In the electron density fit, $\l\,[\mrsun]$ is the density scale height and $N_0\,[\cminvc]$ is the electron density at the footpoint ($h=0$) of the loop. In the electron temperature fit, $a\,[\MK/\mrsun]$ is the slope and $T_0\,[\MK]$ is the electron temperature at the footpoint of the loop. The slope $a$ estimates the radial gradient of the electron temperature along the loop, which we denote as $a = \dr\Tm$ hereafter, {with} $\dr\equiv\er\cdot\nabla$ {being} the radial derivative operator {and} $\er$ the heliocentric radial unit vector.

In the case of the electron density, the fitted function corresponds to the isothermal hydrostatic equilibrium solution, allowing for variation of the solar gravitational acceleration with height. This choice of function provides a straightforward means to directly {compare the observed} coronal thermodynamical state with the hydrostatic solution.

{Coronal magnetic structures for which temperature increases/decreases with height (in the inner corona) were dubbed as “up”/“down” loops by \citet{huang_2012} and \citet{nuevo_2013}, who first reported the presence of down loops. As speculated by the authors of those works, down loops can be expected if the heating deposition is strongly confined near the coronal base of a magnetic loop. Down loops were first predicted by \citet{serio_1981}, and later by \citet{aschwanden_2002}. In a recent study, \citet{schiff_2016} reproduced both down and up loops by means of numerical simulations, using a 1D steady-state model and {considering time-averaged heating rates.}}

{To determine if the leg of a traced field line is} of type up or down, we first determine the Pearson correlation coefficient $\rhoTr$ between the DEMT temperature $\Tm$ and the heliocentric height $r$. We then select field lines for which the temperature {is significantly correlated with height} by requiring $|\rhoTr| > 0.5$. {To test the goodness of the fit we perform a chi-squared test \citep{recipes} considering the uncertainty of the DEMT data, selecting legs for which the fit matches the data with a 90\% confidence level.} In this way, legs for which the DEMT temperature does not show a significant {correlation} with height, or the linear fit to temperature is {weak}, are {excluded from the} analysis. {The test is also applied to the density-height data points, to ensure the trend is reasonably represented by the exponential fit.} Finally, selected legs are then classified as up or down according to $\dr\Tm > 0$ or $\dr\Tm < 0$, respectively. The linear fit allows characterization of the variation of $\Tm$ with height by means of a characteristic temperature gradient $\dr \Tm\,[\MK/\mrsun]$ along each leg. {The chi-squared test to evaluate the quality of the {fit} considers the uncertainty level in the DEMT products due to systematic sources (radiometric calibration and tomographic regularization), that \citet{lloveras_2017} estimated to be $\Delta\Tm\approx 10\%$ and $\Delta\sqravgN\approx 5\%$.} In summary, {a selected leg must meet three conditions:}

\begin{enumerate}
\item 
The leg must go through at least five tomographic grid cells with reconstructed data, and there must be at least one data point in each third of the range of heights spanned by the leg. {This requirement is set to ensure a reasonably spread sample of heights along the leg.}
\item 
The {DEMT temperature and height points must meet} $|\rhoTr| > 0.5$.
\item 
{The confidence level of both the exponential and linear fits must be larger than 90\%.} 
\end{enumerate}

{To characterize the global thermodynamic state of the inner solar corona in distinct magnetic structures, the DEMT and AWSoM results were traced along the magnetic field lines of the latter model. Based on the geometry and size of the loops, as well as on their thermodynamical properties, their legs were classified in four different types:} 

\begin{itemize}
\item Type 0: closed-small-down with footpoints in the range $|{\rm latitude}|<50\mdeg$. 
\item Type I: closed-small-up with footpoints in the range $|{\rm latitude}|<50\mdeg$. 
\item Type II: closed-large-up with footpoints in the range $|{\rm latitude}|>40\mdeg$. 
\item Type III: open with footpoint in the range $|{\rm latitude}|>60\mdeg$.
\end{itemize}

{In the case of closed-small field lines, the population of down and up legs becomes comparable for CR-2082, so we classify them into the two complementary classes of legs of type 0 (down) and legs of type I (up). In the case of closed-large field lines, down legs are $\lesssim 15\%$ of the population for both rotations. In the case of open field lines, down legs are $\lesssim 10\%$ of the population for both rotations. Hence, the requirement of being up for legs of type II and III is included to select the vastly dominating population in each case. On the other hand, the inclusion of latitude limits for the footpoints in the classification of legs from type 0 through III is purposely set to study the streamer belt in progressively outer layers, as well as to separate the field lines of the high latitude CHs (legs of type III). In Section \ref{resu}, the results of both the DEMT and AWSoM models in the four classes of legs are statistically analysed.}

\subsection{{Energy Input Flux}}\label{energia} 

{The high temperature of the corona requires heating mechanisms to compensate for the energy losses. While the vast majority of the existing literature on coronal heating focuses on active regions (ARs), some studies have been dedicated to the heating of quiet-Sun regions. In particular, \citet{maccormack_2017} developed a new application of the DEMT technique to estimate the energy input flux required at the base of quiet-Sun coronal loops to maintain stability. The technique is based on tracing the DEMT results along field lines of a global coronal magnetic model, as described in Section \ref{trace}.}

{Consider} a static energy balance for each magnetic flux tube, {in which the dominating losses of} radiative power ($E_r$) and thermal conduction power ($E_c$) are compensated {by a} coronal heating power ($E_h$) \citep{aschwanden_2004}:

\begin{equation}\label{Balance}
E_h(s) = E_r(s) + E_c(s),
\end{equation}

\noindent
where $s$ is the position along {the flux tube} and the {power quantities} are in units of $[{\rm erg\,sec^{-1}\,cm^{-3}}]$.

{The thermal conduction power $E_c$ equals the divergence of the conductive heat flux $F_c$, \textit{i.e.} $E_c(s)=[1/A(s)]\,{\rm d}[A(s)\,F_c(s)]/{\rm d}s$, where $A(s)$ is the {cross-sectional} area of the magnetic flux tube at position $s$. Under a quiescent solar corona plasma regime, the conductive flux is assumed to be dominated by the electron thermal conduction, described by the usual Spitzer model \citep{spitzer_1962}

\begin{equation}\label{Fc}
F_c(s)=-\kappa_0\,{T(s)}^{5/2}\,\frac{dT}{ds}(s),
\end{equation}
where $\kappa_0 = 9.2 \times 10^{-7}  {\rm erg\,sec^{-1}\,K^{-7/2}}$ is the Spitzer thermal conductivity.}

{In the corona, EUV emission is dominated by collisions of the emitting ions with free electrons, so that the radiative power scales as $N_e^2$. The radiative power of an isothermal plasma at temperature $T$ is then computed as $E_r=N_e^2\,\Lambda(T)$, where the radiative loss function $\Lambda(T)$ is calculated by means of an emission model. In this work we used the latest version of the atomic database and plasma emission model CHIANTI \citep{delzanna_2015}. The radiative power in terms of the LDEM is then given by:}

\begin{equation}\label{Er}
E_r = \int\,\textrm{d}T\,\LDEM(T)\,\Lambda(T).
\end{equation}

{The energy balance given by Equation \ref{Balance} is then integrated in the volume of any given coronal magnetic flux tube. Dividing the result by the flux tube area at the coronal base, and making use of the soleidonal condition of the magnetic field, a {field line integrated} version of that energy balance is found,

\begin{equation}\label{FluxBalance}
\phi_h = \phi_r + \phi_c,
\end{equation}

\noindent
where the {line-integrated} flux quantities $\phi_{r,c}\,[\rm{erg\,sec^{-1}\,cm^{-2}}]$ are given by \citep{maccormack_2017}:}

\begin{equation}\label{phi_r}
\phi_r = \left( \frac{B_0 \ B_L}{B_0 + B_L} \right) \int_{0}^{L} ds \ \frac{E_r(s)}{B(s)},
\end{equation}

\begin{equation}\label{phi_c}
\phi_c = \left( \frac{B_0 \ F_{c,L} - B_L \ F_{c,0}}{B_0 + B_L} \right) ,
\end{equation}

\noindent 
{where $L$ is the length of the loop, and $B_0$ and $B_L$ are the values of the magnetic field at the footpoint locations of the loop in the coronal base, namely $s=0$ and $s=L$.} {{For} any given field line, all quantities in these two expressions can be computed {from the DEMT results traced along field lines, and the AWSoM magnetic field model,} through Equations \ref{Fc}-\ref{Er}. Once computed, the quantity $\phi_h$ can be calculated, which is the energy input flux required at the coronal base of each coronal field-line to maintain a stable coronal structure.}

\section{Results}\label{resu} 

\subsection{Tomographic Results}\label{demt_res} 

{As described in Section \ref{demt}, we carried out DEMT reconstructions of the coronal structure for rotations CR-2082 and CR-2208 using STEREO/EUVI and SDO/AIA data, respectively.} {Once the LDEM was determined for each rotation, the square root of the mean value of the electron density squared ($\sqravgN$) and the electron mean temperature ($\Tm$) were computed at each voxel of the tomographic computational grid by means of Equations \ref{momento1} and \ref{momento2}{, and the measure $R$ was calculated by means of Equation \ref{R}.}} 

\begin{figure}[h!]
\begin{center}
\includegraphics[width=0.495\textwidth]{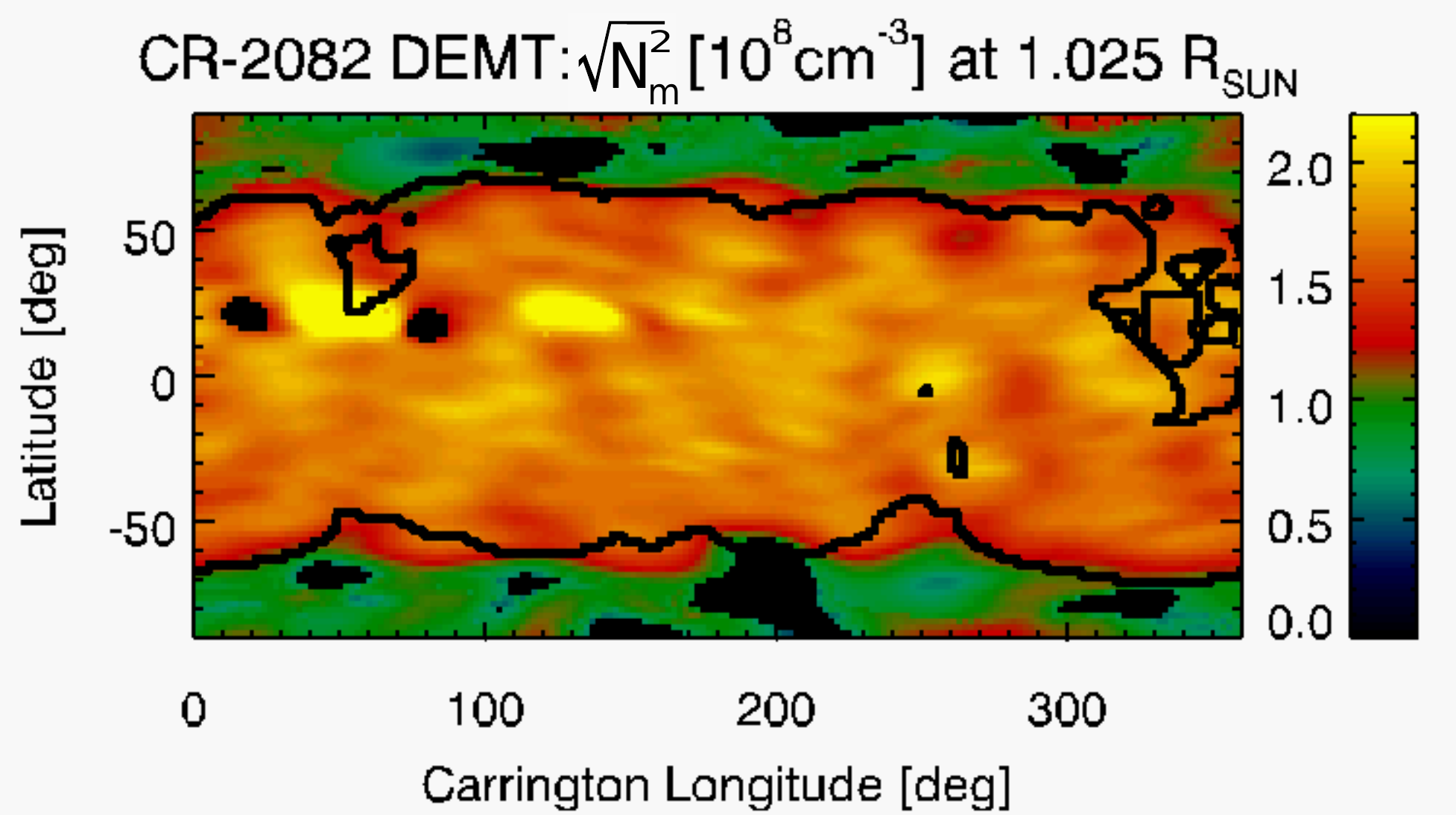}
\includegraphics[width=0.495\textwidth]{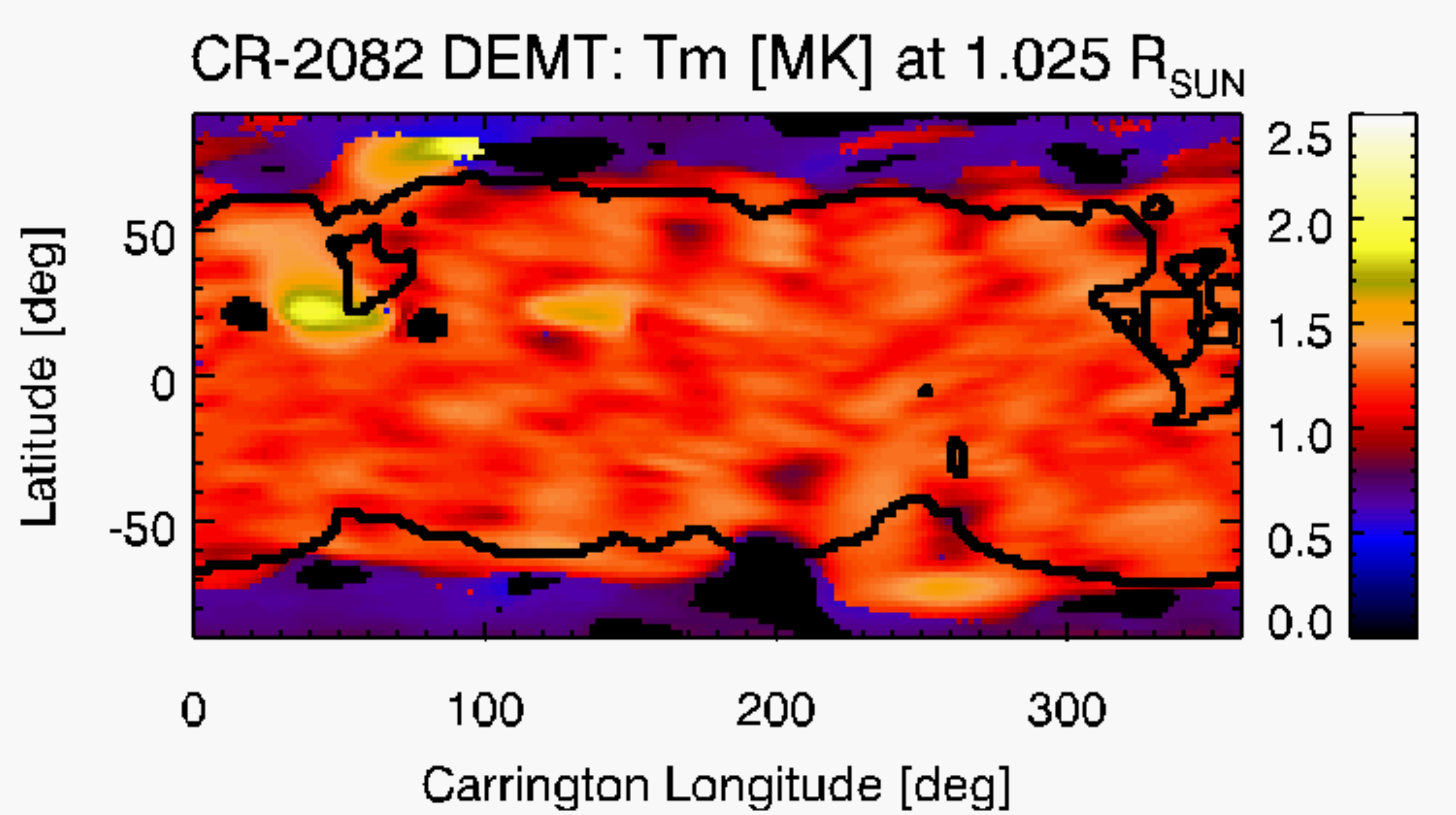}
\includegraphics[width=0.495\textwidth]{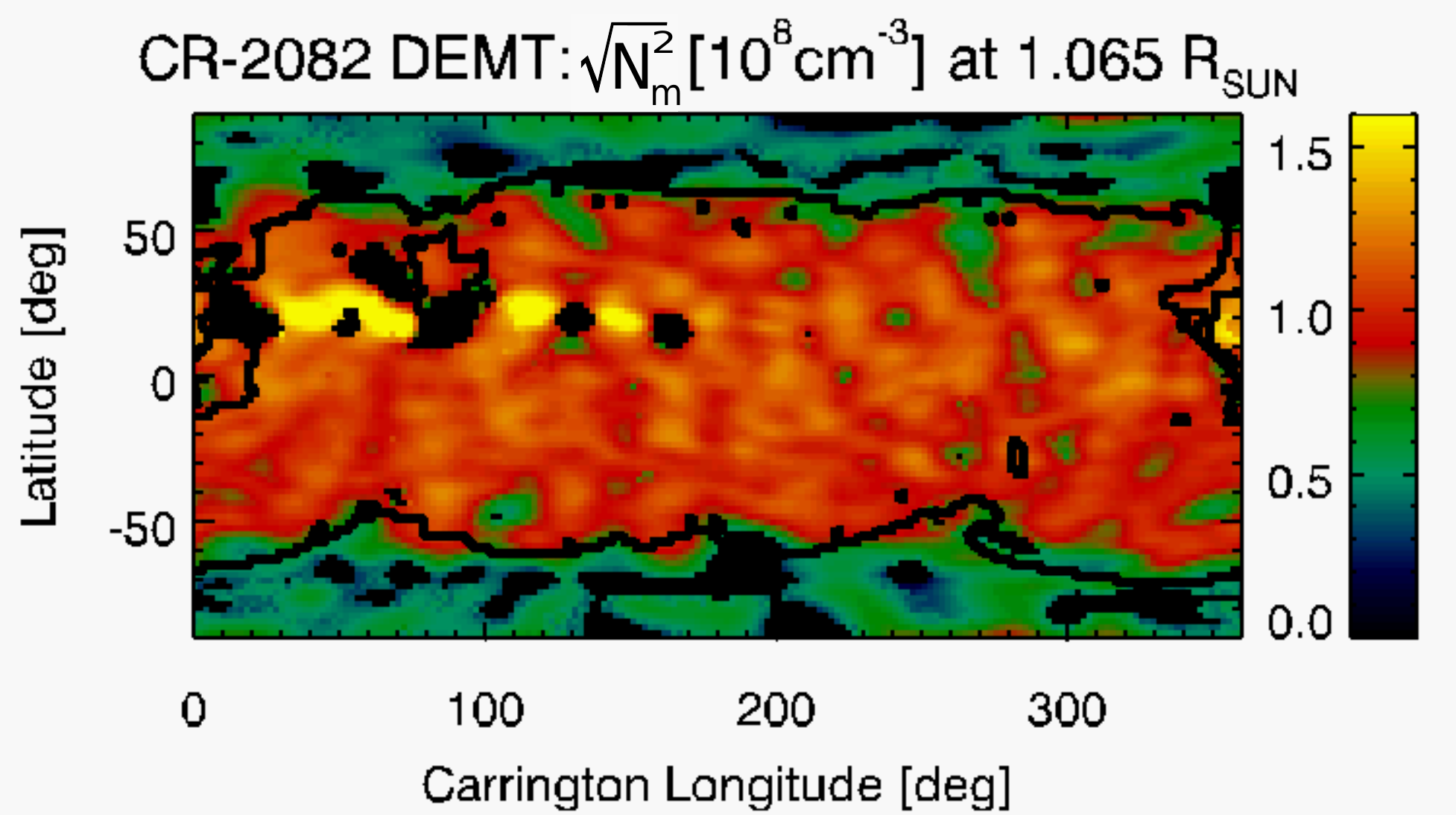}
\includegraphics[width=0.495\textwidth]{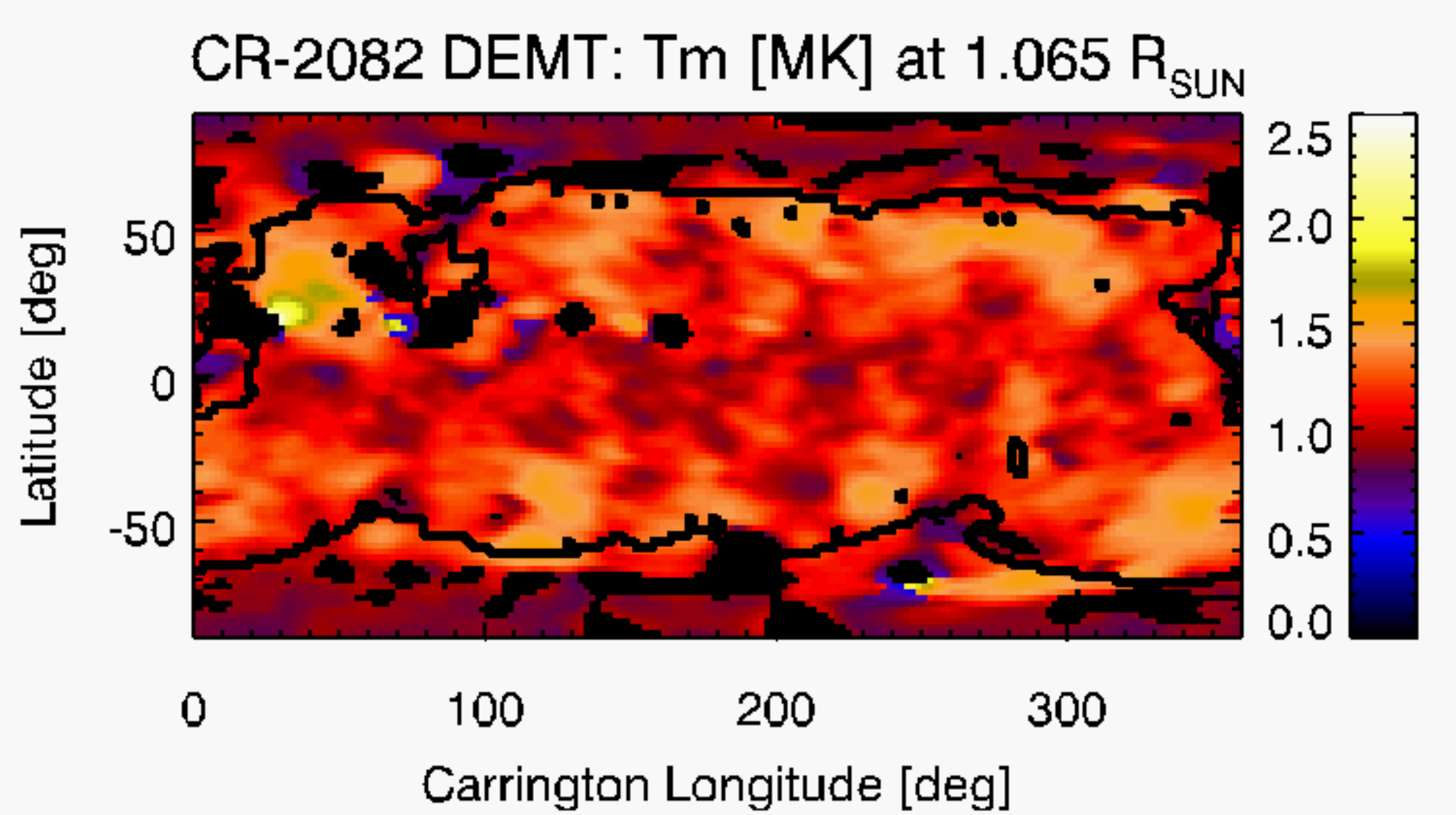}
\includegraphics[width=0.495\textwidth]{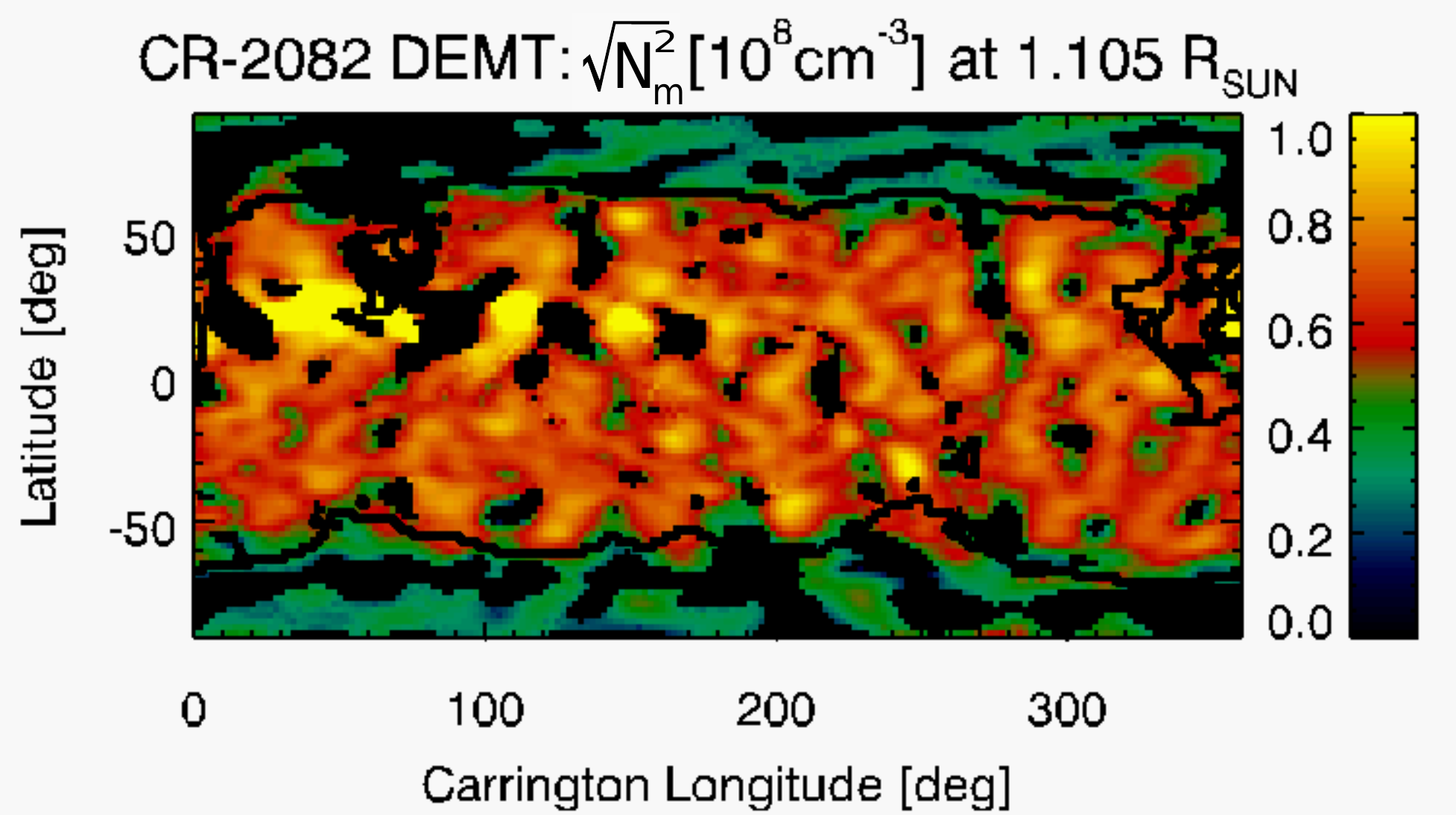}
\includegraphics[width=0.495\textwidth]{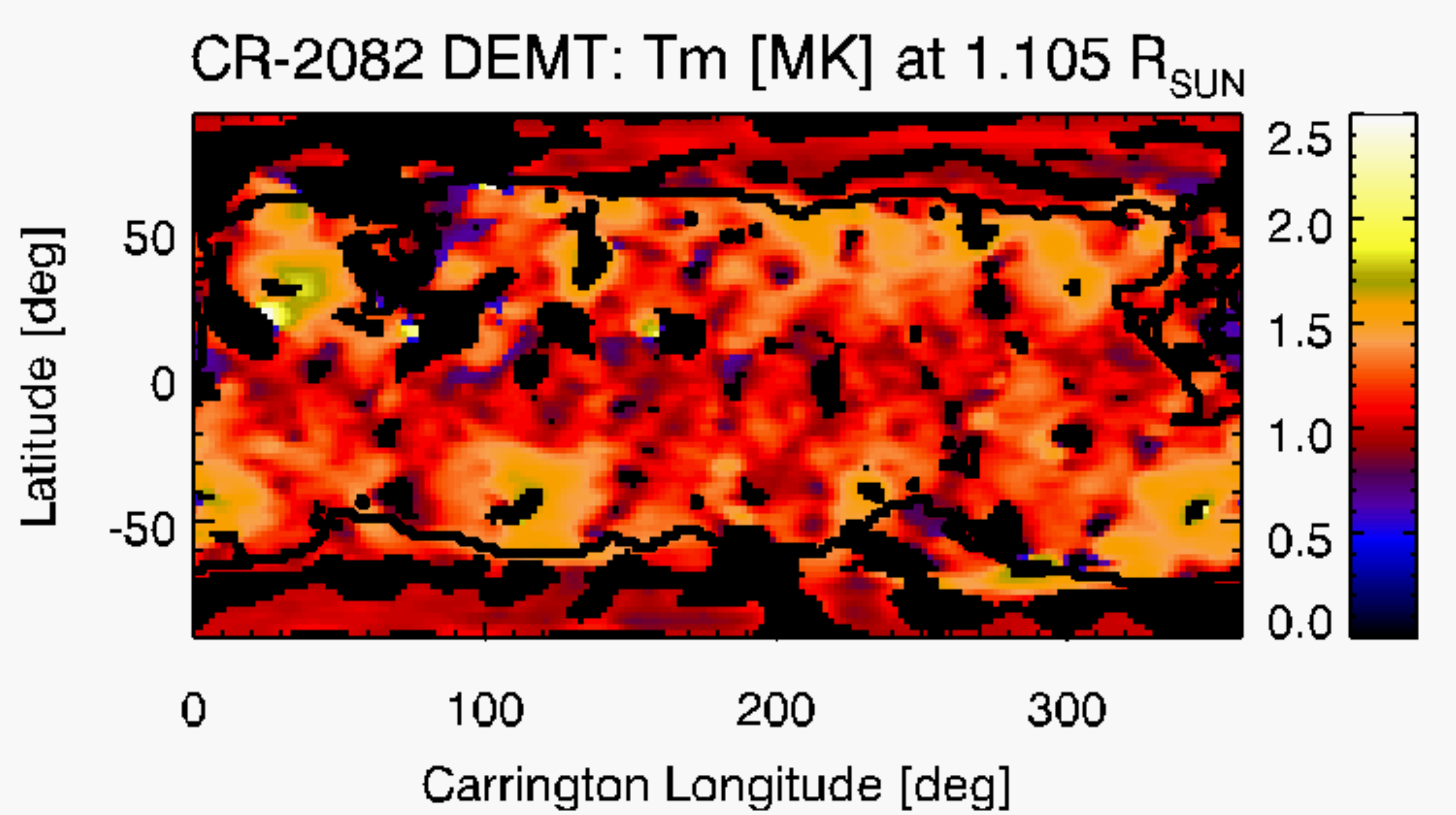}
\caption{Carrington maps of DEMT {products $\sqravgN$} (left panels) and $\Tm$ (right panels) for CR-2082. Top, middle and bottom panels show the results at three heliocentric heights, $1.025$, $1.065$ and $1.105\,\mrsun$ respectively. Black voxels correspond to non-reconstructed regions (see text in Section \ref{demt_res}) and thick-black curves indicate the open/closed boundaries.}
\label{carmaps_demt_2082}
\end{center}
\end{figure}

\begin{figure}[h!]
\begin{center}
\includegraphics[width=0.495\textwidth]{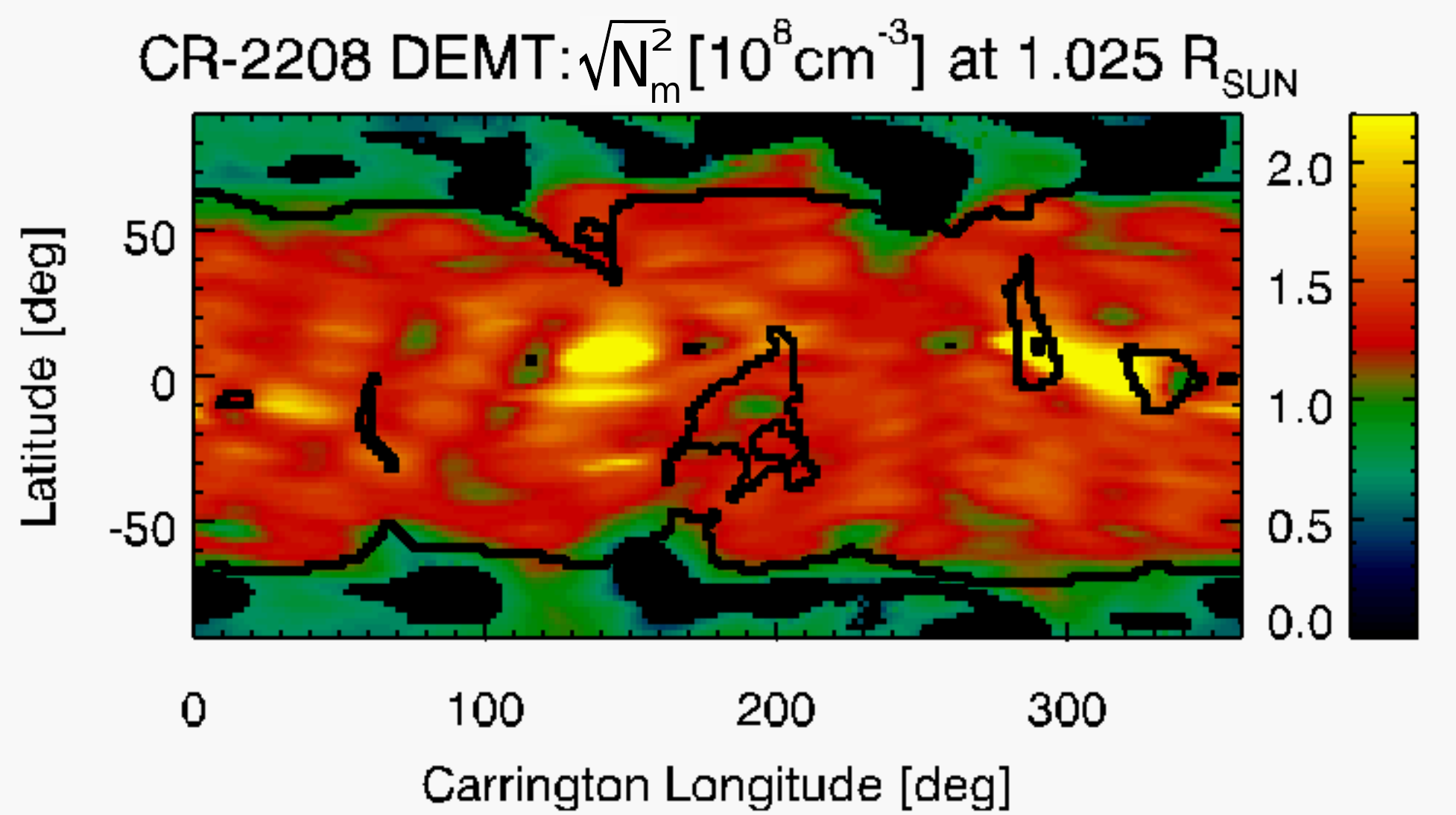}
\includegraphics[width=0.495\textwidth]{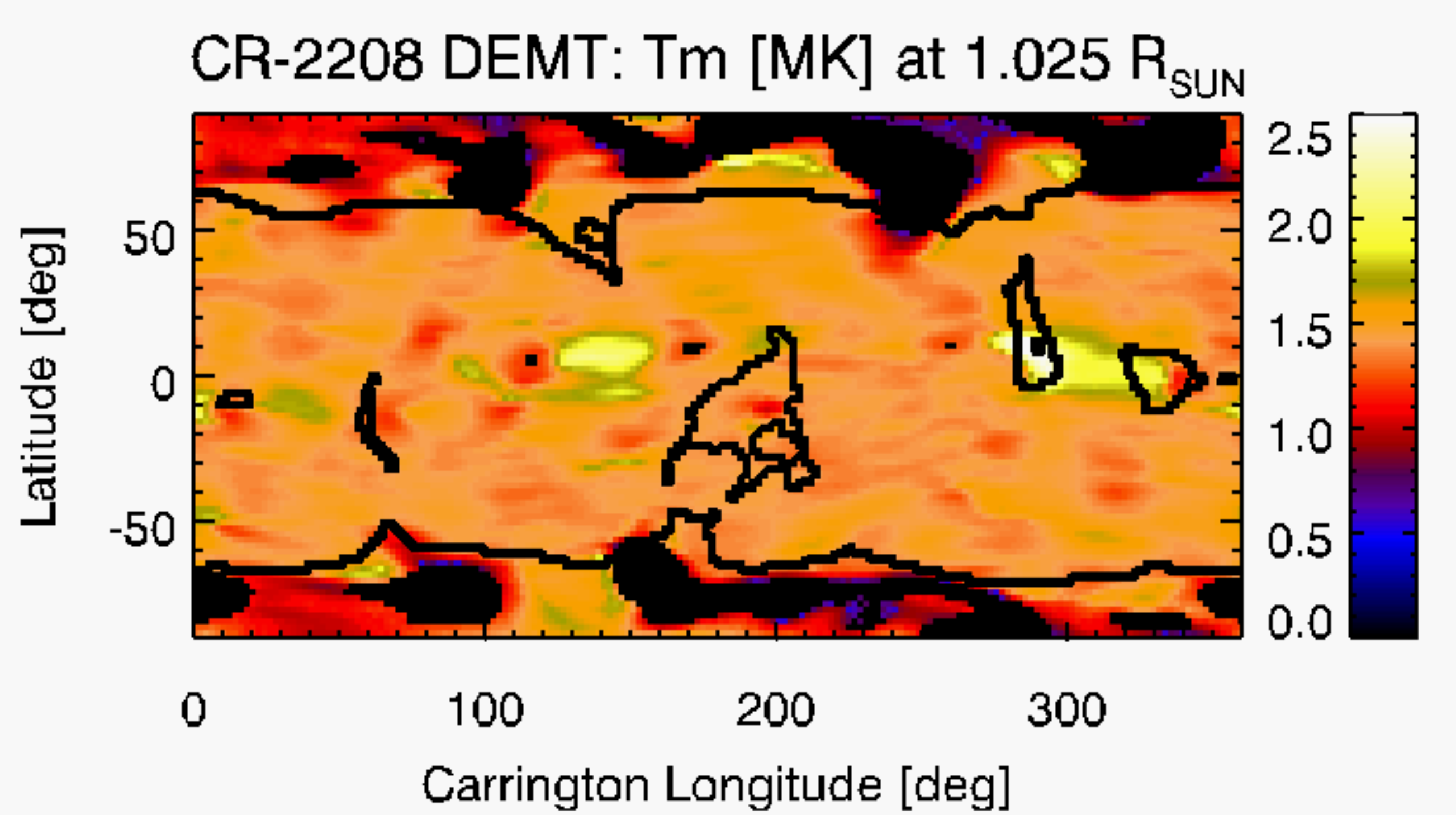}
\includegraphics[width=0.495\textwidth]{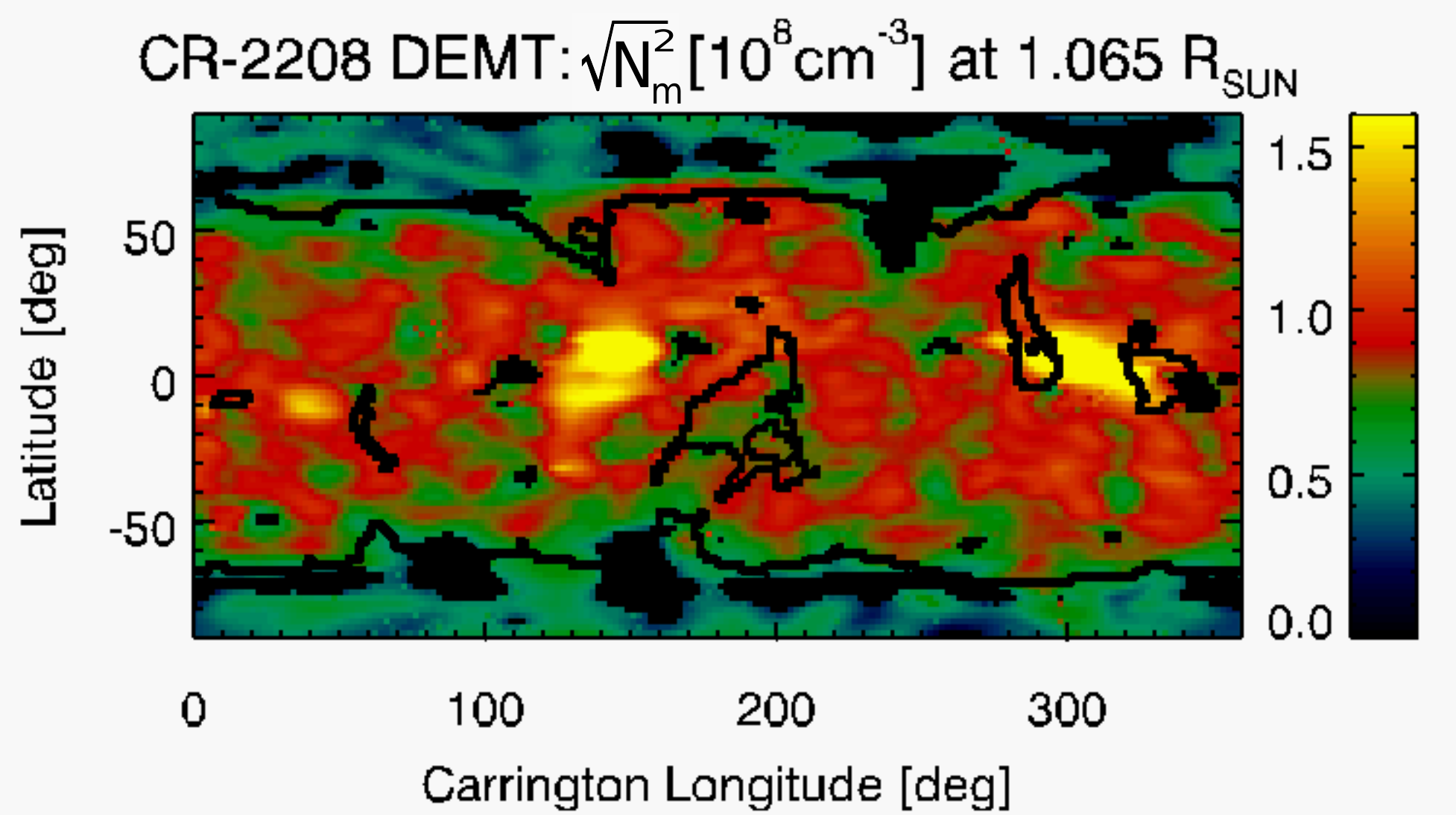}
\includegraphics[width=0.495\textwidth]{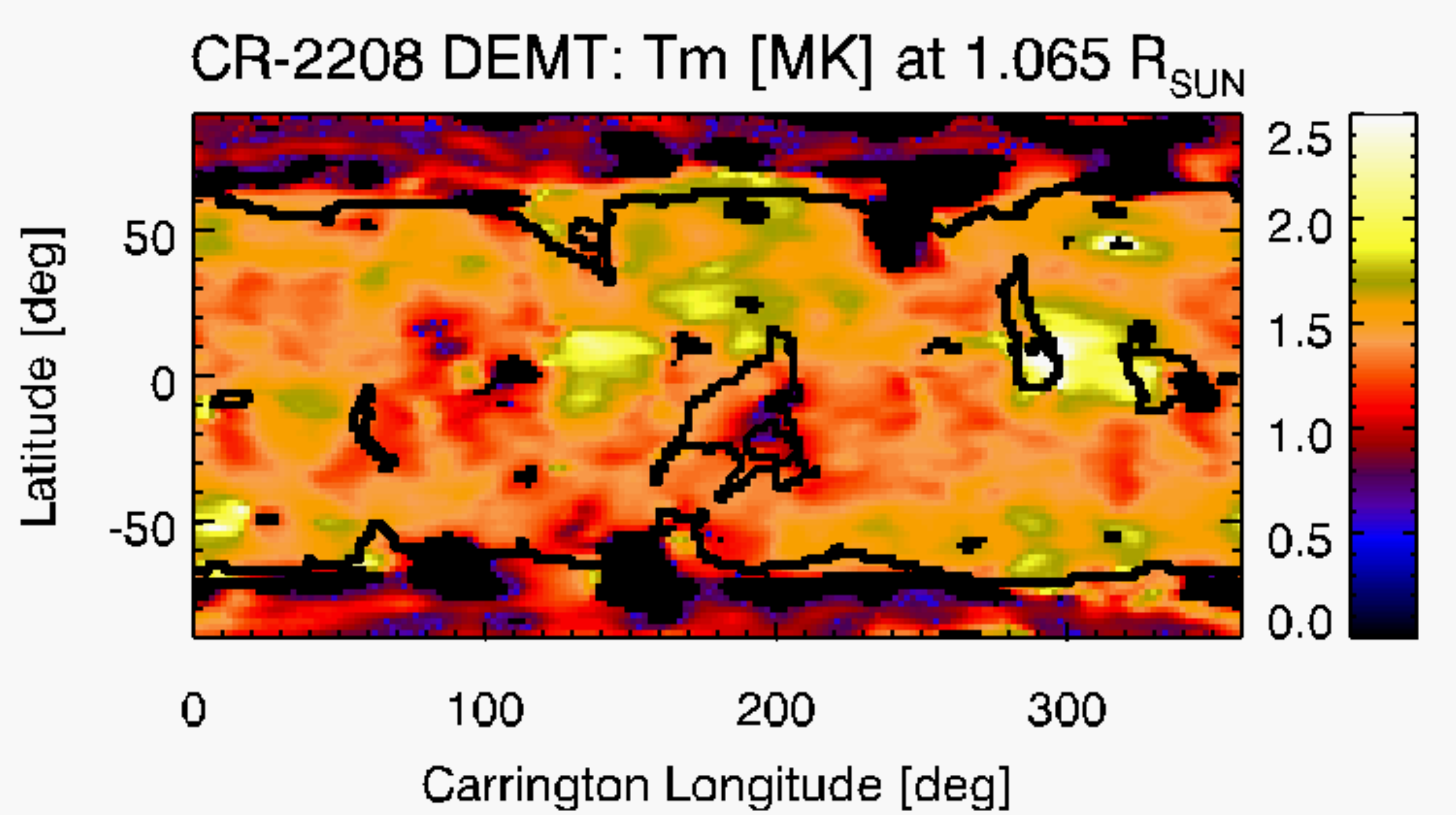}
\includegraphics[width=0.495\textwidth]{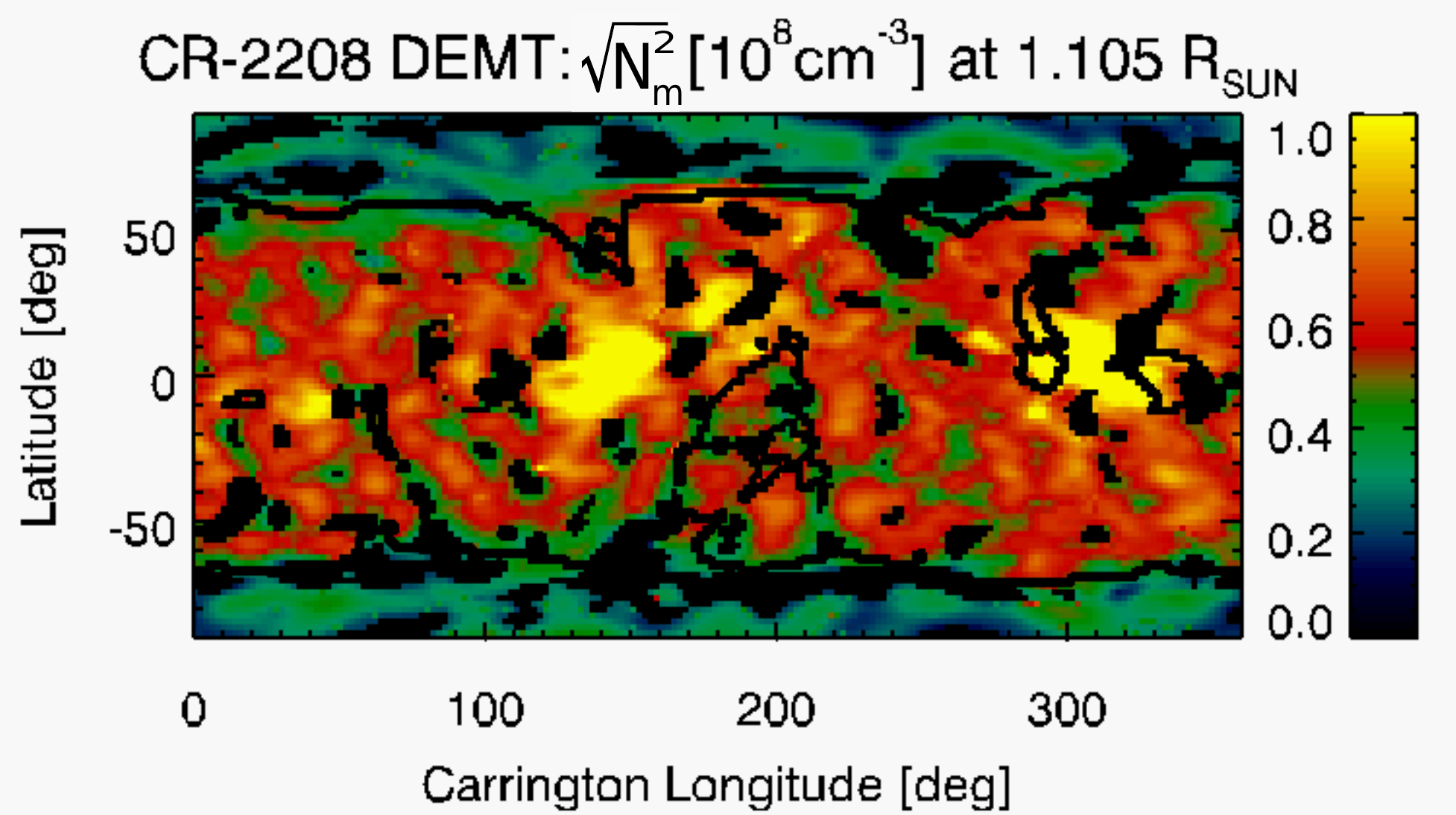}
\includegraphics[width=0.495\textwidth]{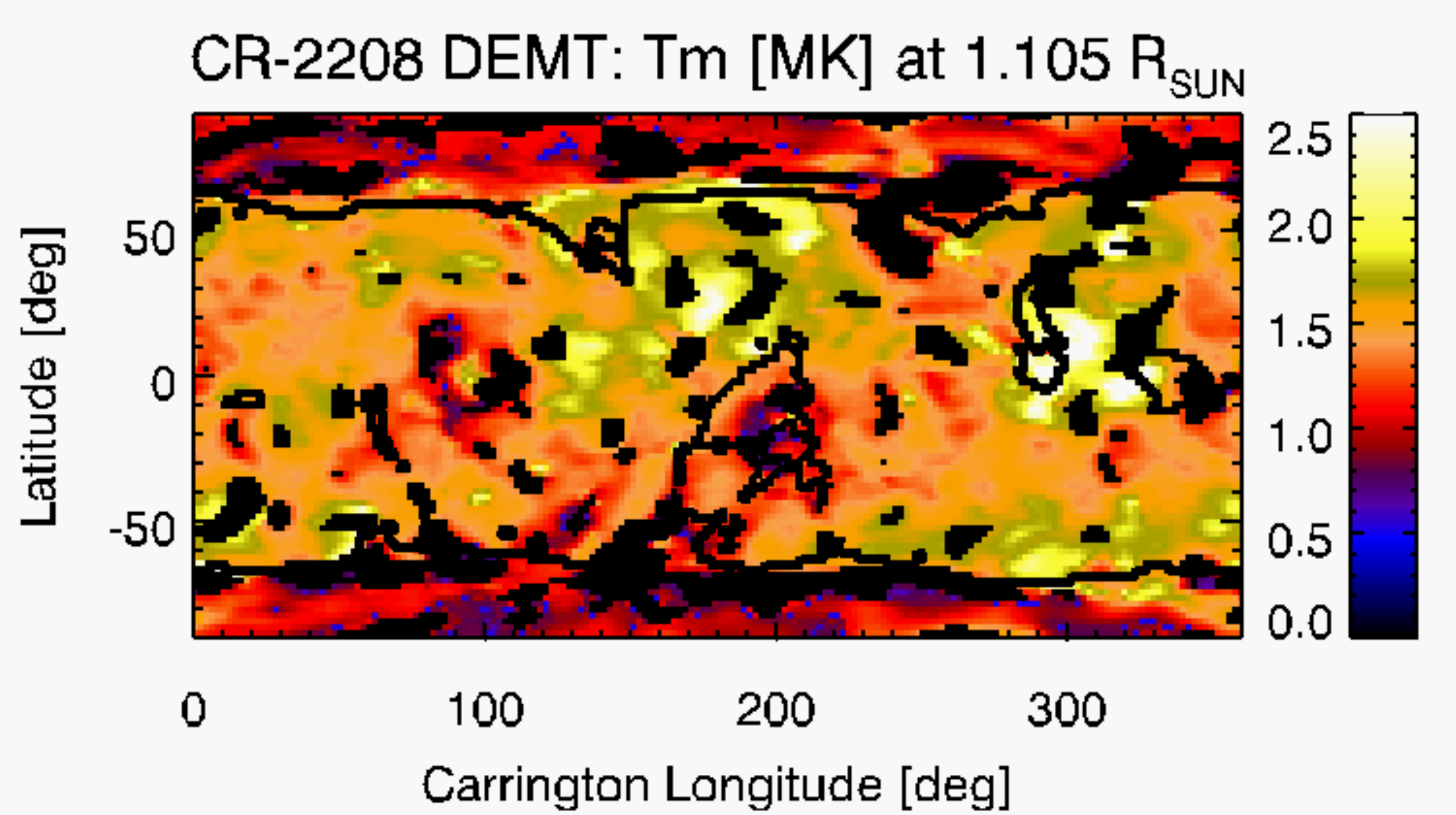}
\caption{Same as Figure \ref{carmaps_demt_2082} but for CR-2208.}
\label{carmaps_demt_2208}
\end{center}
\end{figure}

Figures \ref{carmaps_demt_2082} and \ref{carmaps_demt_2208} show {latitude-longitude} maps of DEMT results for both {rotations. Three different heights {of interest are selected from the tomographic grid}, providing also a detailed 3D view of the tomographic results: the lowest height of {the} tomographic grid ($1.025\,\mrsun$), {the lowest height where the AWSoM results are fully} consistent with coronal conditions ($1.065\,\mrsun$), and a middle height of the tomographic grid ($1.105\mrsun$)}. Black voxels correspond to non-reconstructed voxels ({see} Section \ref{demt}). Thick-black {curves indicate} the {open/closed boundaries of the magnetic field of the} AWSoM model{, detailed in Section \ref{awsom}.} 

{Both target rotations are highly axisymmetric, \textit{i.e.} characterised by a high azymuthal symmetry. Rotation CR-2082} {has} two {small} ARs, both near latitude $+30\mdeg$ and {around longitudes} $50\mdeg$ and $120\mdeg$  {(not identified in the NOAA catalog)}. {Rotation CR-2208} {has} two {ARs, both near latitude} $+5\mdeg$ and {around longitudes $140\mdeg$ and $300\mdeg$ {(NOAA 12722, 12721)}.} 

{The magnetically open and closed regions of the AWSoM model are associated {with CHs} and the equatorial streamer belt, respectively. The {location of the} open/closed boundaries derived from the respective AWSoM model quite accurately matches {the regions of the DEMT maps, which exhibit the strongest latitudinal gradient of both the electron density and temperature.}}

{{Figures \ref{carmaps_demt_2082} and \ref{carmaps_demt_2208} show that} the DEMT reconstruction of the streamer belt is characterised by relatively higher densities and temperatures in comparison to the CHs. They also show that the streamer belt region of CR-2082 was denser and colder than  that of CR-2208.} In the case of CR-2082, which belongs to the deep minimum epoch between {SCs 23 and 24}, the {low latitudes of the streamer belt are characterised by lower electron temperature than in its mid-latitudes.} A similar behavior is seen in CR-2208, belonging to the {end of the declining} phase of {SC 24}, but having a {somewhat less axisymmetric} structure this characteristic is not so obvious. This thermodynamic structure of the streamer has been reported for other solar minimum rotations in previous DEMT works \citep{lloveras_2017,nuevo_2013,vasquez_2010}.

Latitude-longitude maps of the score $R$ defined by Equation \ref{R} show that {the agreement between the tomographic and synthetic FBEs is 5\% or better in more than 90\% of the reconstructed coronal volume (i.e. where FBEs are non zero), and of order 10\% in the rest of the volume. This implies that the LDEM found in each voxel accurately predicts the reconstructed FBEs.}

\begin{figure}[h!]
\begin{center}
\includegraphics[width=0.495\textwidth,clip=]{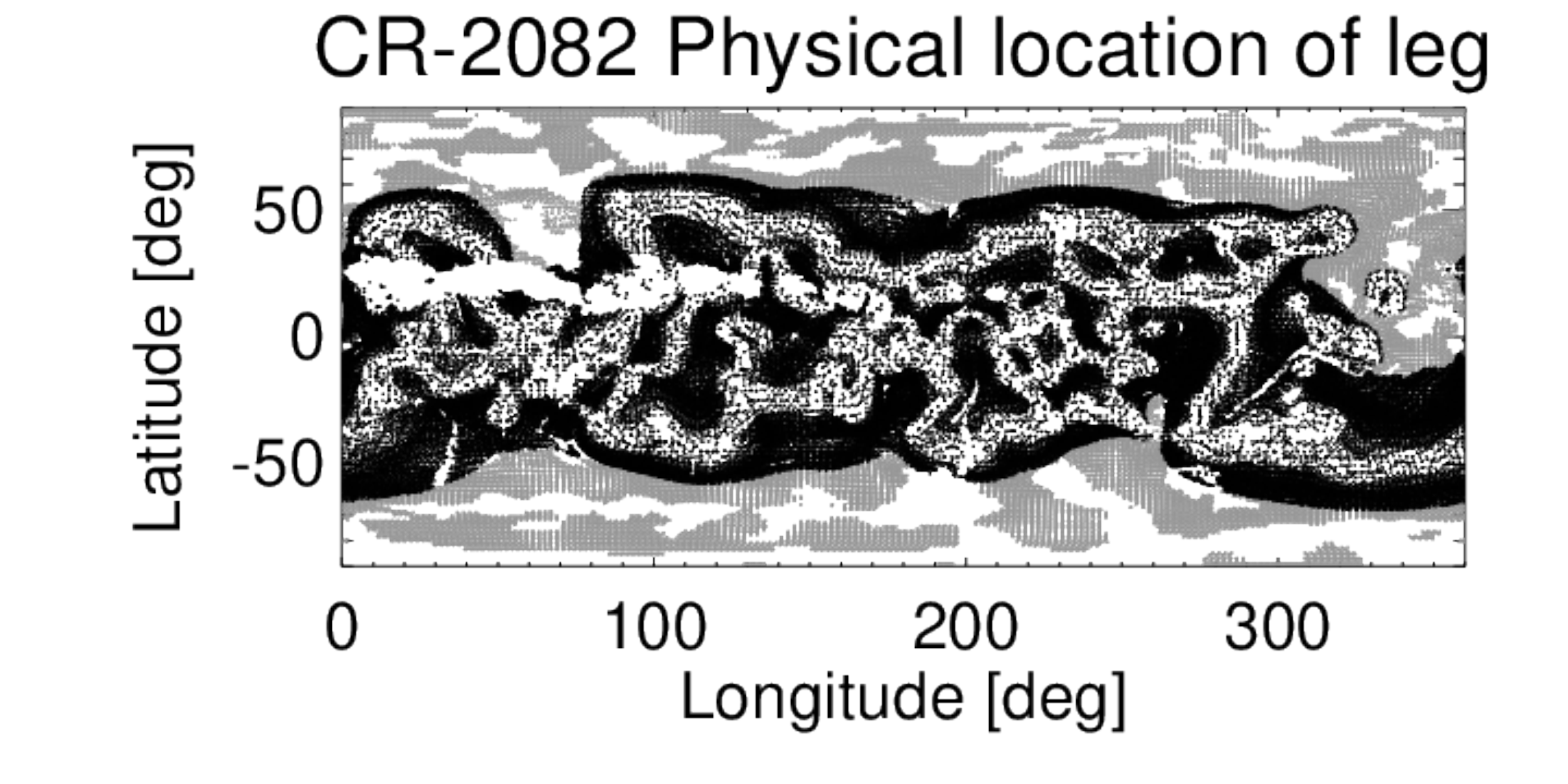}
\includegraphics[width=0.495\textwidth,clip=]{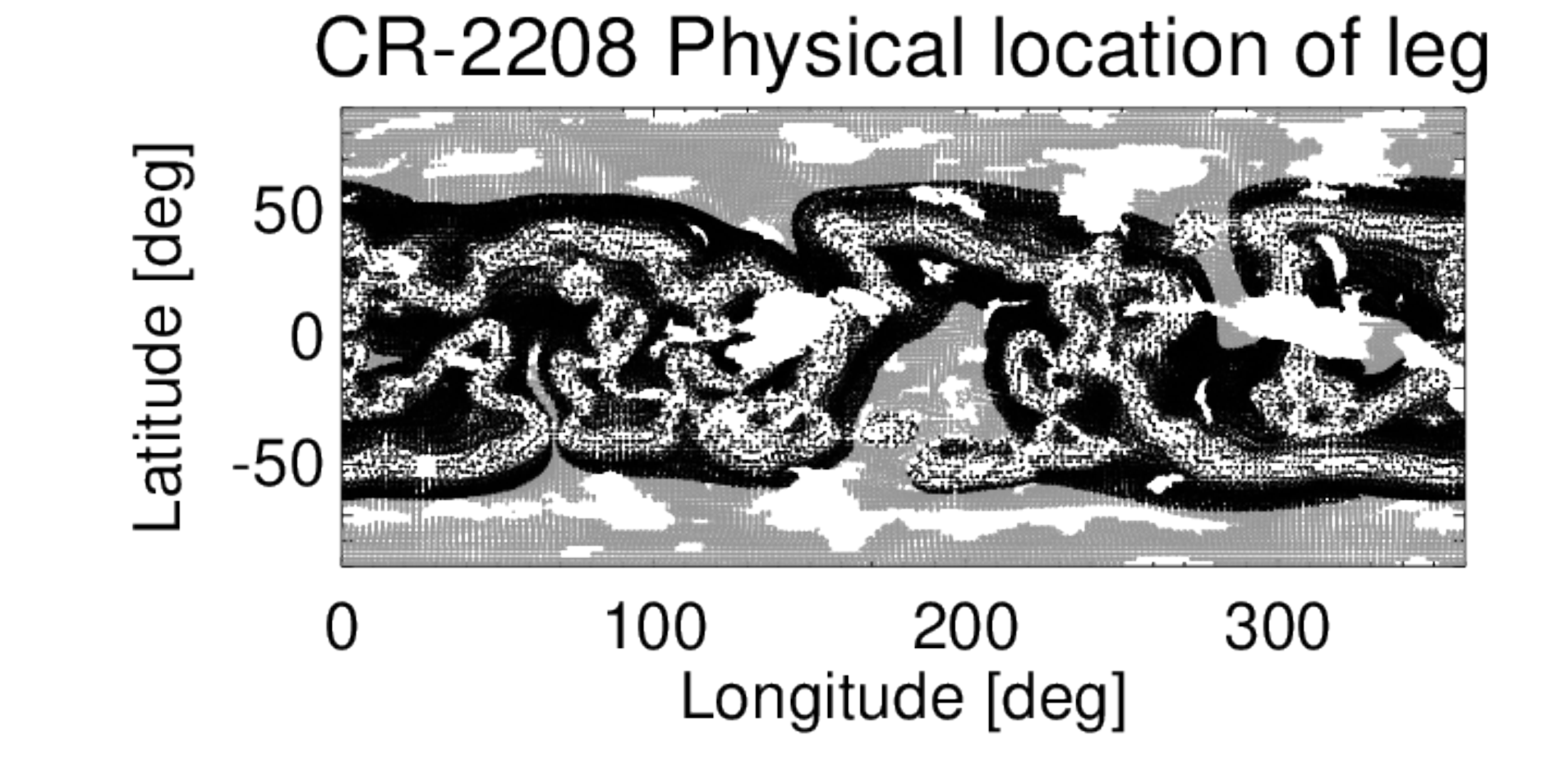}
\includegraphics[width=0.495\textwidth,clip=]{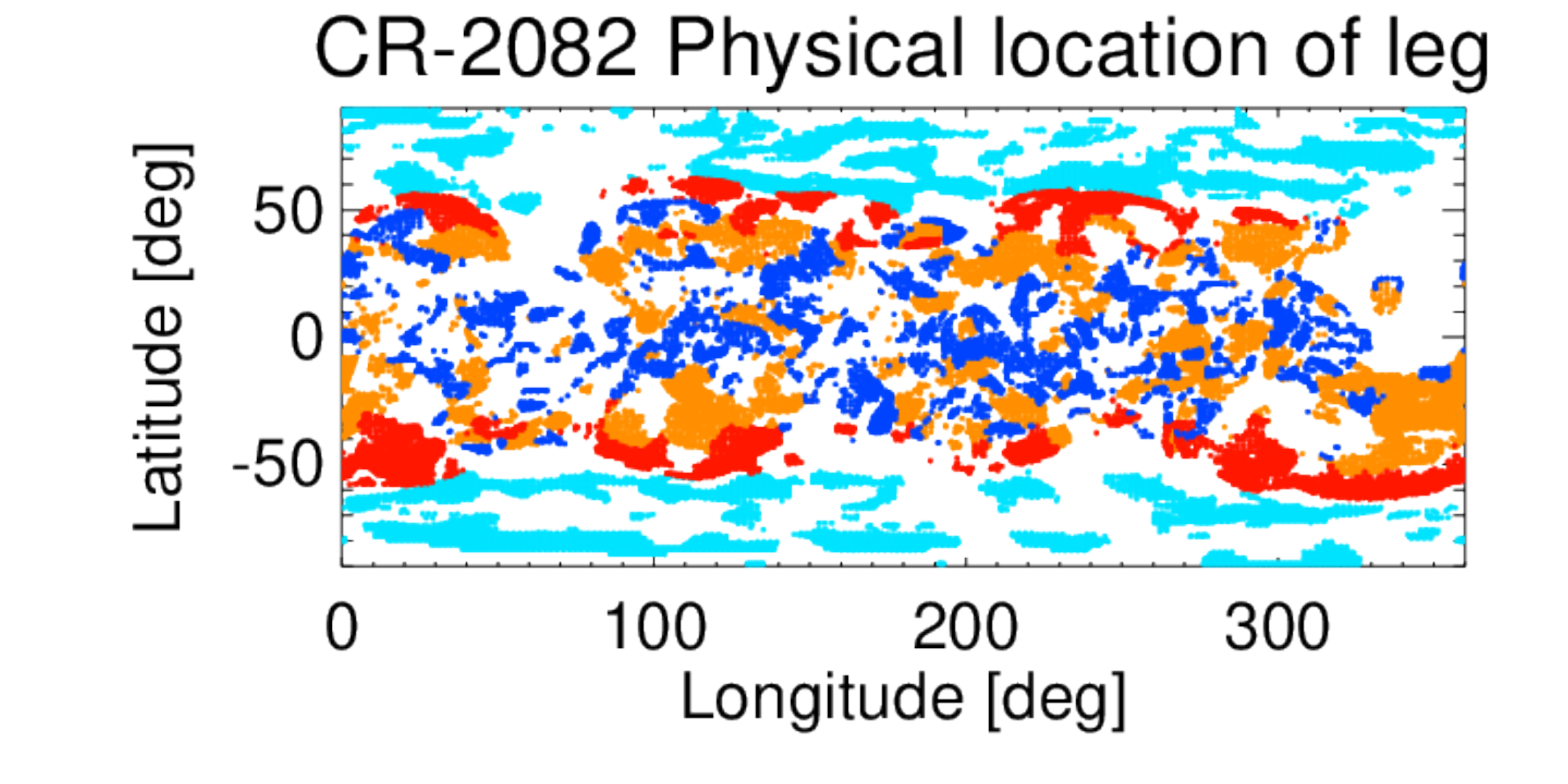}
\includegraphics[width=0.495\textwidth,clip=]{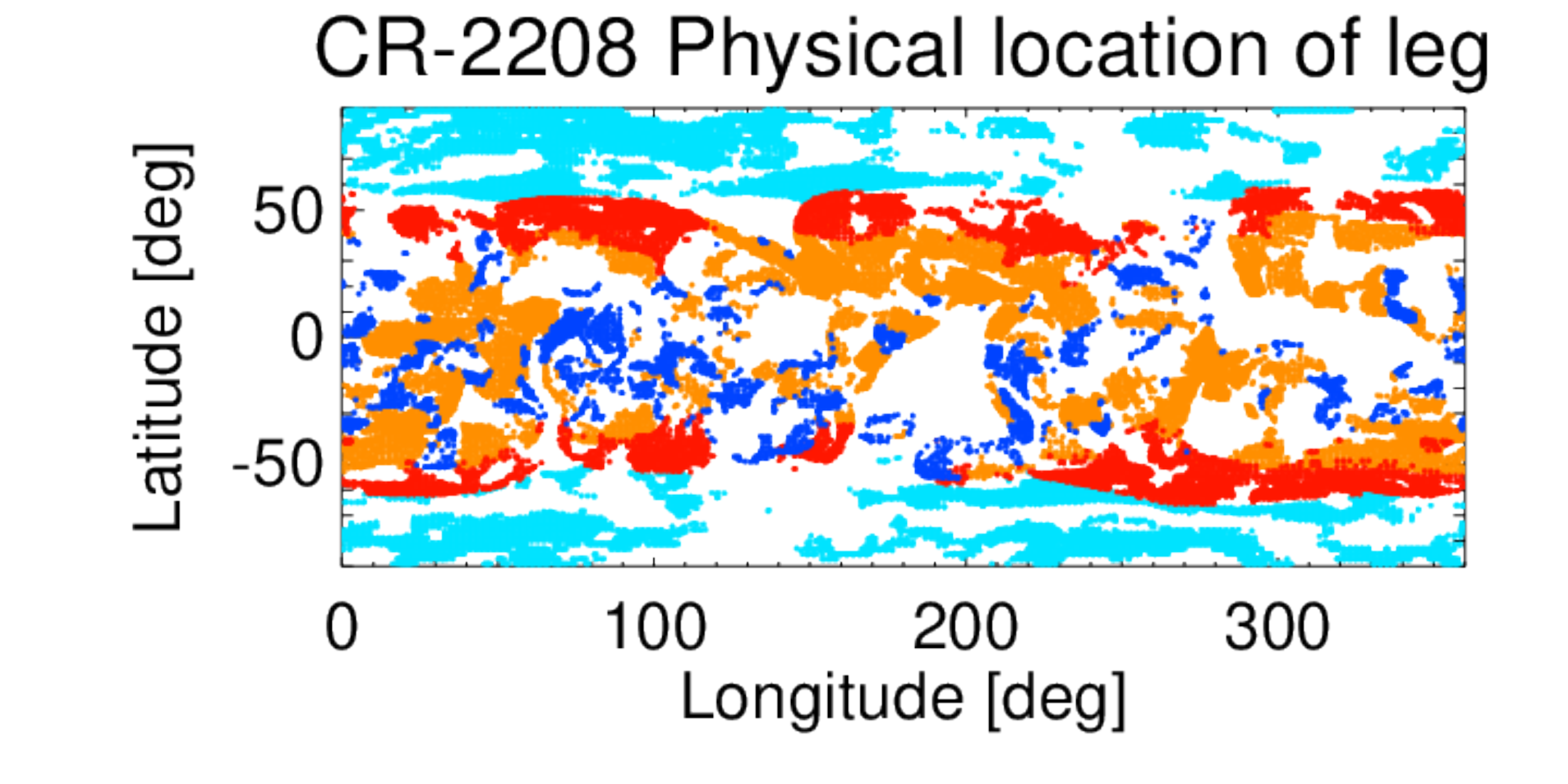}
\includegraphics[width=0.495\textwidth,clip=]{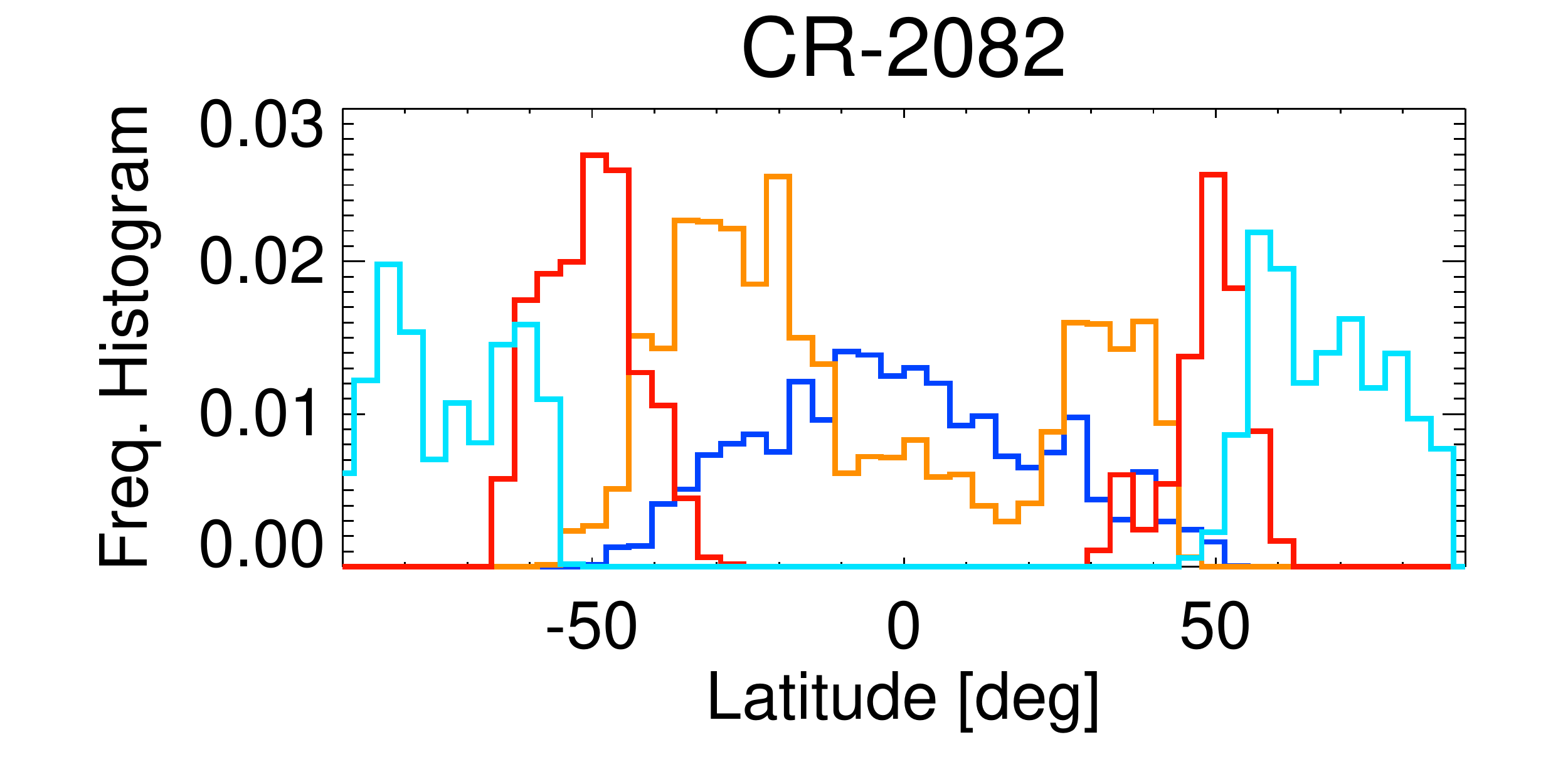}
\includegraphics[width=0.495\textwidth,clip=]{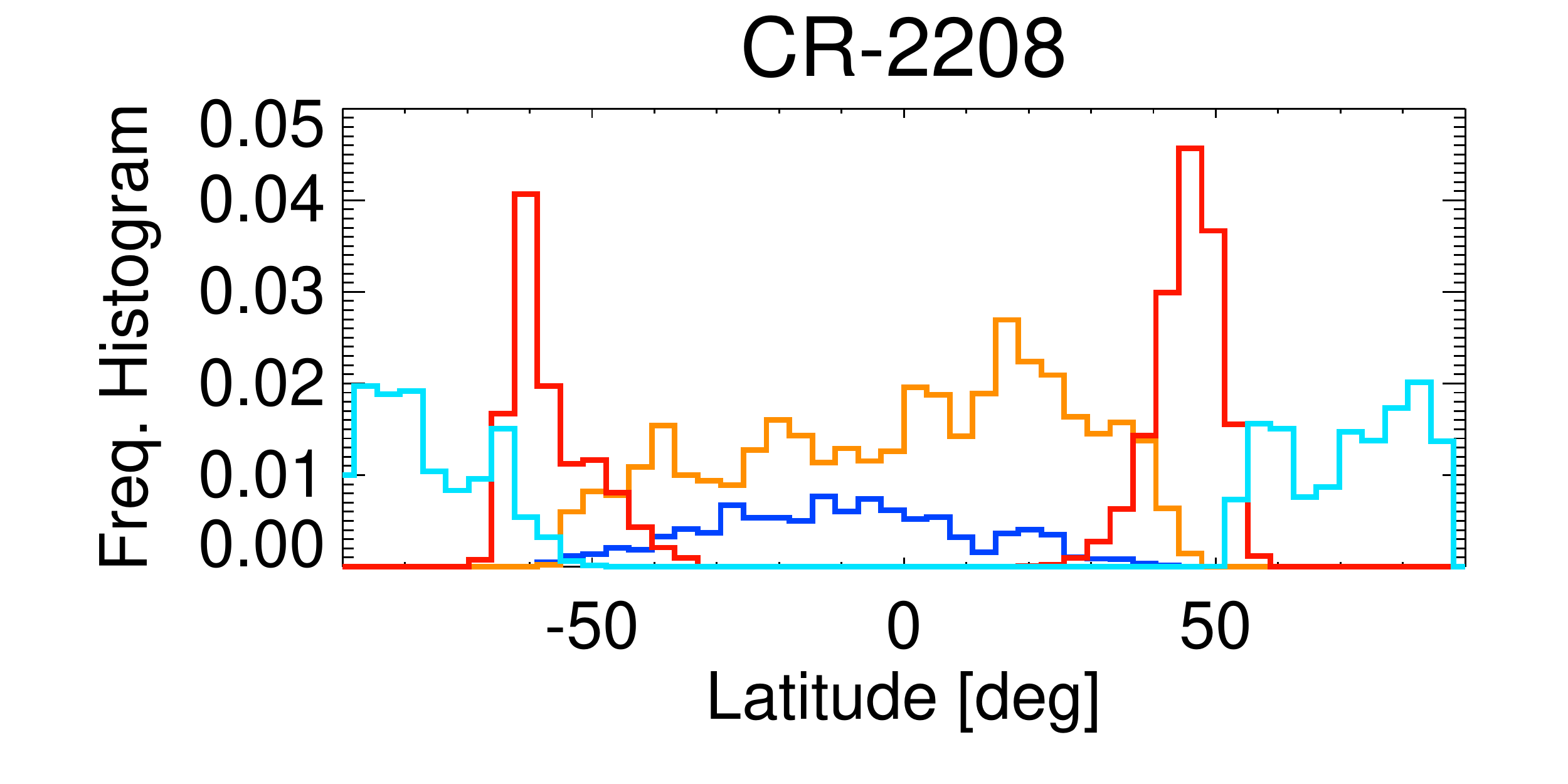}
\caption{{Top panels:} {latitude-longitude location at heliocentric height $r=1.105\,\mrsun$ of all open (grey color) and closed (black color) traced field line legs for which criterion (i) of Section \ref{trace} is met, for both CR-2082 (left) and CR-2208 (right). {Middle panels}: latitude-longitude location of the subset for which also both criteria (ii) and (iii) of Section \ref{trace} are met. {The location of} type 0, I, II and III legs {is shown} in blue, {orange, red, and cyan} color, respectively.} {Bottom panels: frequency histograms of the latitude of the four types of legs of the middle panel.}}
\label{rpoint_demt}
\end{center}
\end{figure}

For both rotations, the top panels of Figure \ref{rpoint_demt} show the latitude-longitude location (at heliocentric height $r=1.105\,\mrsun$) of all traced field line legs for which criterion (i) of Section \ref{trace} is met. Open legs are indicated in gray color and closed ones in black color. Considering the DEMT data points and the resulting fits along each leg, the {middle} panels of Figure \ref{rpoint_demt} show the latitude-longitude location of the subset for which also both criteria (ii) and (iii) of Section \ref{trace} are met. Using a four-color code, type 0, I, II and III legs are shown in blue, {orange, red, and cyan} color, respectively. {The bottom panels show histograms of the latitude distribution of the legs of the middle panel, using the same color code.} Of the {$\approx 44000$} legs selected for CR-2082, 20\% are type 0, 31\% are type I, 23\% are type II and 26\% type III. On the other hand, of the {$\approx 50000$} legs selected for CR-2208, 10\% are type 0, {38}\% are type I, {27}\% are type II and {25}\% type III.

{For both rotations, the population of type 0 (small-closed-down) legs peaks at the core of the streamer belt, around the equator in the case of CR-2082, and towards the southern hemisphere in the case of CR-2208.} This kind of structure was originally found by \citet{huang_2012}, and their existence was shown to be anti-correlated with the solar activity level around the solar minimum between {SCs 23 and 24} by \citet{nuevo_2013}. Later on, \citet{lloveras_2017} showed {that} equatorial down loops in streamers were also to be found in the deep minimum between SCs {23 and 24}. Here, we verify the existence of this type of structure for the two rotations. The relatively smaller population of down legs seen in CR-2208, as compared to CR-2082, is consistent with the aforementioned results by \citet{nuevo_2013}. {Type I (small-closed-up) legs are present at all latitudes within the streamer belt. Their population peaks in the mid-latitudes of both hemispheres for CR-2082, and in the mid-latitudes of the northern hemisphere for CR-2208. The latitude distribution of legs of type 0 and I in CR-2082 is consistent to the distribution of down and up loops of CR-2081 in the analysis by \citet{nuevo_2013}, which did not place any limits on the latitude location of the analysed structures.}

{Type II (large up) legs are mostly very large trans-equatorial field lines forming the envelope of the streamer belt (the requirement of footpoints located beyond mid-latitudes was included precisely to select this kind of loop). Finally, type III (open) legs populate the high latitude CHs.}

\begin{figure}[h!]
\begin{center}
\includegraphics[width=0.495\textwidth,height=0.25\textwidth,clip=]{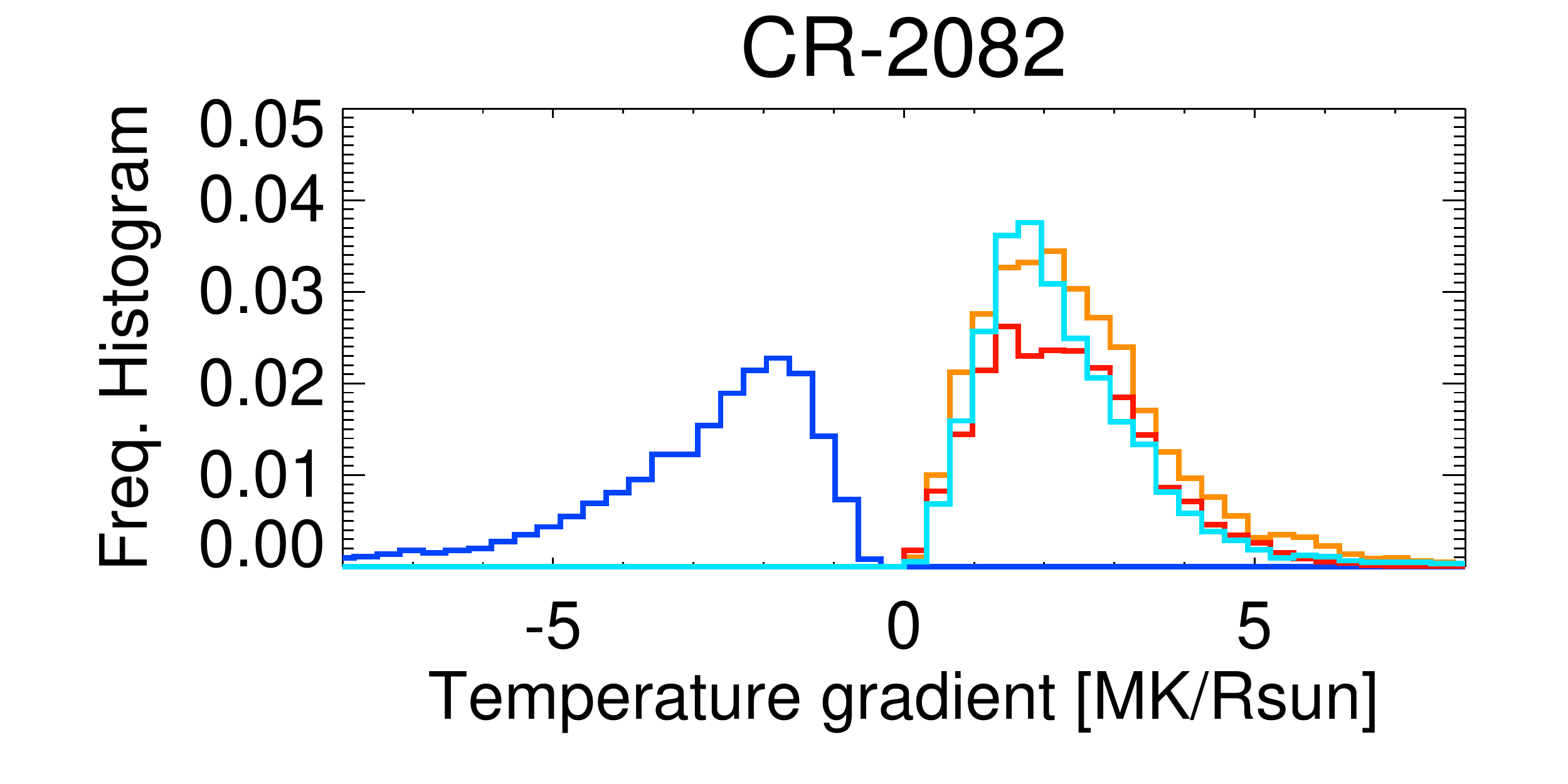}
\includegraphics[width=0.495\textwidth,height=0.25\textwidth,clip=]{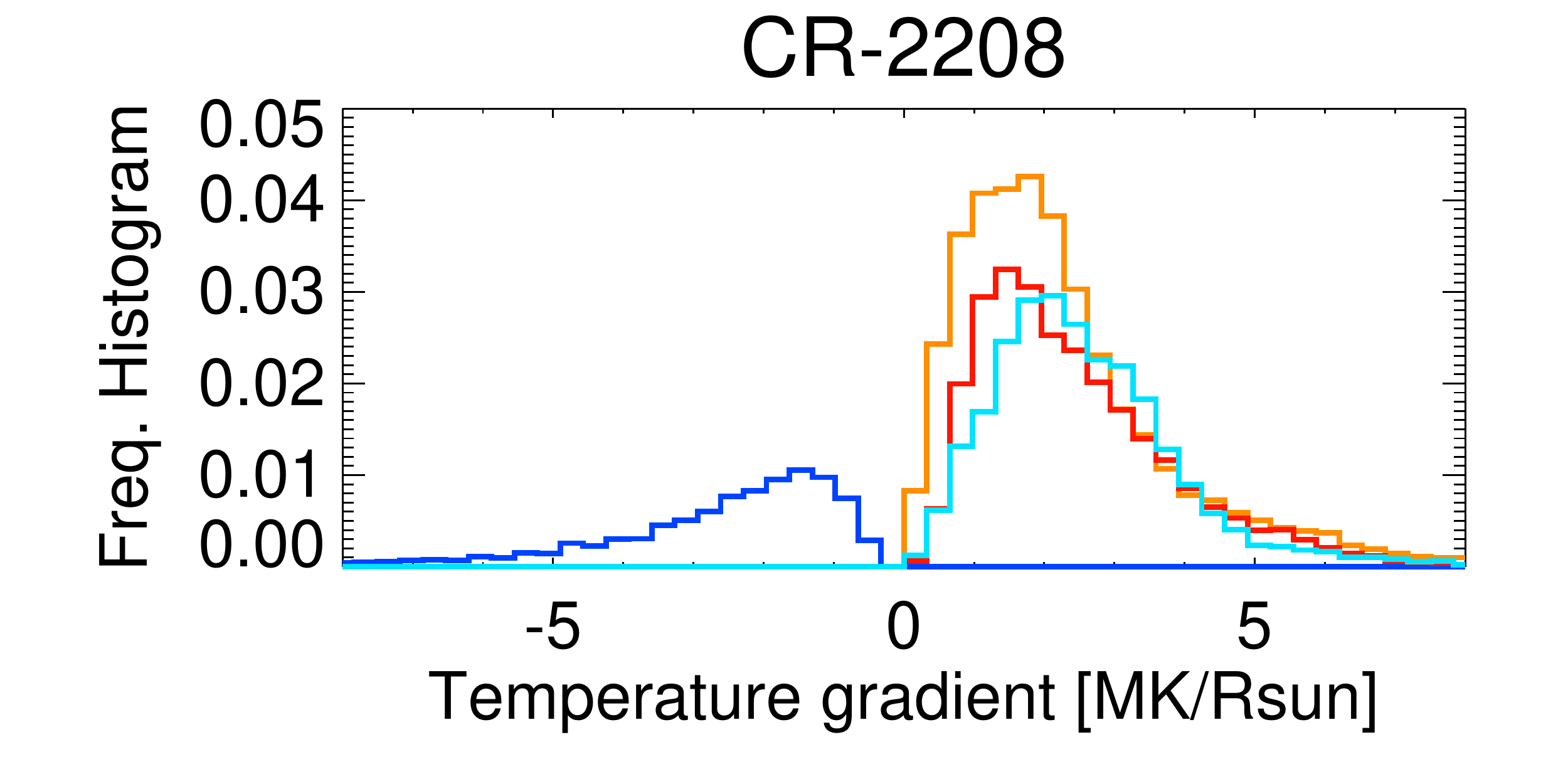}
\caption{Frequency histograms of the temperature radial gradient for {the four types of legs} in Figure \ref{rpoint_demt} (using the same color code) for CR-2082 (left panel) and CR-2208 (right panel).}
\label{gradt_demt}
\end{center}
\end{figure} 

Figure \ref{gradt_demt} shows frequency histograms of $\dr\Tm$ for legs of type 0, I, II and III. The lack of population around values close to zero is due to the requirement $|\rhoTr| > 0.5$ which discards quasi-isothermal legs. For both rotations, the median value of the temperature radial gradient is {${\rm Md}\left(\dr\Tm\right) \approx -2.2$, $+2.3$, $+2.4$ and $+2.3\,\MK/\mrsun$, for legs of type 0, I, II and III, respectively.}

\begin{figure}[h!]
\begin{center}
\includegraphics[width=0.31\textwidth,clip=]{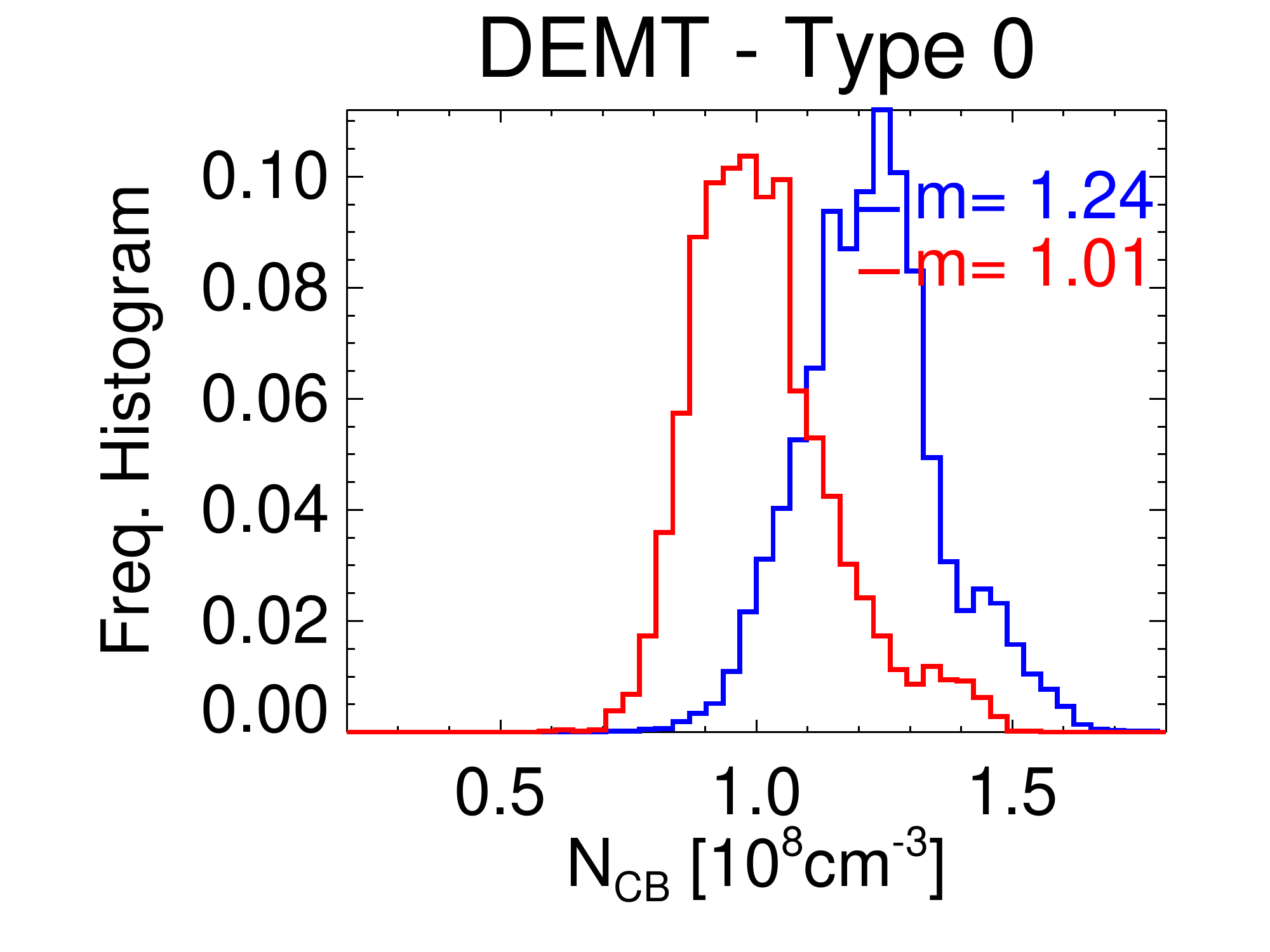}
\includegraphics[width=0.31\textwidth,clip=]{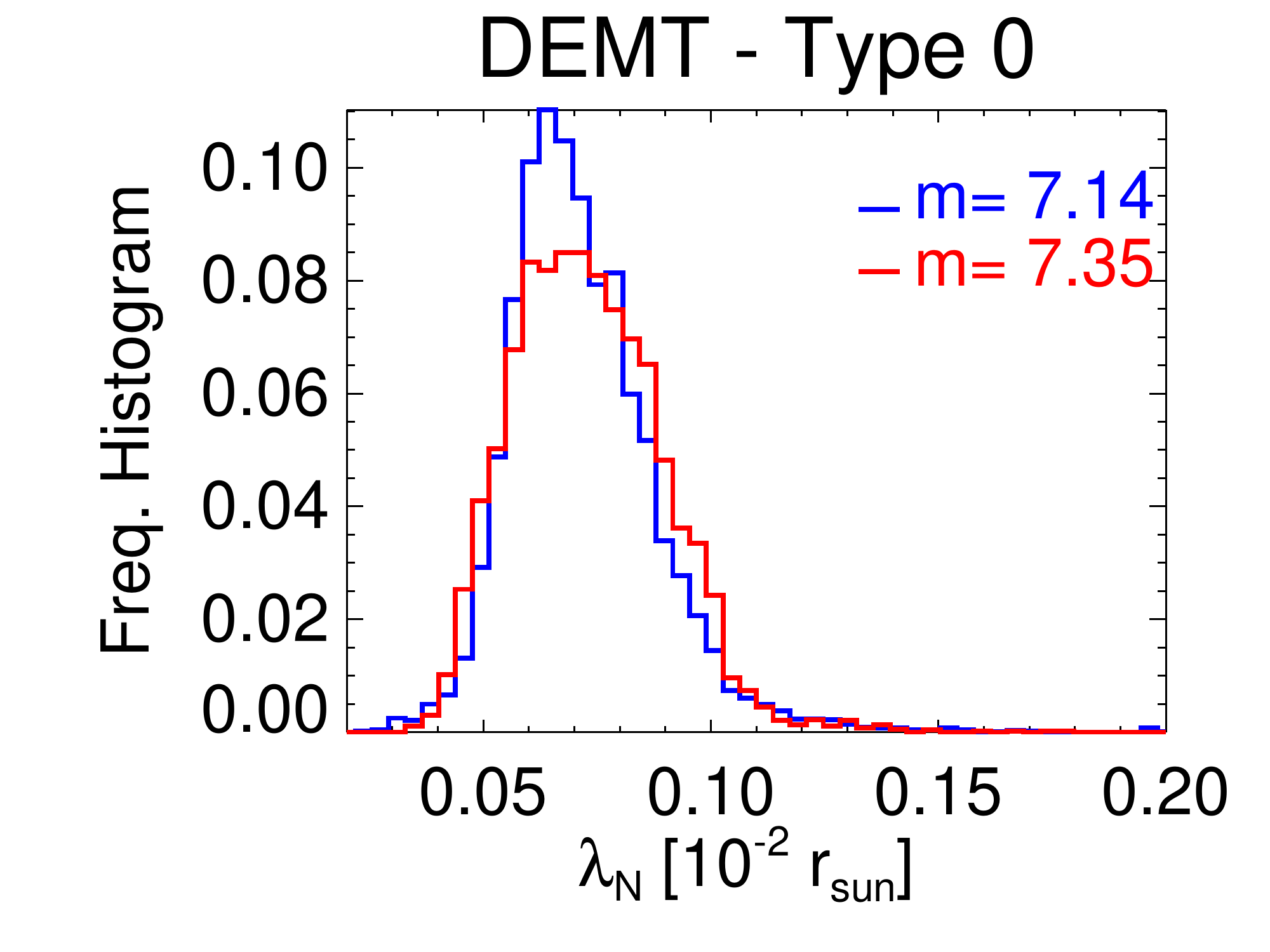}
\includegraphics[width=0.31\textwidth,clip=]{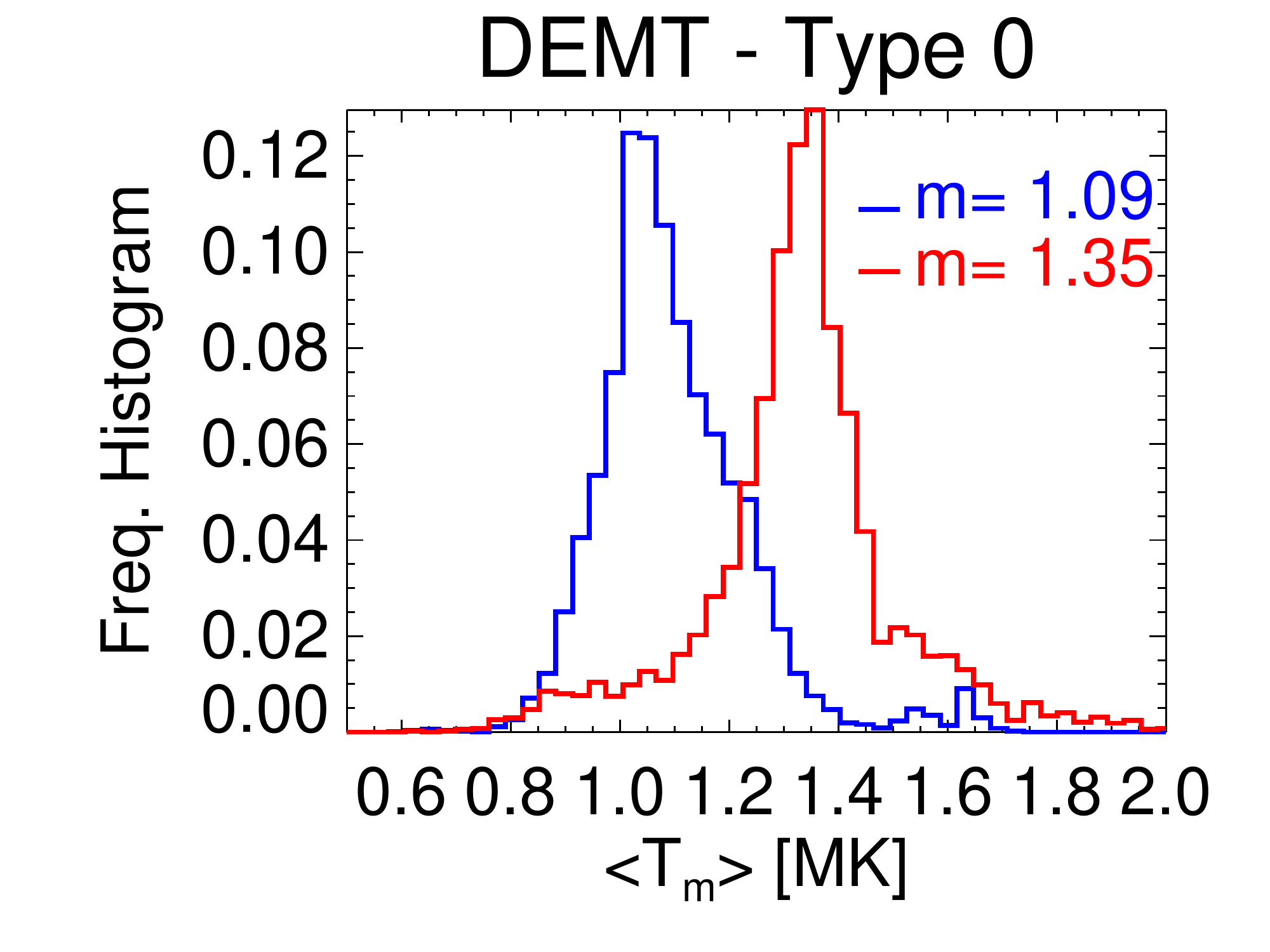}
\includegraphics[width=0.31\textwidth,clip=]{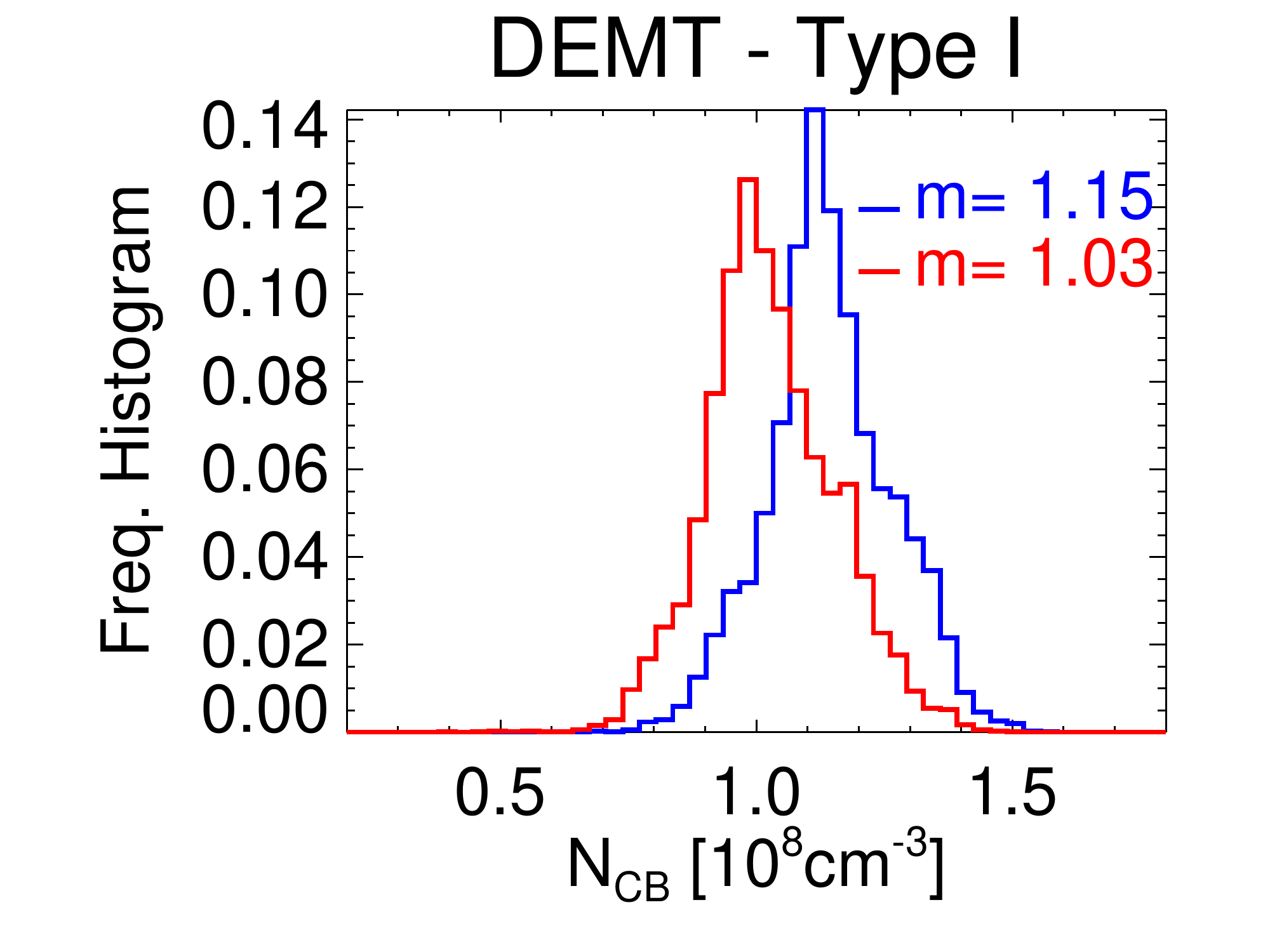}
\includegraphics[width=0.31\textwidth,clip=]{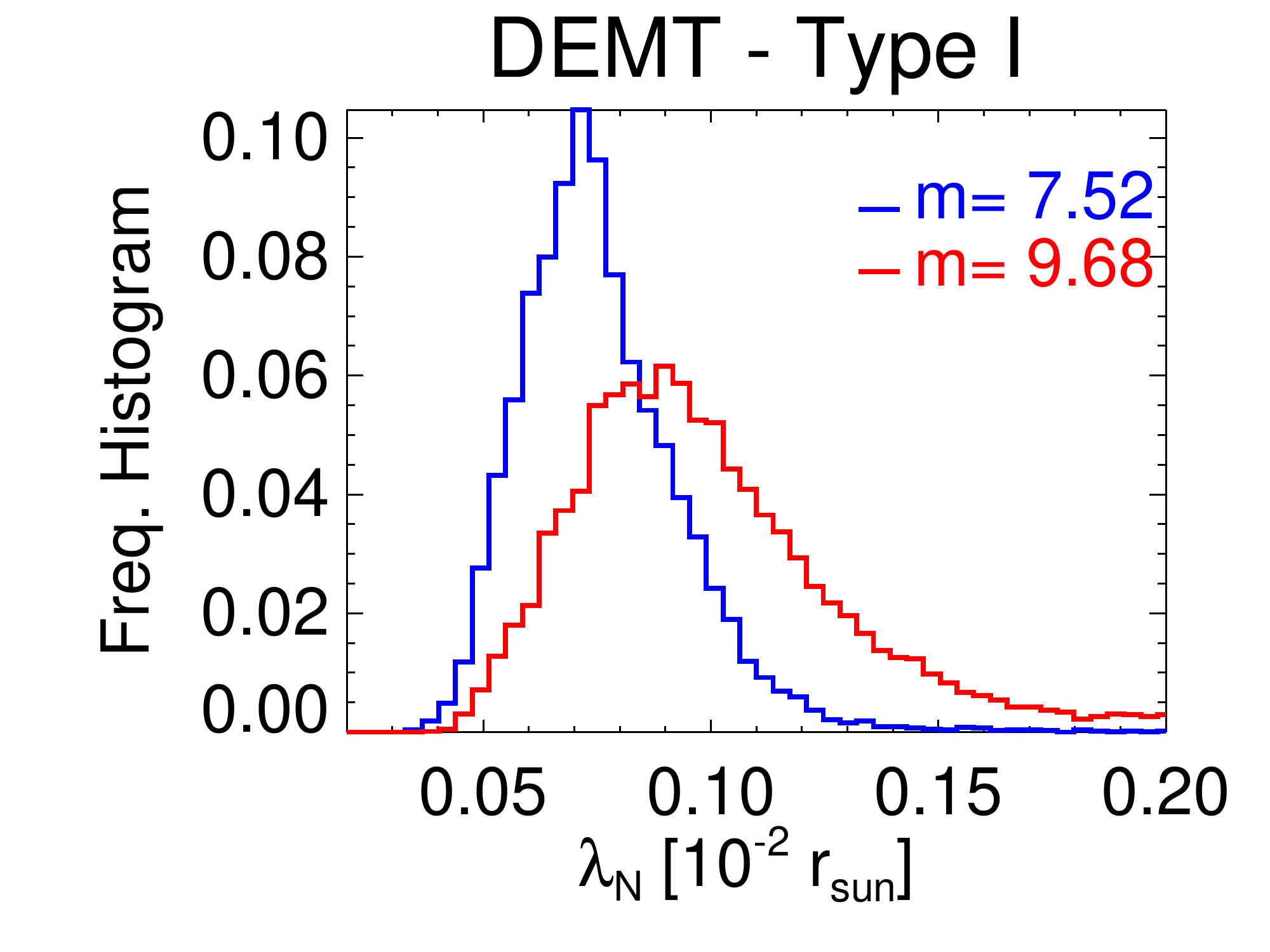}
\includegraphics[width=0.31\textwidth,clip=]{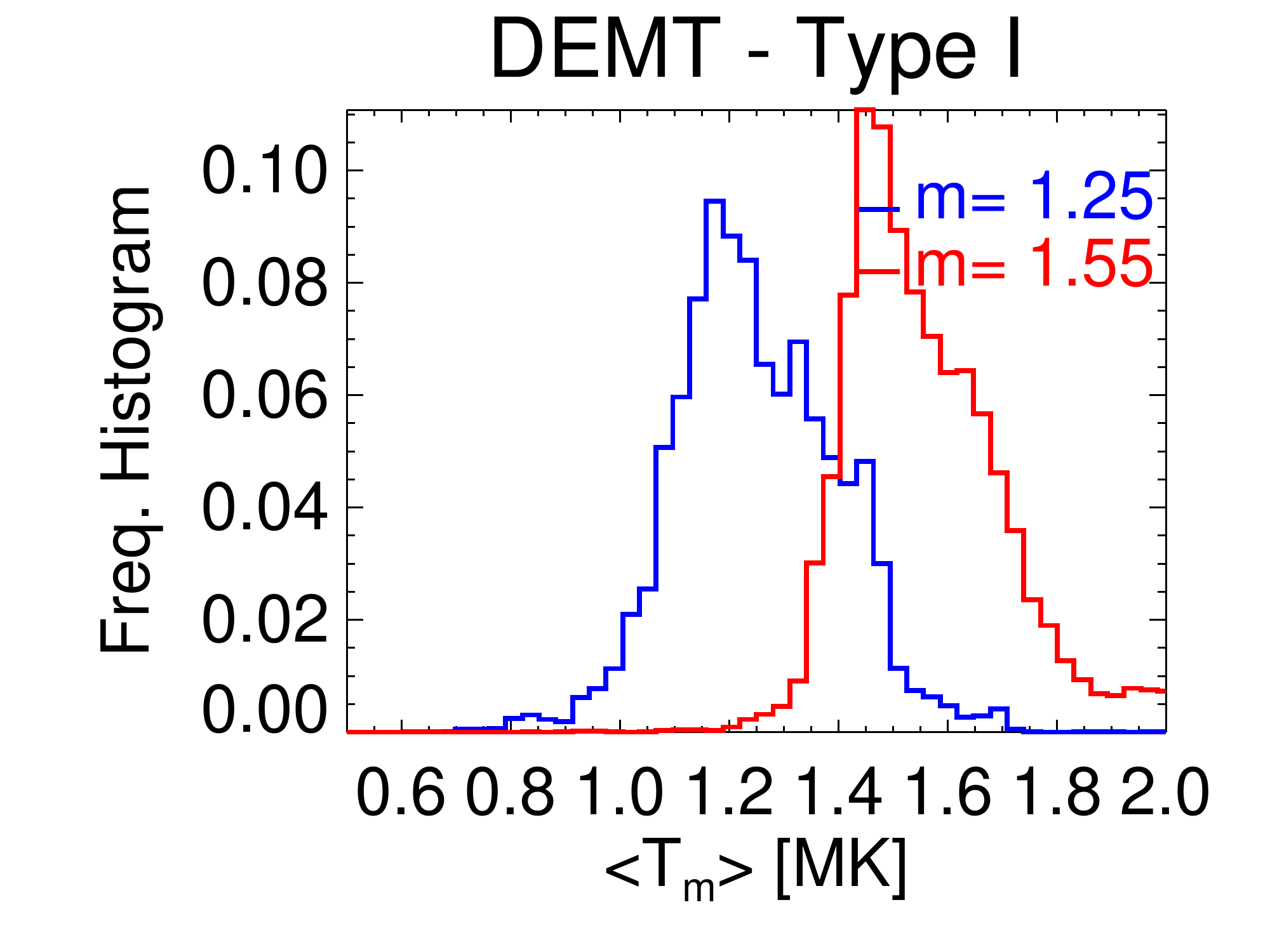}
\includegraphics[width=0.31\textwidth,clip=]{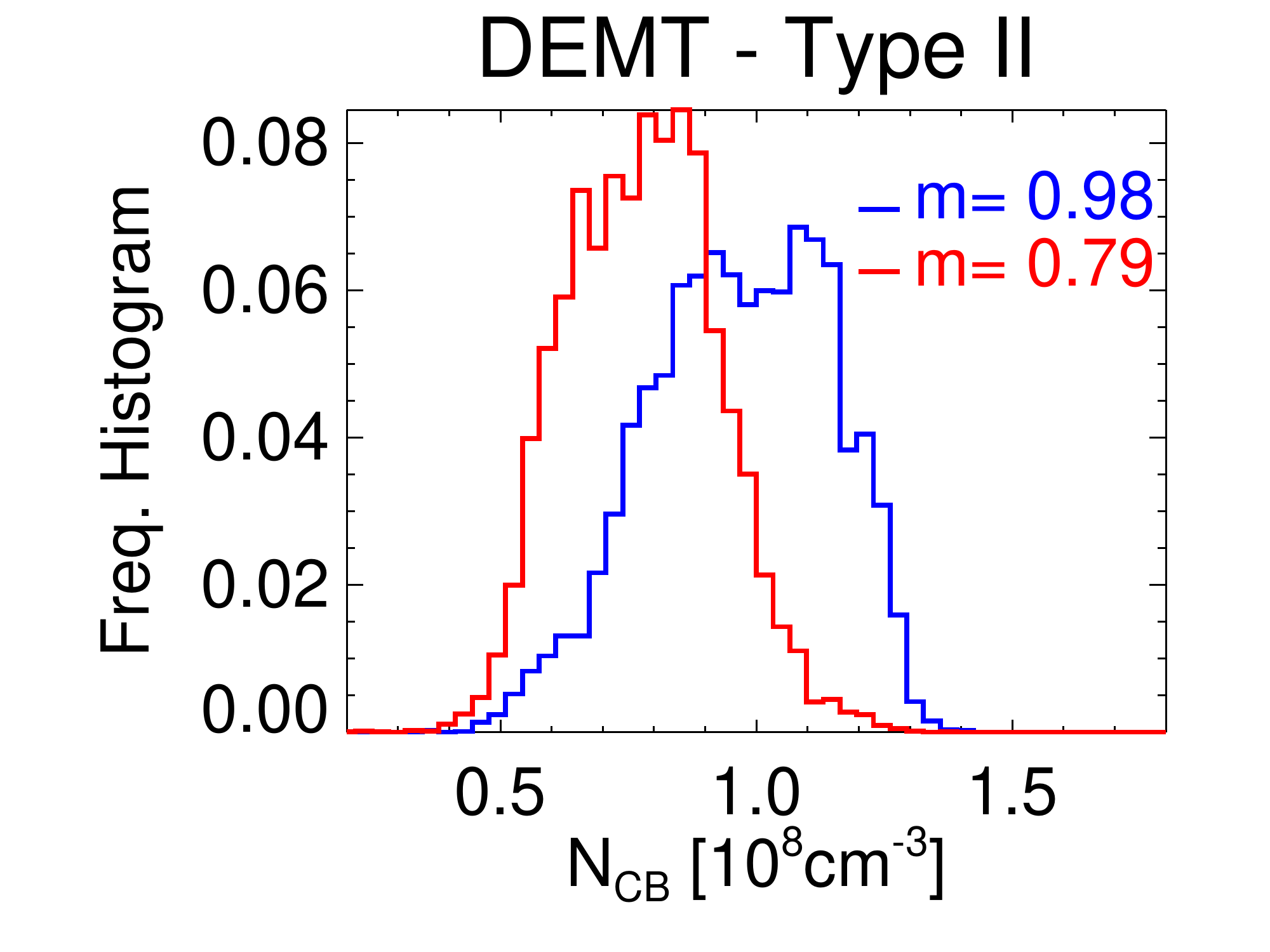}
\includegraphics[width=0.31\textwidth,clip=]{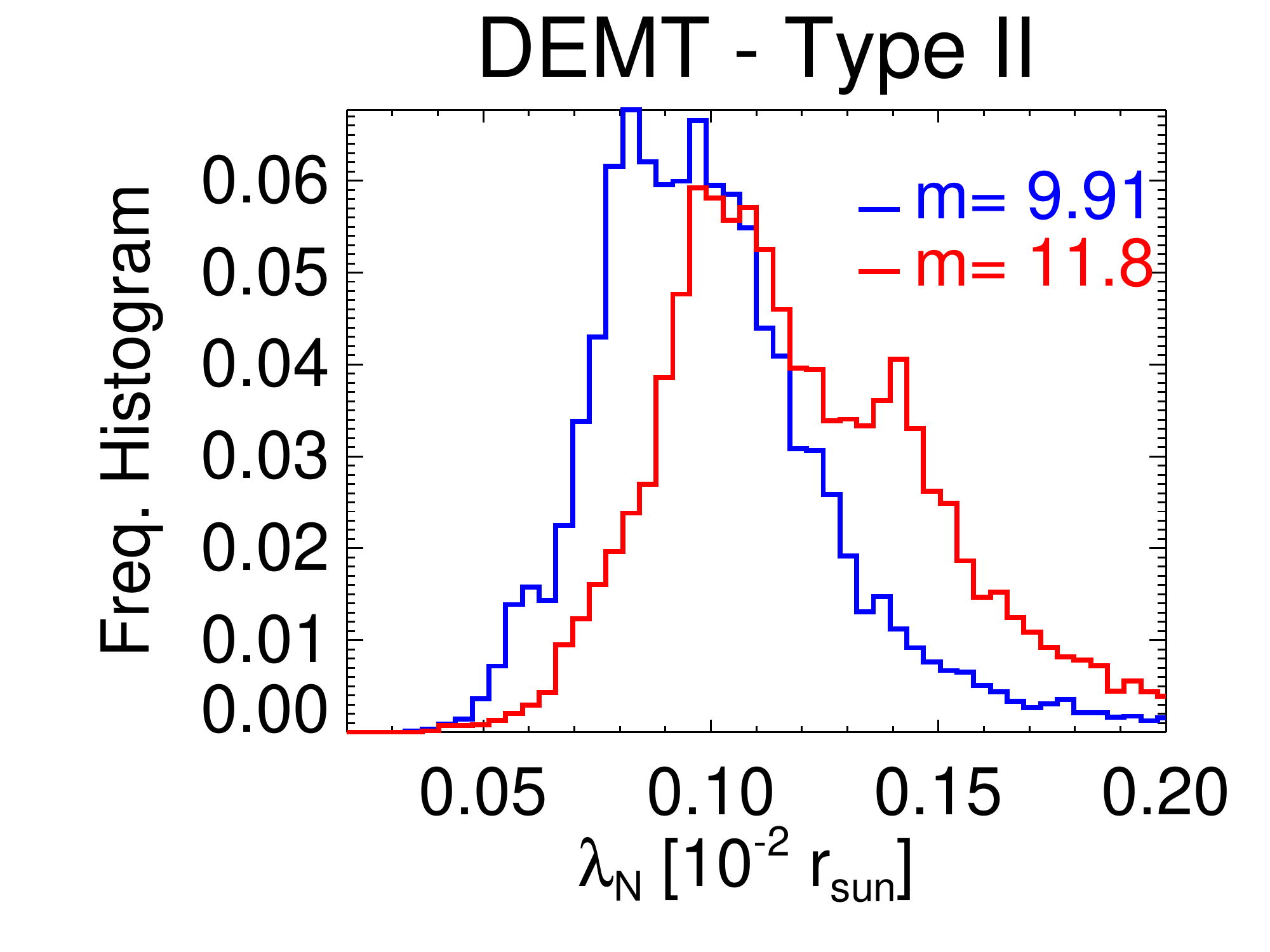}
\includegraphics[width=0.31\textwidth,clip=]{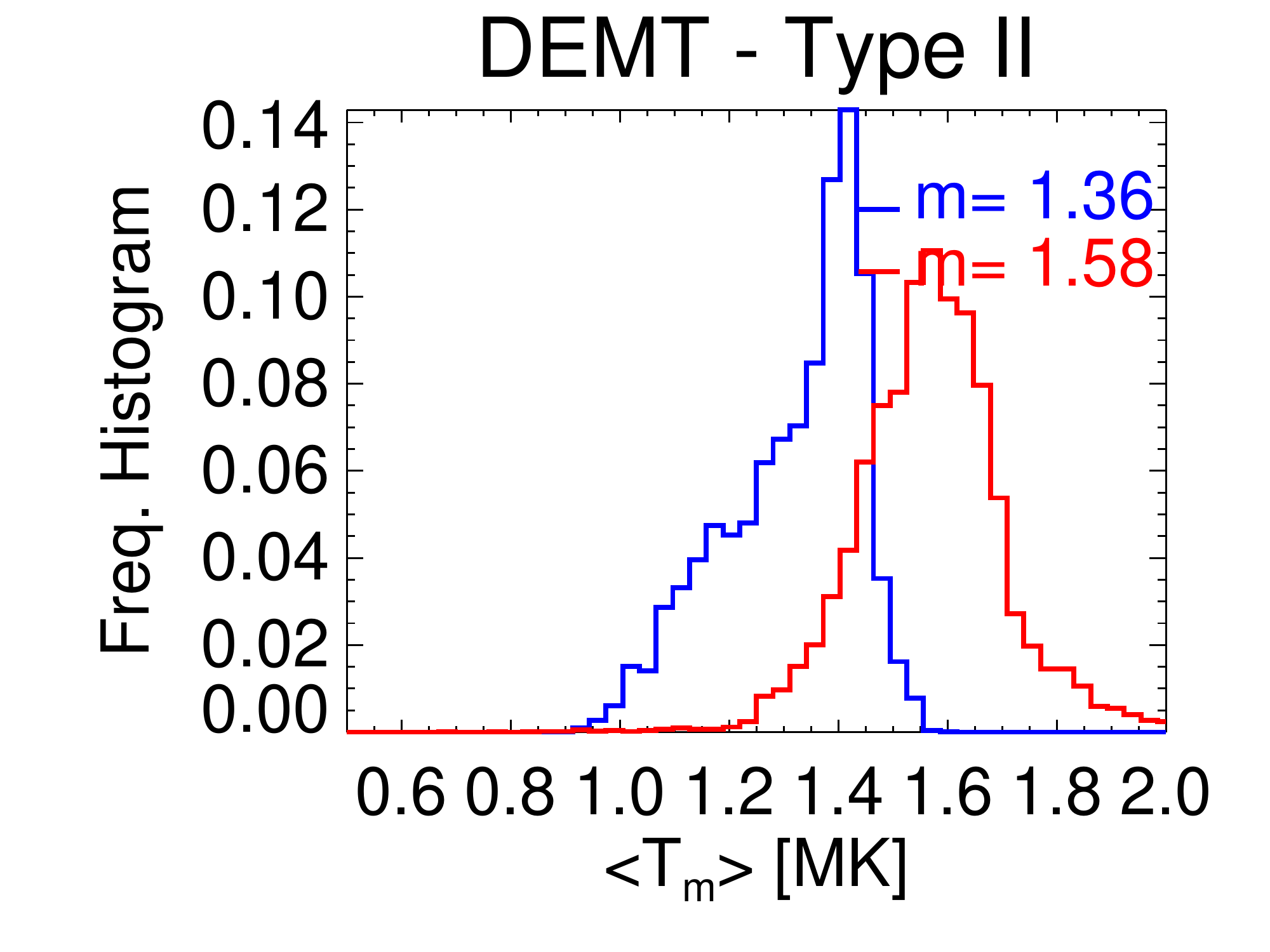}
\includegraphics[width=0.31\textwidth,clip=]{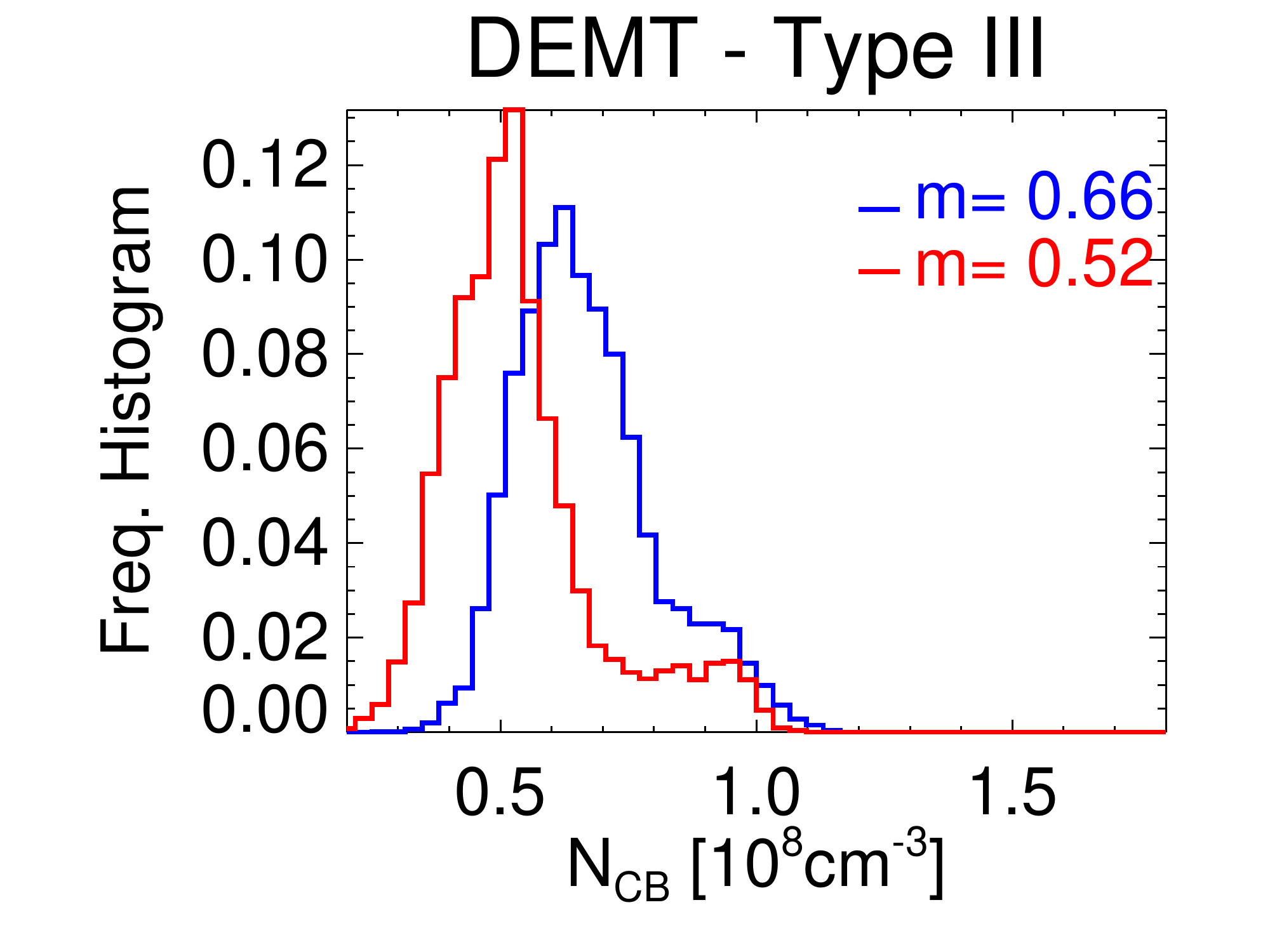}
\includegraphics[width=0.31\textwidth,clip=]{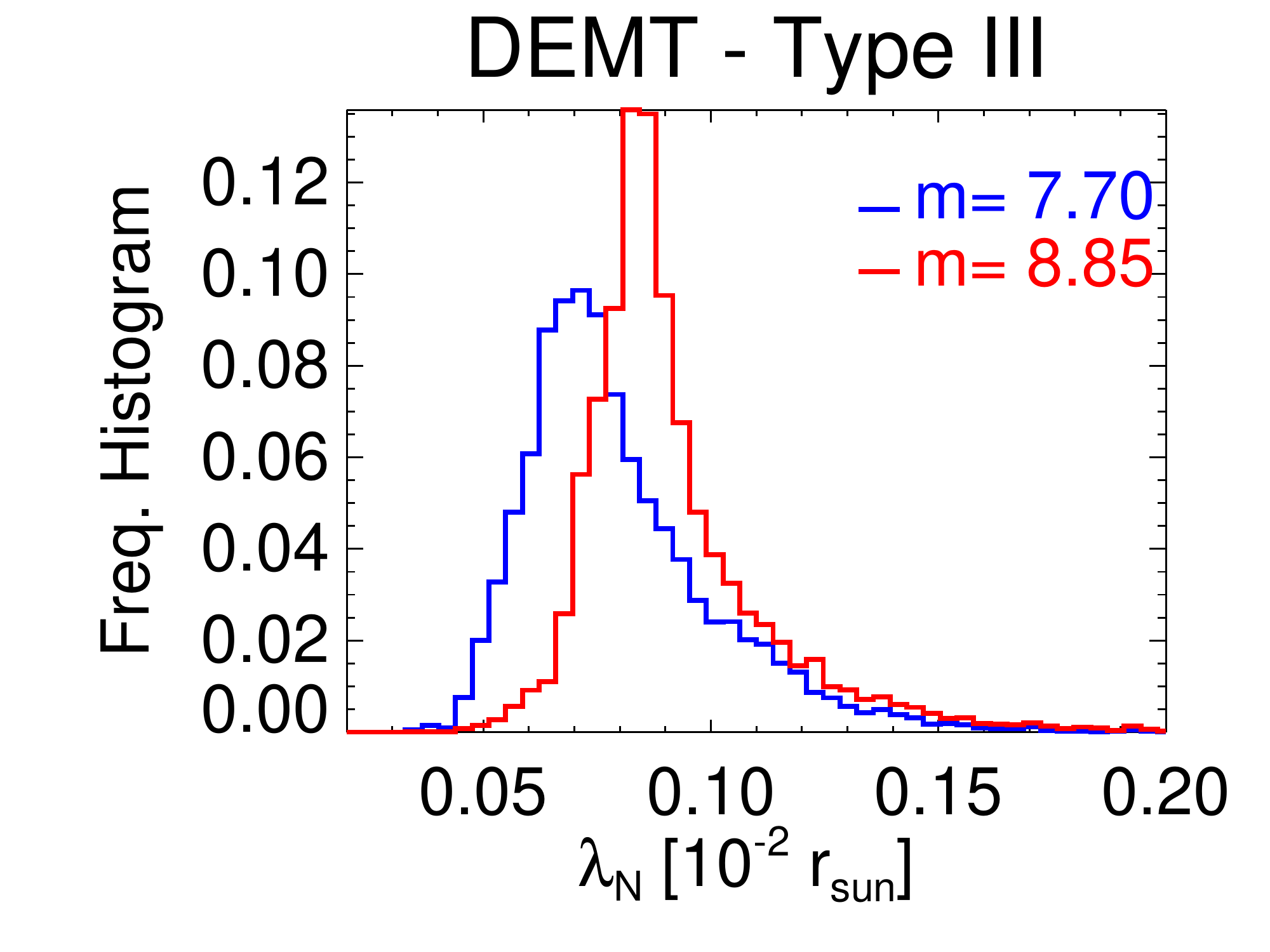}
\includegraphics[width=0.31\textwidth,clip=]{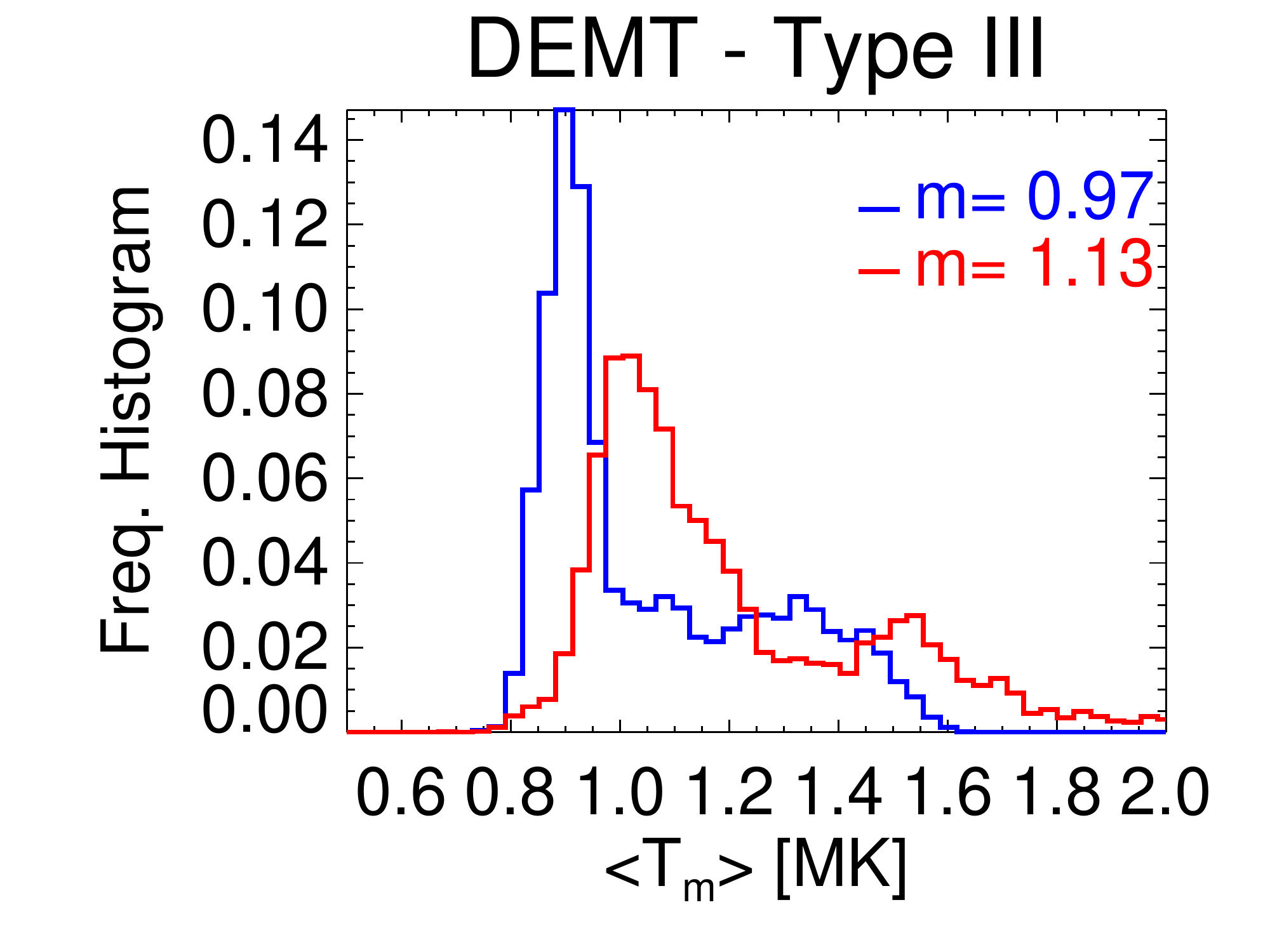}
\caption{{Statistical distribution of {DEMT results for rotations CR-2082 (blue) and CR-2208 (red)} traced along legs of type 0, I, II and III (from top to bottom), as defined in Section \ref{trace}. From left to right: {electron density $\NCB\equiv\sqravgN(r=1.055\,\mrsun)$,} electron density scale height $\lN$, {and loop-averaged temperature $\aTm$. In each panel} the median value $m$ is indicated.} }
\label{histos_fulldemt}
\end{center}
\end{figure} 

{For both rotations, Figure \ref{histos_fulldemt} shows, in a statistical fashion, the DEMT results traced along field lines discriminated by leg type}. From top to bottom {results are shown for} type 0 to type III {legs}{, respectively}. From left to right the panels show the statistical distribution of the electron density value $\NCB \equiv \sqravgN(r=1.055)$, {the scale height $\lN$, and the height-averaged (along the leg) electron temperature $\aTm$}, {with the median value $m$ indicated in each plot}.

\begin{table}[h!]
\begin{tabular}{l r@{.}l@{\hskip 0.05in} r@{\hskip 0.01in} r  r@{.}l@{\hskip 0.05in} r@{\hskip 0.01in} r r@{.}l@{\hskip 0.05in} r@{\hskip 0.01in} r }
\hline
Type    & \multicolumn{4}{c}{$\med(\NCB)$}             & \multicolumn{4}{c}{$\med(\lN)$} & \multicolumn{4}{c}{$\med(\avgTe)$} \\
        & \multicolumn{4}{c}{$[10^8\,{\rm cm}^{-3}]$}  & \multicolumn{4}{c}{$[{\rm 10}^{-2}\,\mrsun]$} & \multicolumn{4}{c}{$[\MK]$} \\
\hline
0    & 1&24 &(\Mi&19\%)  &   7&1 &(\Pl&~3\%) &   1&09 &(\Pl&24\%) \\
I    & 1&15 &(\Mi&10\%)  &   7&5 &(\Pl&29\%) &   1&25 &(\Pl&24\%) \\
II   & 0&98 &(\Mi&20\%)  &   9&9 &(\Pl&19\%) &   1&36 &(\Pl&16\%) \\
III  & 0&66 &(\Mi&{21}\%)  &   7&7 &(\Pl&15\%) &   0&97 &(\Pl&{17}\%) \\
\hline          
\end{tabular}
\caption{Median value (indicated as ``Md'') of the statistical distribution of $\NCB$, $\lN$, and $\aTm$ for each coronal type of legs defined in Section \ref{trace}. For CR-2082 values are expressed in absolute terms, while for CR-2208 they are {given} as a percentual variation relative to the CR-2082 value {in the parentheses.}}
\label{tabla_demt}
\end{table}

Table \ref{tabla_demt} summarizes {the results of the quantitative comparative analysis between the two {target rotations}}. For CR-2082 quantities are expressed as absolute values, {while} for CR-2208 they are expressed relative to the corresponding results for CR-2082.

{Throughout the magnetically closed region of both rotations, type 0, I and II legs, associated to  increasingly outer layers of the equatorial streamer belt, exhibit progressively decreasing coronal base density, increasing density scale height, and increasing electron temperature. In both rotations also, type III legs in the CHs are characterised by sub-MK temperatures, and electron density values of order $\approx 1/2$ of those observed for the type 0 and type I legs in the core of the equatorial streamer.} {A comparison of the results between the two rotations shows that, compared to CR-2082, rotation CR-2208 was characterised by {$\approx 10-20\%$} lower values of the electron density at the coronal base, $\approx 5-30\%$ larger values of density scale height, and {$\approx 15-25\%$} larger values of the electron temperature.} 

\begin{figure}[h!]
\begin{center}
\includegraphics[width=0.495\textwidth]{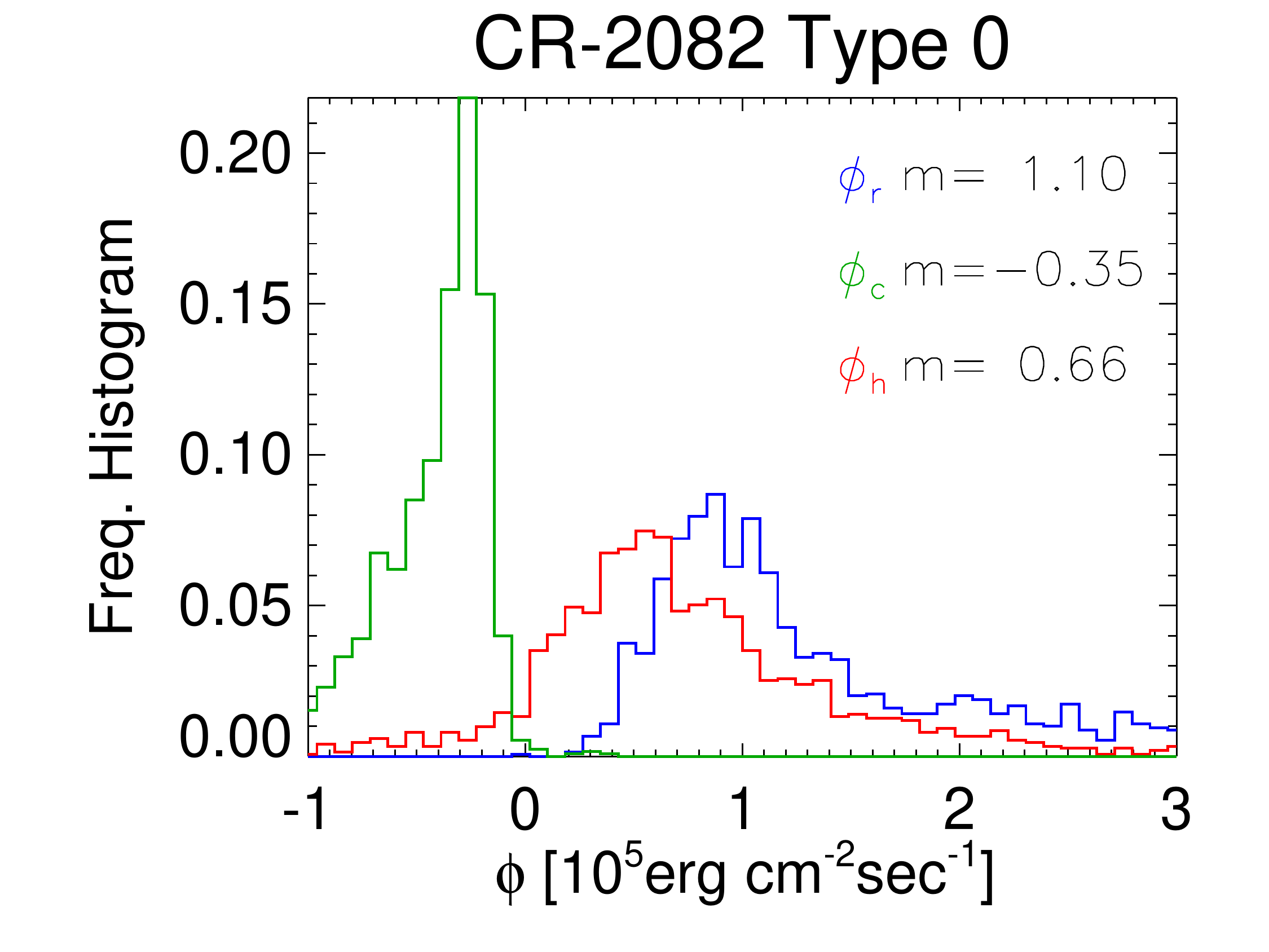}
\includegraphics[width=0.495\textwidth]{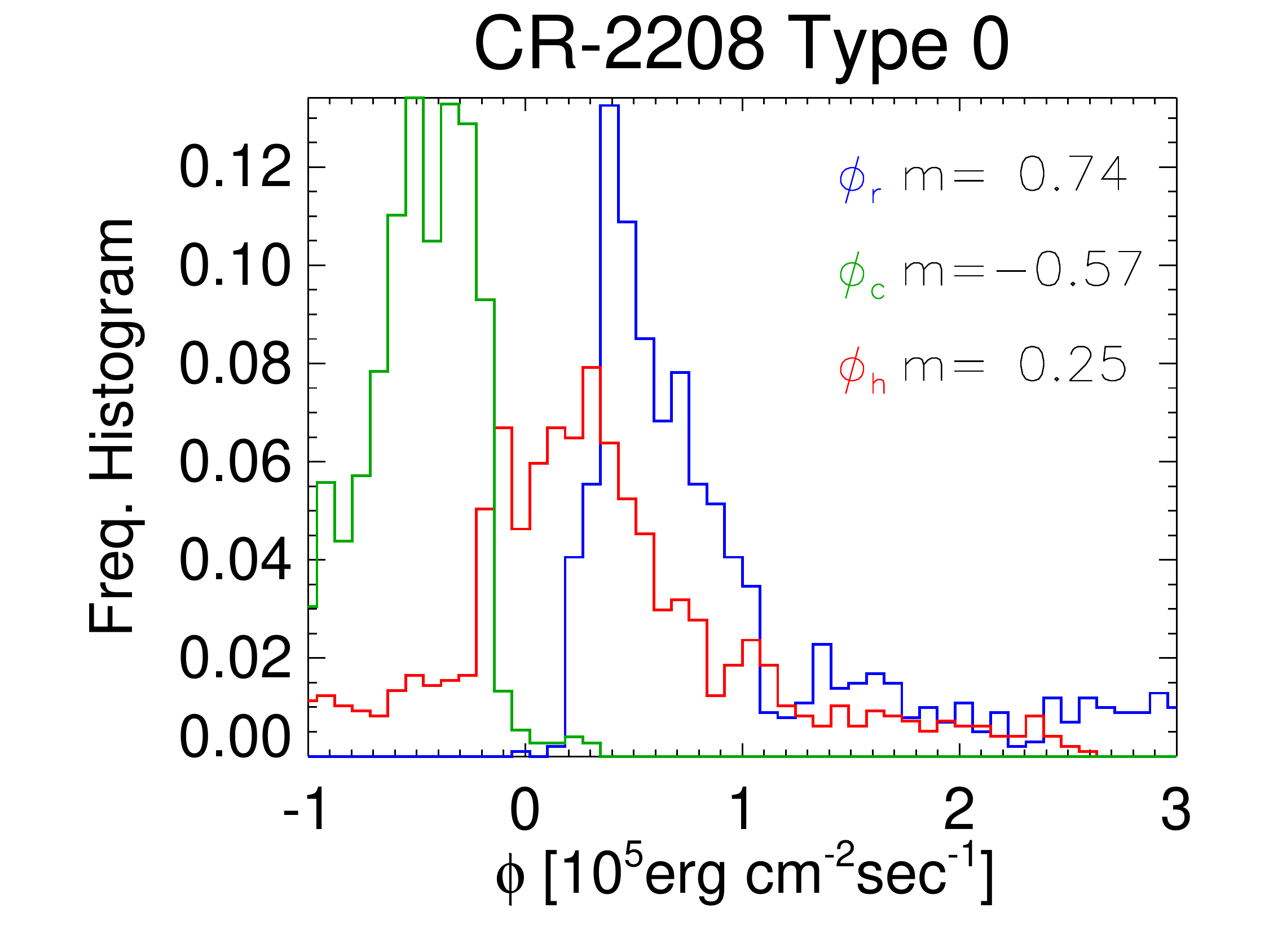}
\includegraphics[width=0.495\textwidth]{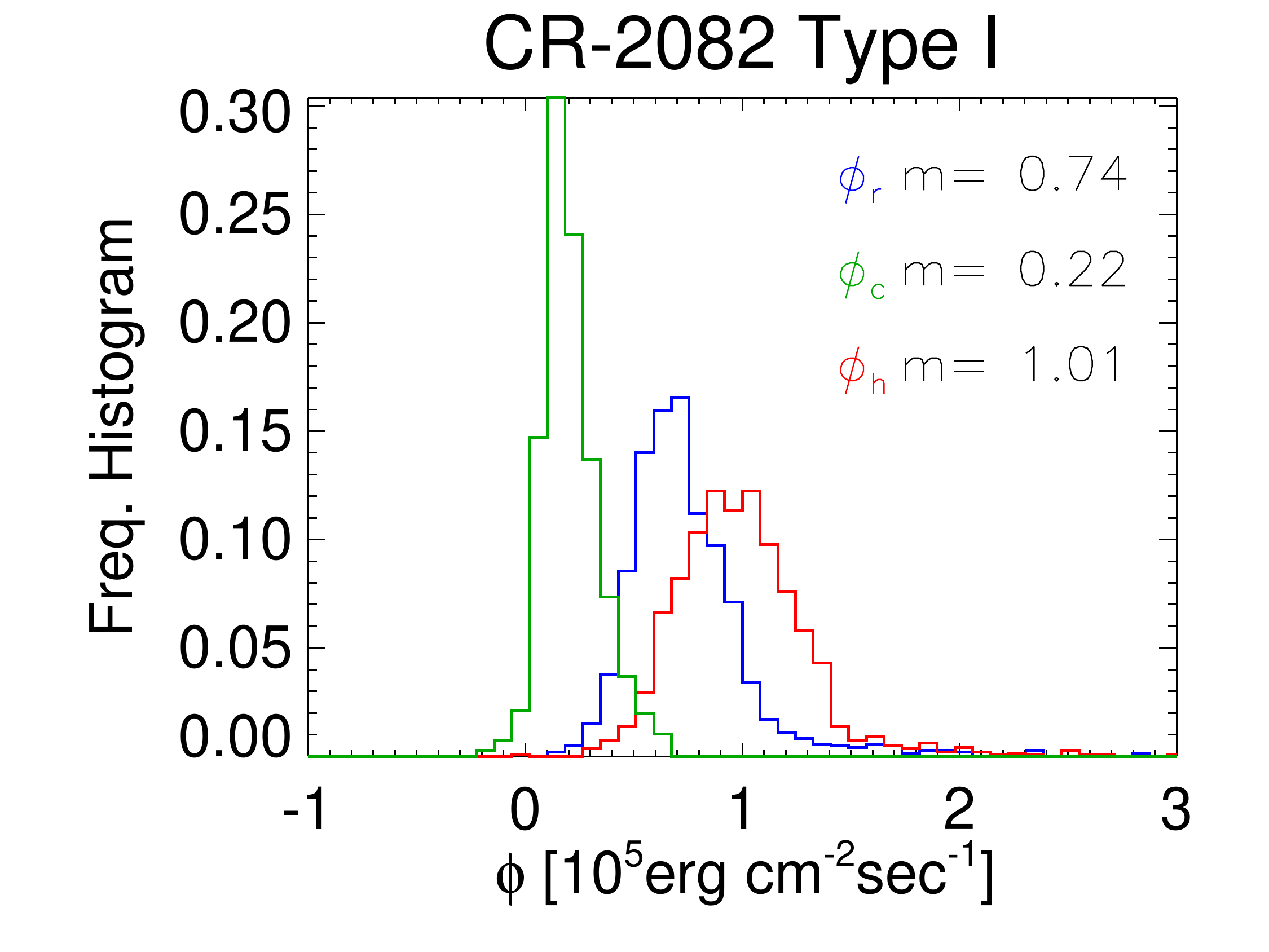}
\includegraphics[width=0.495\textwidth]{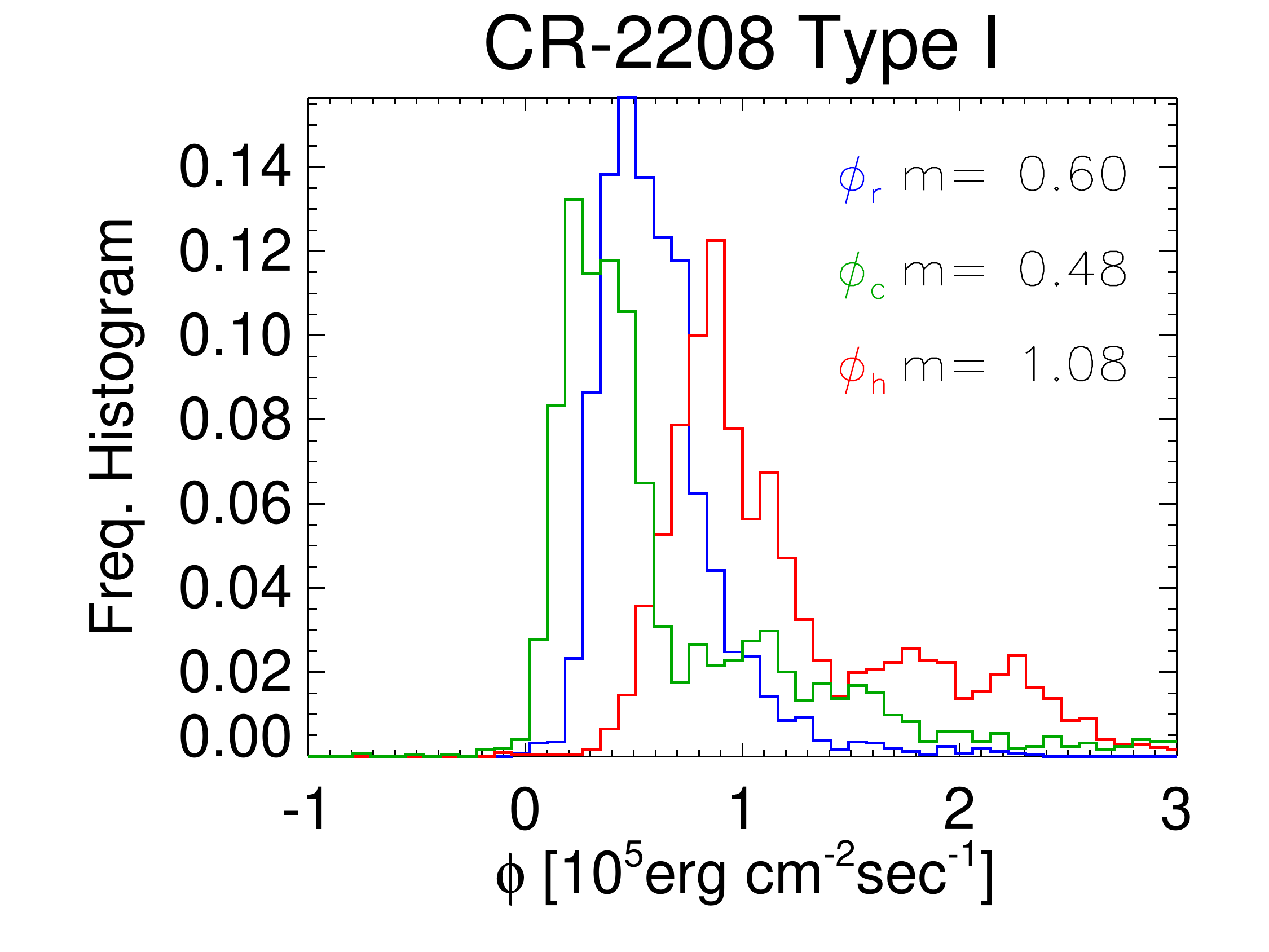}
\includegraphics[width=0.495\textwidth]{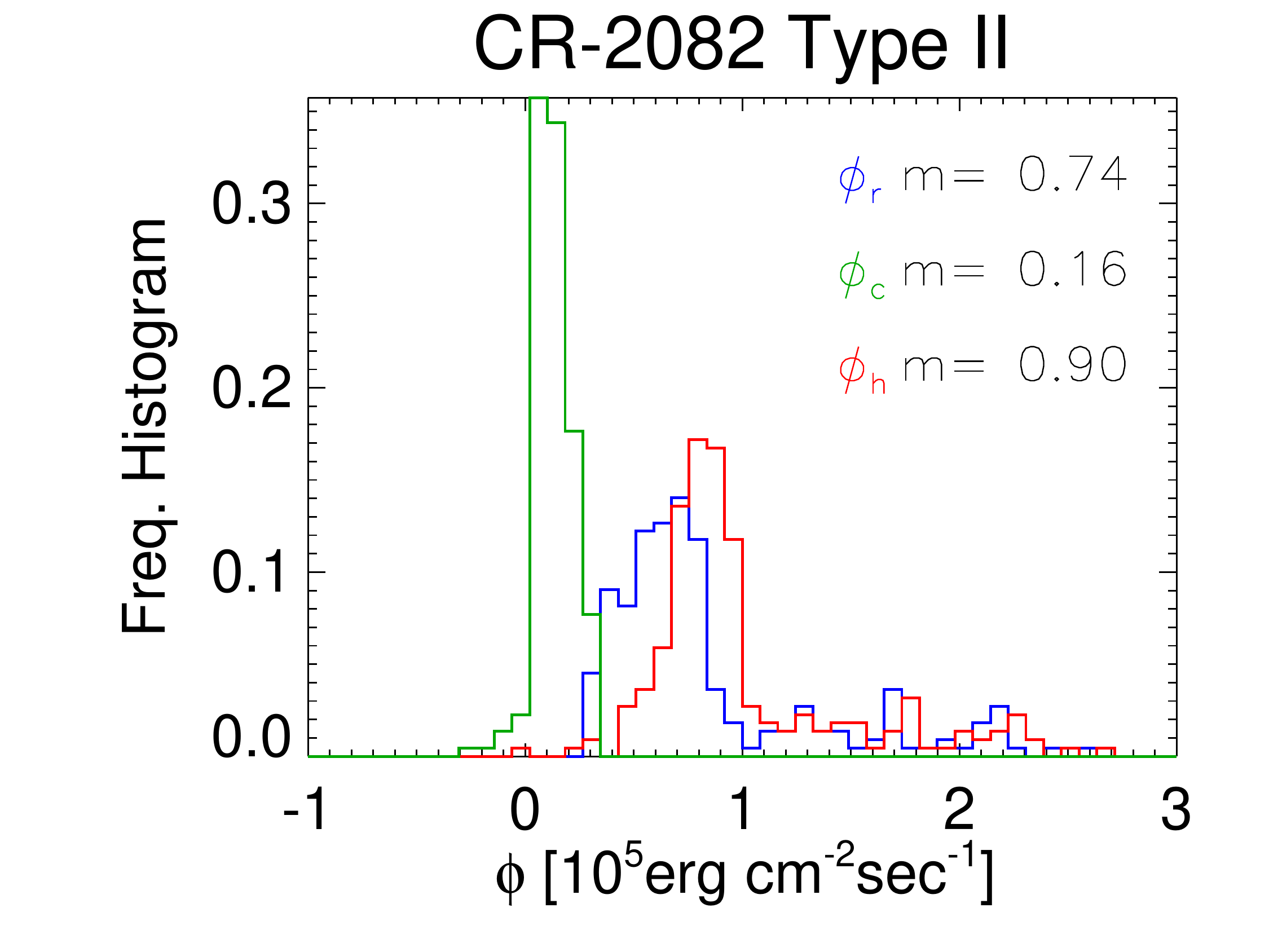}
\includegraphics[width=0.495\textwidth]{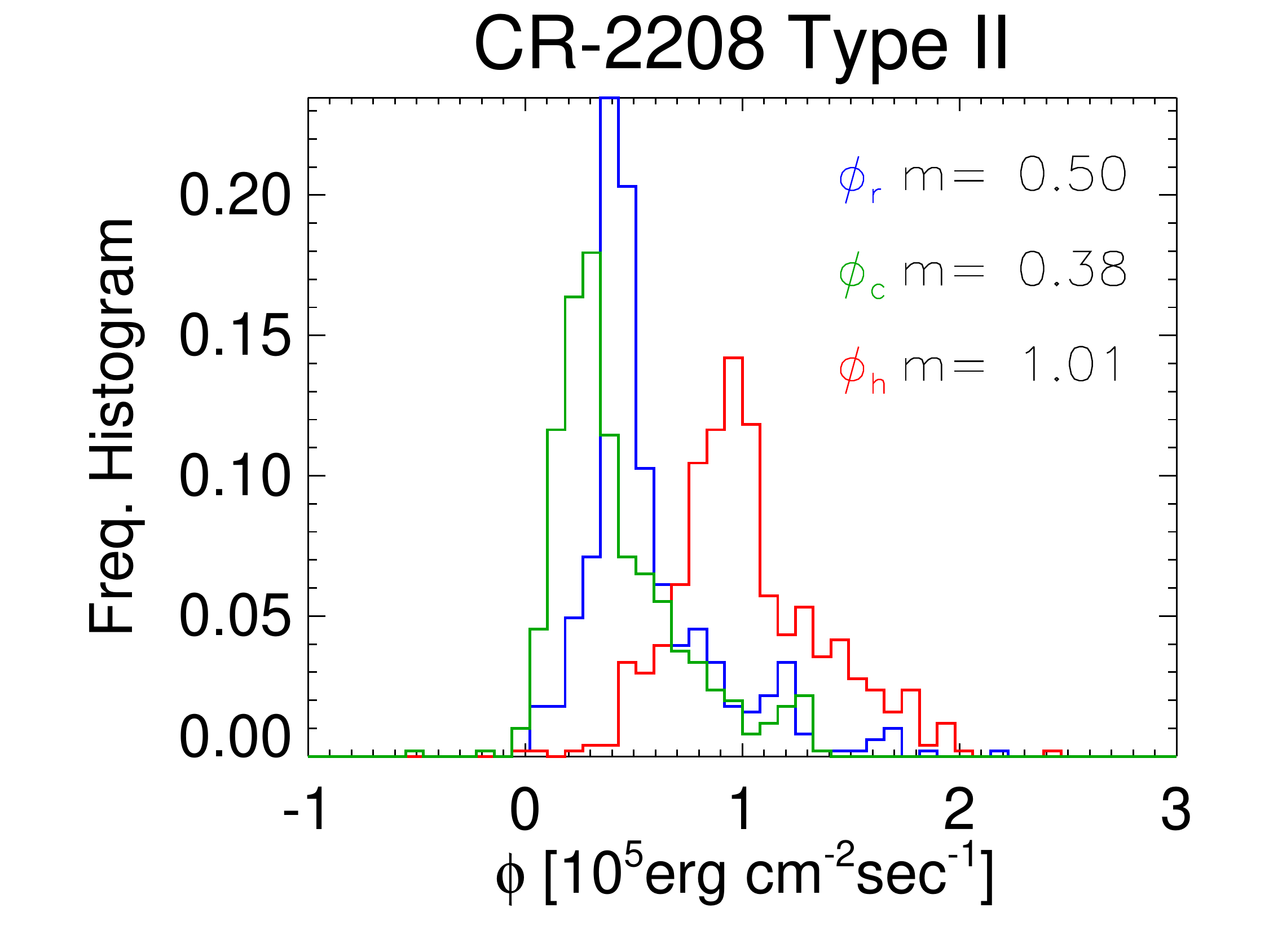}
\caption{{Statistical results of the loop-integrated energy flux quantities $\phi_r,\phi_c,$ and $\phi_h$ in colors blue, red and green, respectively for CR-2082 (left) and CR-2208 (right). From top to bottom, panels show the results for loops of type 0, I and II, which are loops for which both legs meet the criteria from Section \ref{trace}.}}
\label{energia_demt}
\end{center}
\end{figure} 

In comparing the DEMT results obtained for the two selected targets, {we highlight} they rely on data provided by two different instruments, {namely EUVI and AIA for CR-2082 and CR-2208, respectively.} To quantify the systematic difference of the DEMT products {due to the different filter sets of} both instruments, \citet{nuevo_2015}, who were the first to apply DEMT to AIA data, {analysed} a single target using both instruments independently. They found that while the density {products are} essentially equal, the temperature {based on AIA data} is systematically 8\% larger than the one based on EUVI data, \textit{i.e.} $\Tm^{\rm(AIA)}/\Tm^{\rm(EUVI)} \approx 1.08$. Considering {this error}, Figure \ref{histos_fulldemt} and Table \ref{tabla_demt} indicate that CR-2208 was {characterised by temperatures $\approx 10-15\%$ larger relative to CR-2082} throughout the streamer belt region. As for the electron density products, CR-2208 was found to be $\approx 10-20\%$ less dense than CR-2082 throughout the streamer belt region. These systematic differences are {around or beyond the uncertainty level in the DEMT products due to systematic sources (radiometric calibration and tomographic regularization), that \citet{lloveras_2017} {estimated to be $\Delta\Tm\approx 10\%$ and $\Delta\sqravgN\approx 5\%$}.}

{To analyse the loop-integrated energy flux quantities introduced in Section \ref{trace}, we selected closed loops for which both legs have the same sign of the radial gradient of the electron temperature $\dr\Tm$.} In this way, according to the classification of both its legs, each loop was classified as of type 0 (small down loop), I (small up loop), or II (large up loop). {For both target rotations, and for loops of type 0, I and II, Figure \ref{energia_demt} shows the frequency histogram of the loop-integrated energy flux quantities {$\phi_r$, $\phi_c$ and $\phi_h$ in blue, green and red color}, respectively.}

{For both rotations, the value of the {loop-integrated} radiative power $E_r$, measured by the quantity $\phi_r$, is largest for loops of type 0. This is due to $E_r\propto N_e^2\,\Lambda(T_e)$, with both factors contributing to maximize $E_r$ for loops of type 0. As shown in Figure \ref{histos_fulldemt} and Table \ref{tabla_demt}, loops of type 0 are {characterised} by {the largest} values of electron density. Also, in the range of sensitivity of the EUVI and AIA instruments, namely 0.5–3.0 MK \citep{nuevo_2015}, the radiative loss function $\Lambda(T)$ has a local maximum at $\Tc\approx 1\,\MK$. According to Figure \ref{histos_fulldemt}, loops of type 0, I and II are {characterised by temperatures that are are progressively larger, and farther from $\Tc$, for both rotations.}}

The sign of the quantity $\phi_c$ depends on that of the conductive flux $F_c$. Equations \ref{Fc} and \ref{phi_c} imply that, by definition, down loops (type 0) and up loops (type I and II) are characterised by $\phi_c<0$ and $\phi_c>0$, respectively, {as verified} in Figure \ref{energia_demt}.

{{Adding the} radiative and conductive terms, the characteristic energy input flux at the coronal base is in the range $\phi_h\approx 0.5-1.5 \times 10^5\,\erg\,\cminvs\,\s^{-1}$, depending on the rotation and the type of {loop, matching the values reported by} \citet{maccormack_2017}. {For} type 0 loops there is a marginal population characterised by the unphysical result $\phi_h<0$. As shown by \citet{maccormack_2017}, this affects only the smallest sized loops of the type 0, and it is likely due {to emission out of the instrumental sensitivity range}. Though accounting for most of the coronal plasma, there surely is additional emission out of the instrumental sensitivity range. As a result,} the positive term $\phi_r$ is most likely underestimated, leading to values $\phi_h<0$ in loops of type 0, {being characterised by $\phi_c<0$.}

\subsection{{Comparison of the DEMT and AWSoM Models}}\label{awsom_res} 

\begin{figure}[h!]
\begin{center}
\includegraphics[width=0.495\textwidth,clip=]{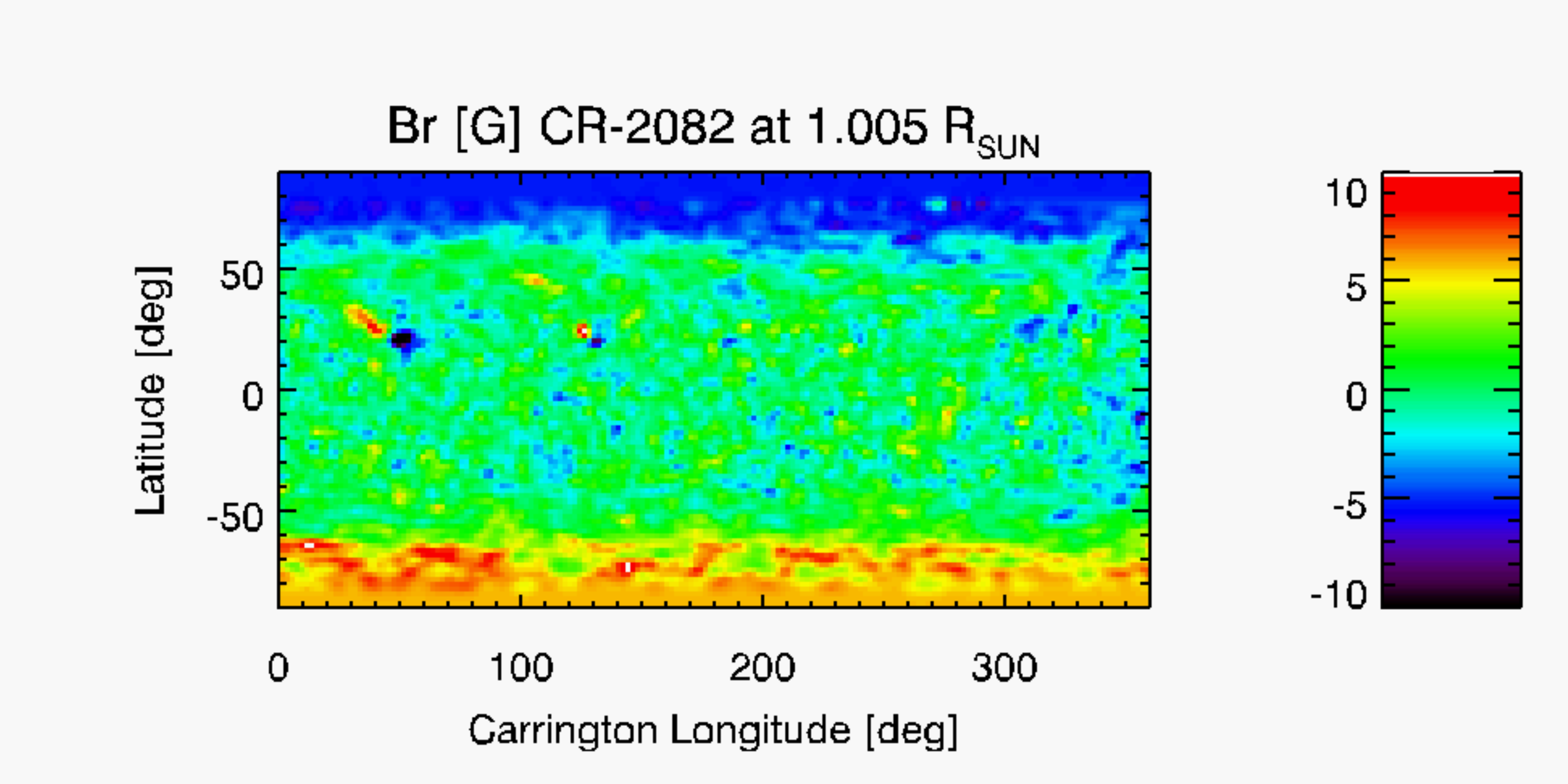}
\includegraphics[width=0.495\textwidth,clip=]{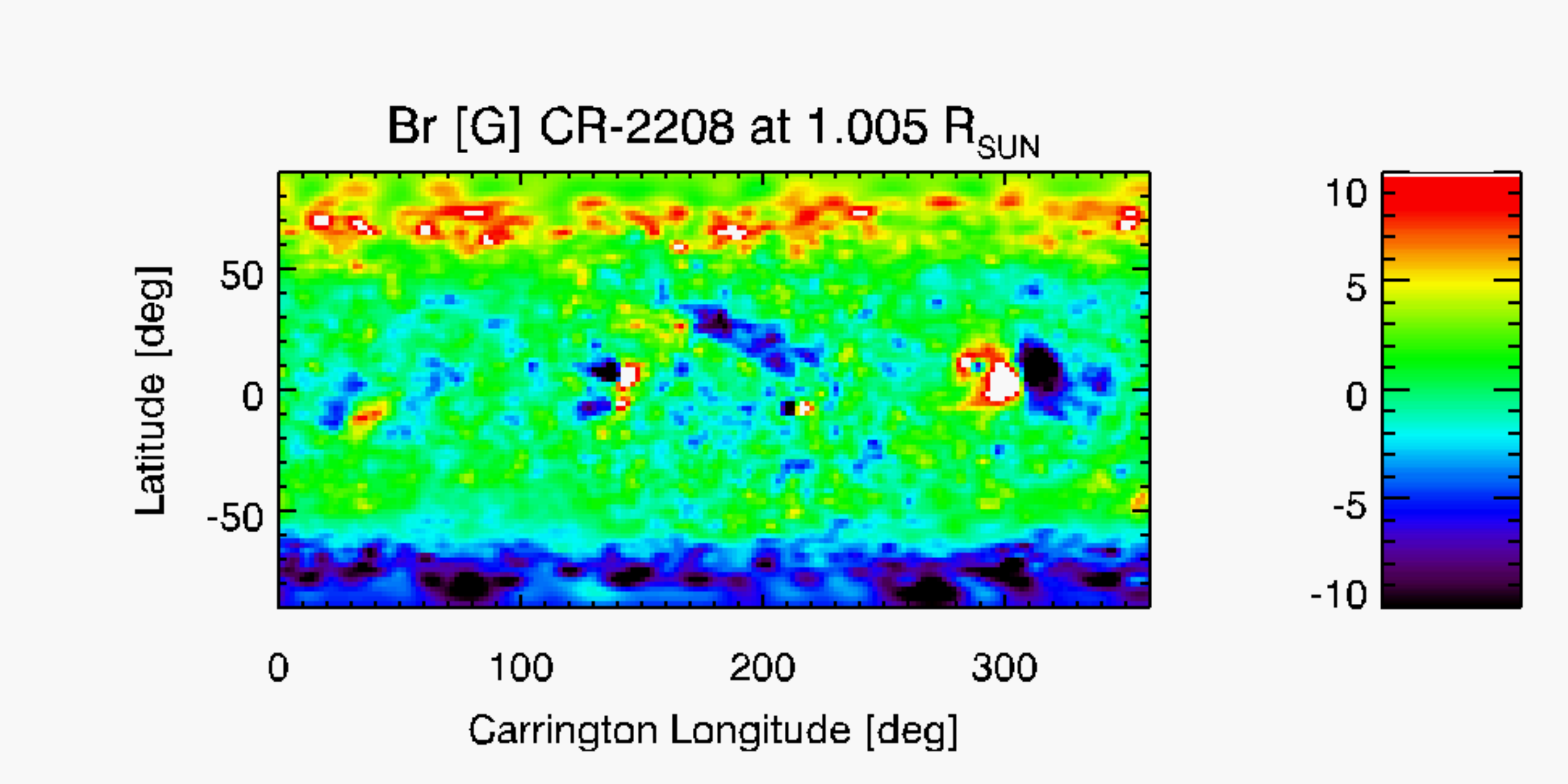}
\caption{Carrington maps of {the radial magnetic field $B_r$ of the AWSoM model} at $1.005 \, \mrsun$ for CR-2082 (left) and CR-2208 (right). }
\label{Br}
\end{center}
\end{figure}

{Figure \ref{Br} shows carrington maps of {the radial} magnetic {field ($B_r$)} for both rotations at $1.005\, \mrsun$. {Both maps clearly show the large-scale dipolar field, characteristic of solar minimum conditions. Differences between both maps are observed in the sub-polar latitudes, due to the different treatments applied there by the GONG (CR-2082) and the ADAPT-GONG (CR-2208) maps.}}

\begin{figure}[h!]
\begin{center}
\includegraphics[width=0.495\textwidth]{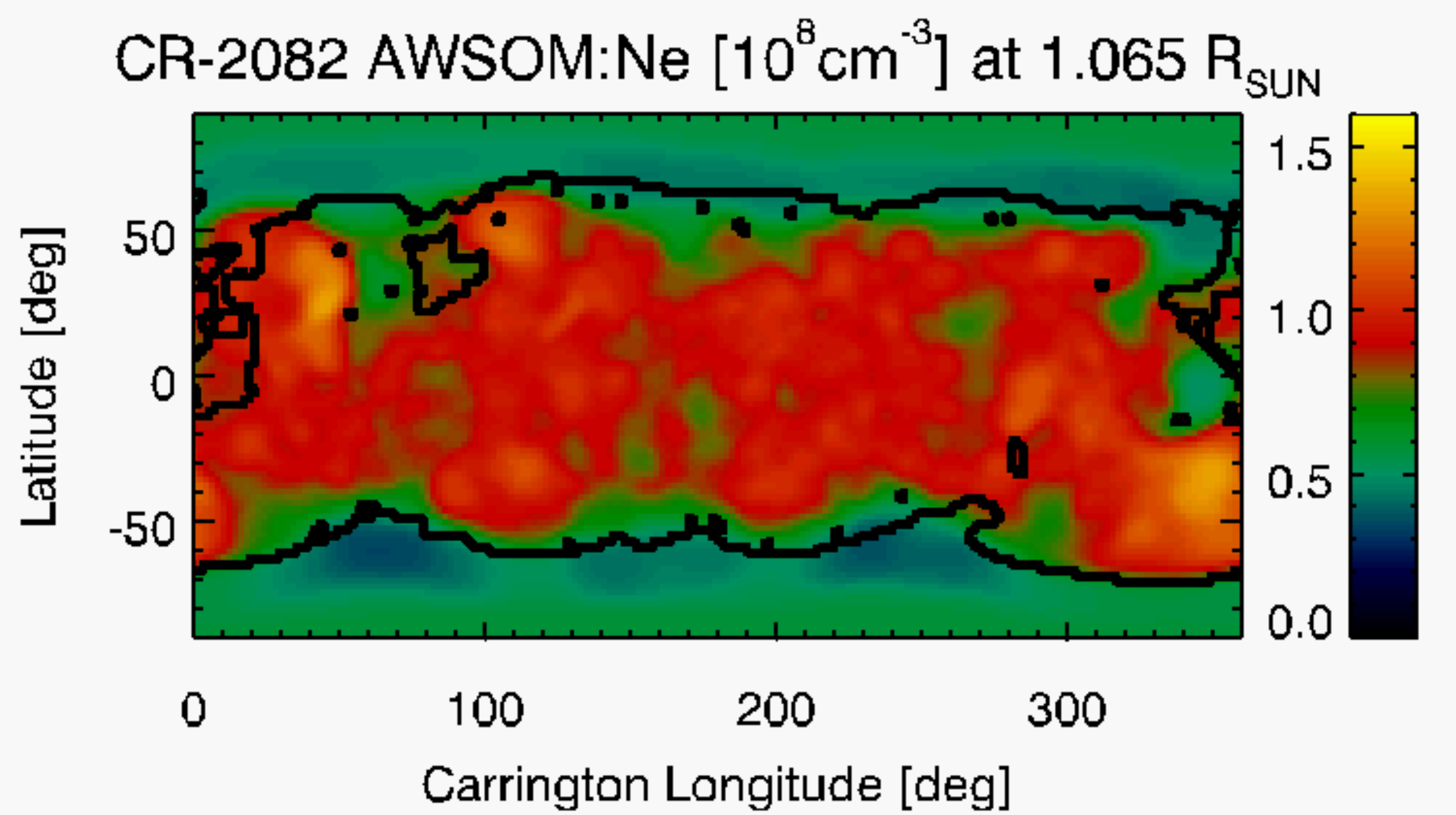}
\includegraphics[width=0.495\textwidth]{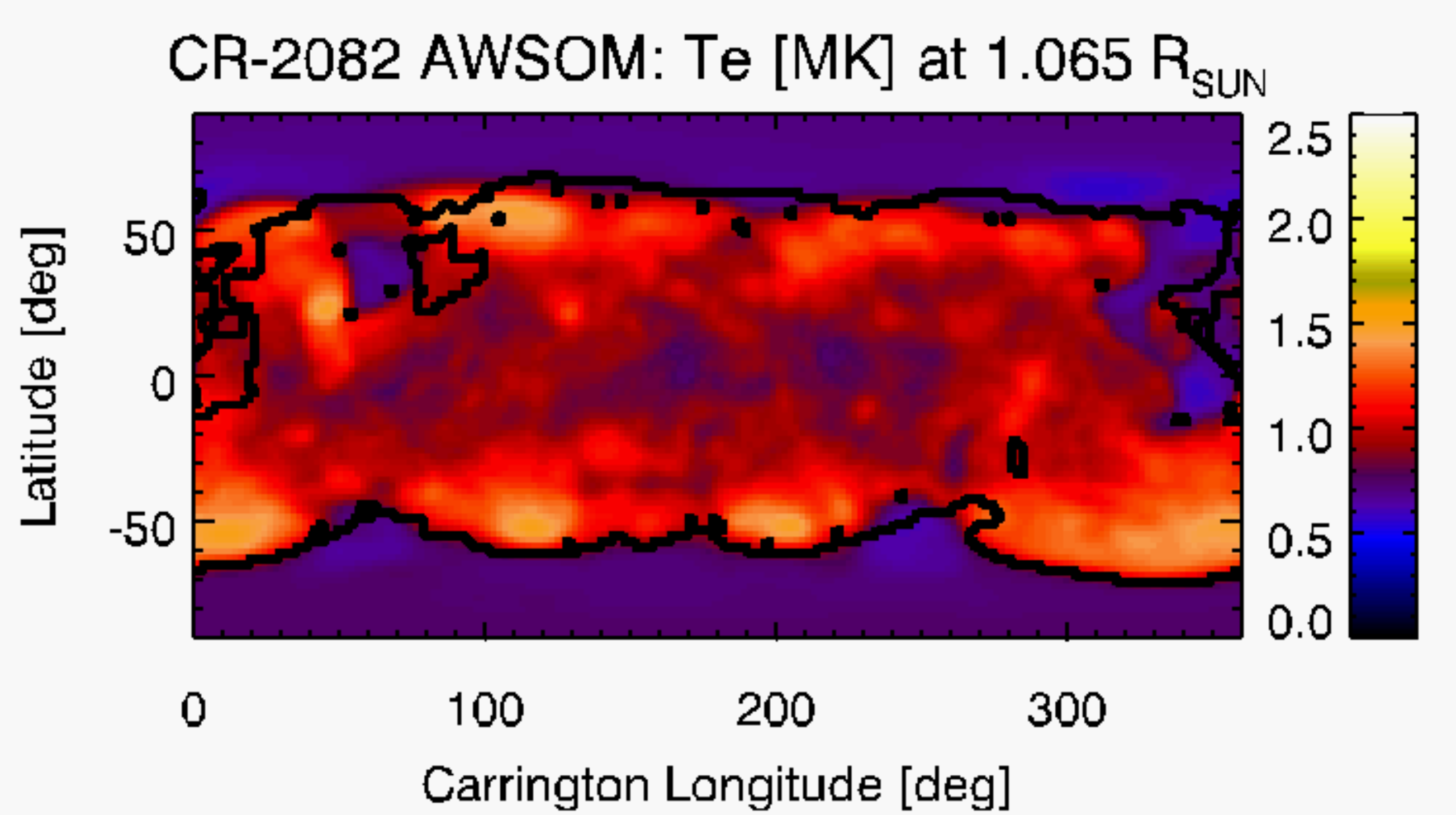}
\includegraphics[width=0.495\textwidth]{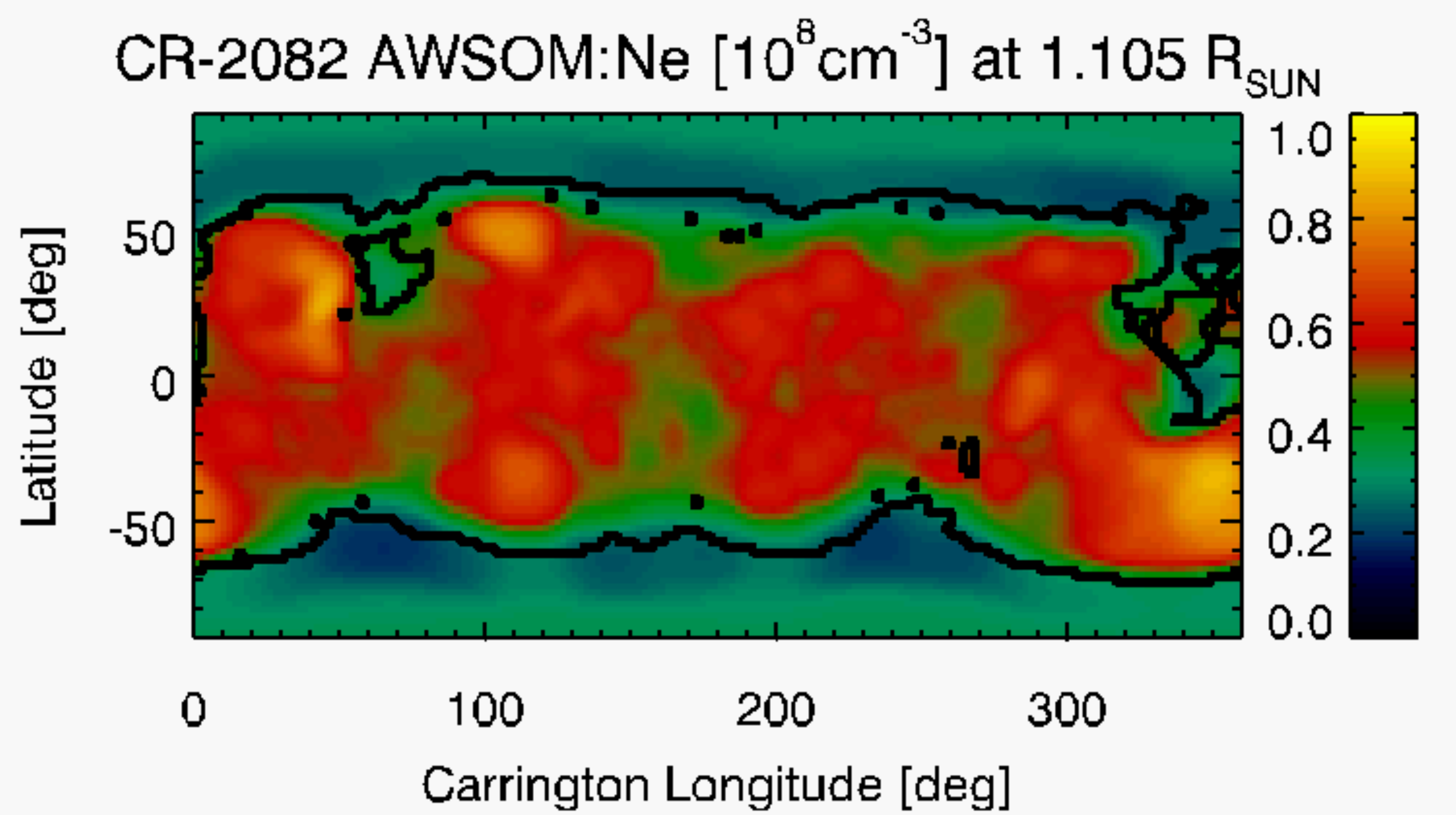}
\includegraphics[width=0.495\textwidth]{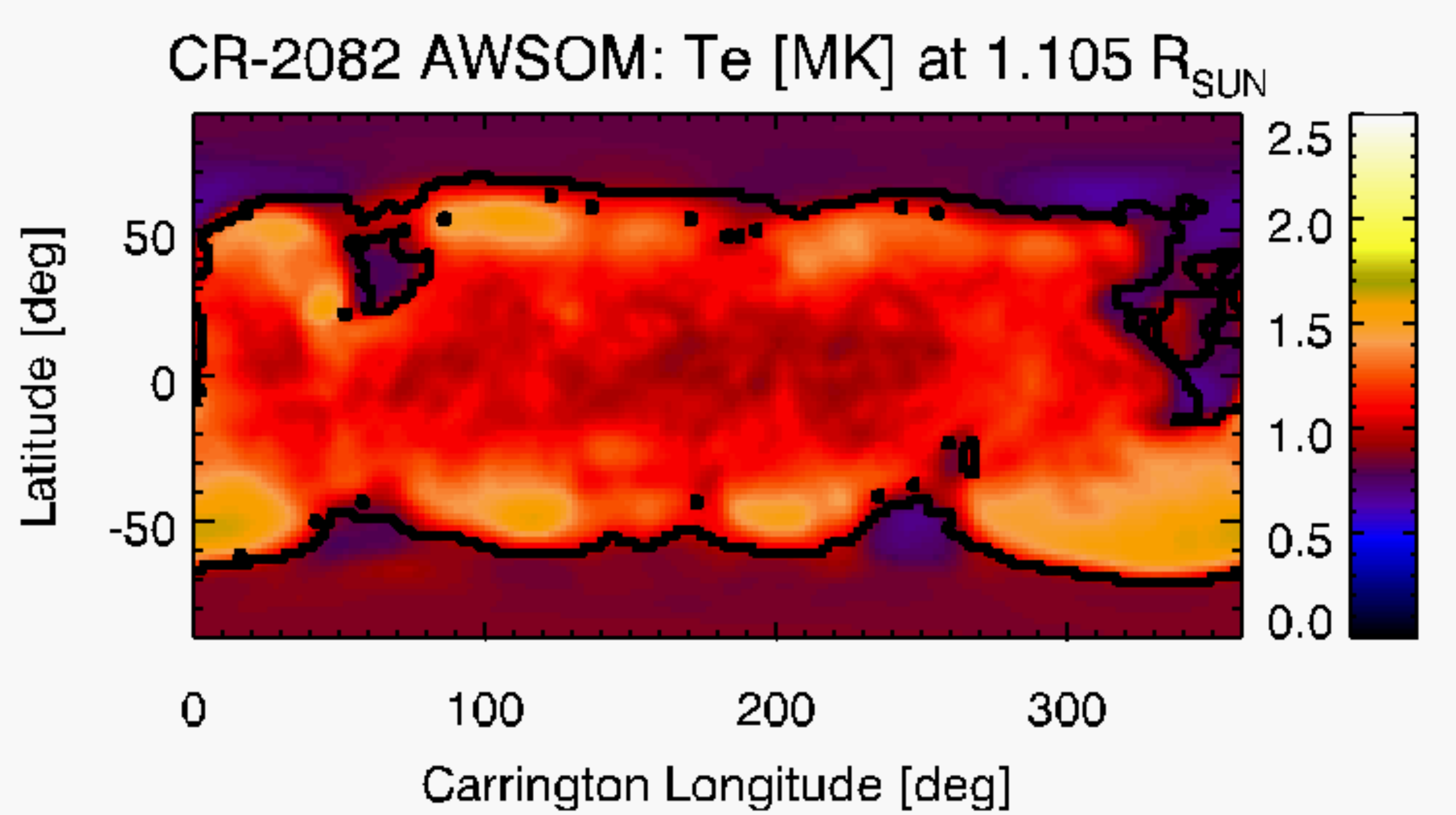}
\caption{Carrington maps of density (left panels) and temperature (right panels) obtained with AWSoM model at heliocentric heights $1.065$ (top panels) and $1.105\,\mrsun$ (bottom panels).}
\label{carmaps_awsom_2082}
\end{center}
\end{figure}

\begin{figure}[h!]
\begin{center}
\includegraphics[width=0.495\textwidth]{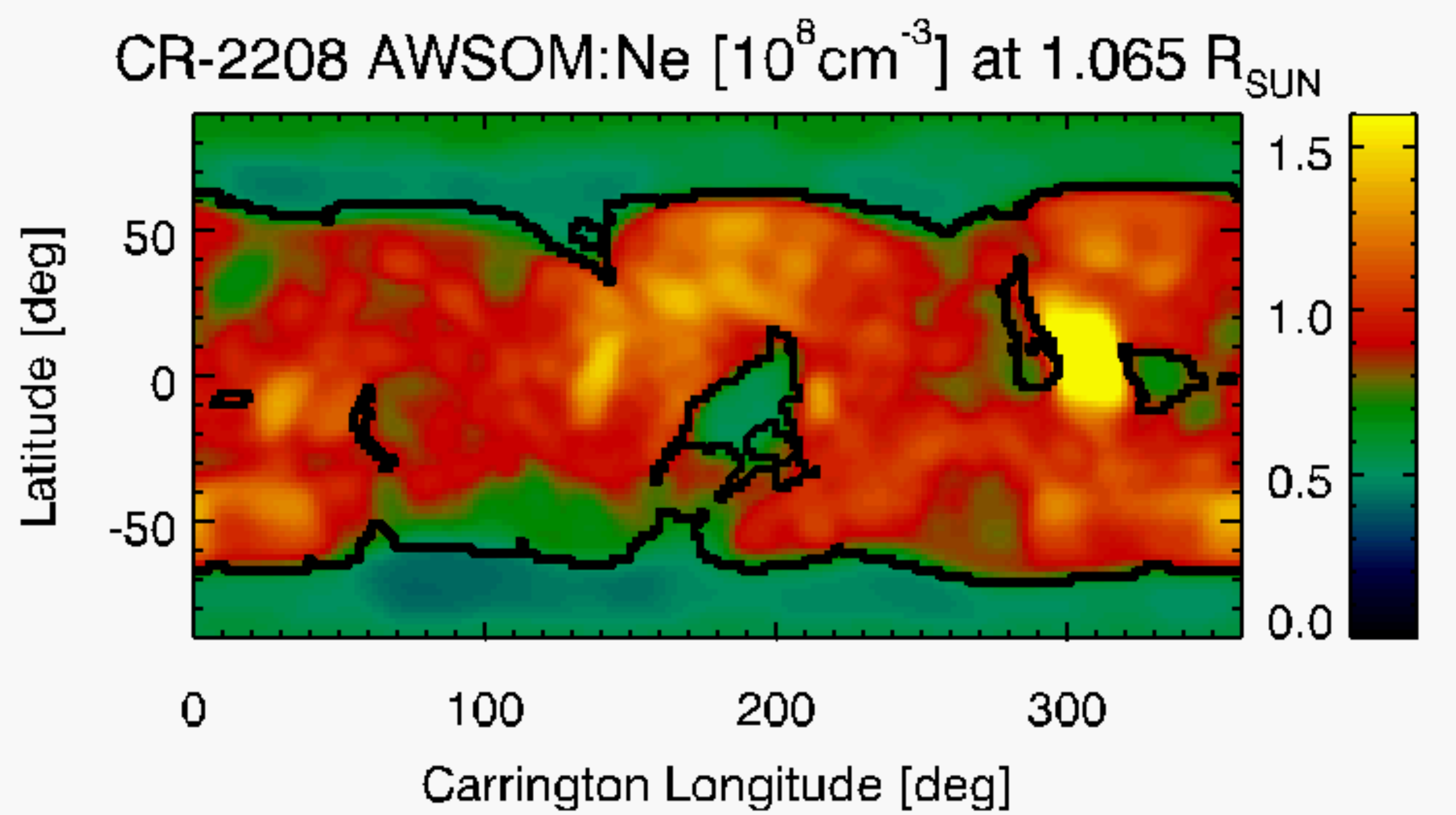}
\includegraphics[width=0.495\textwidth]{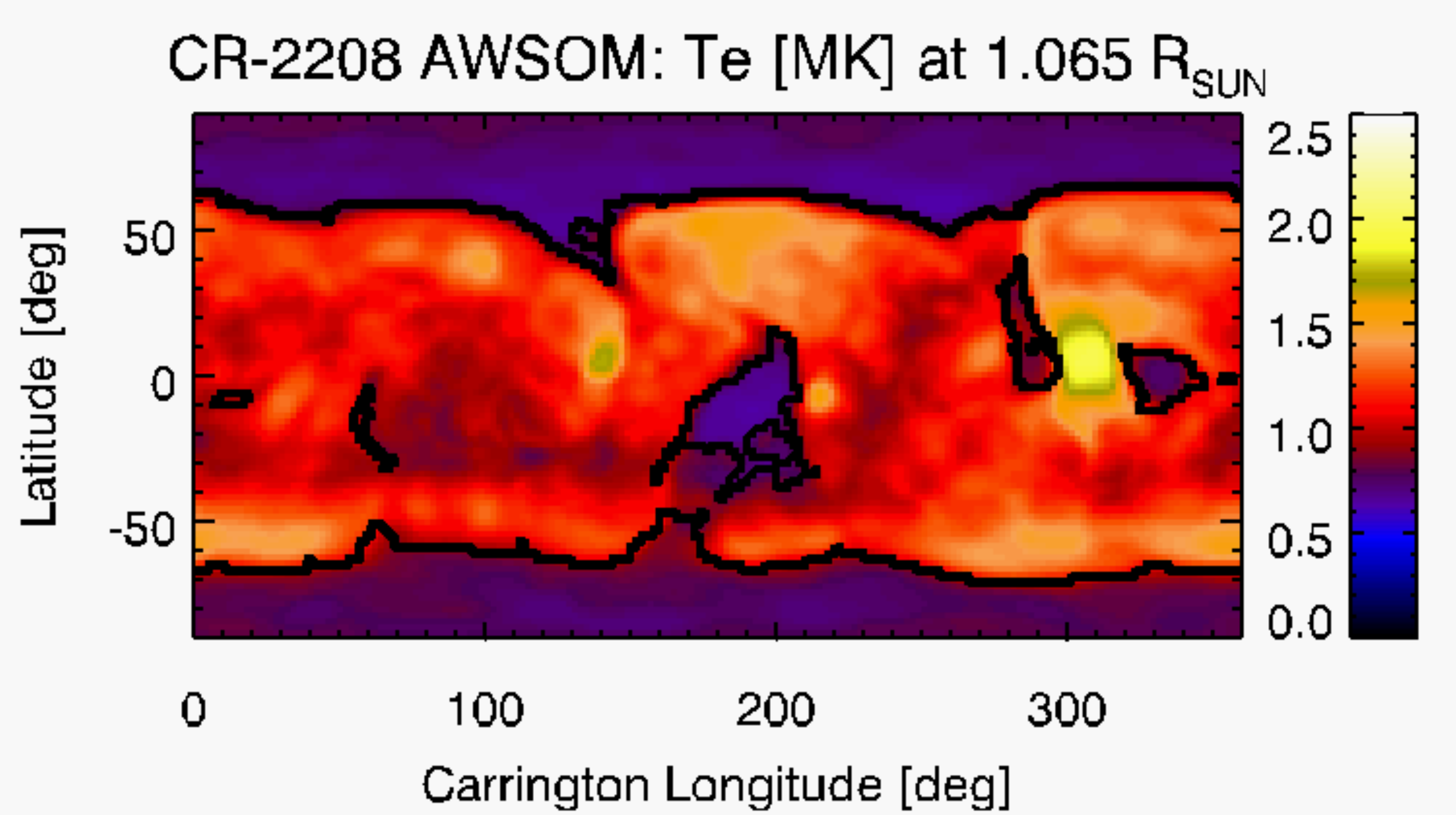}
\includegraphics[width=0.495\textwidth]{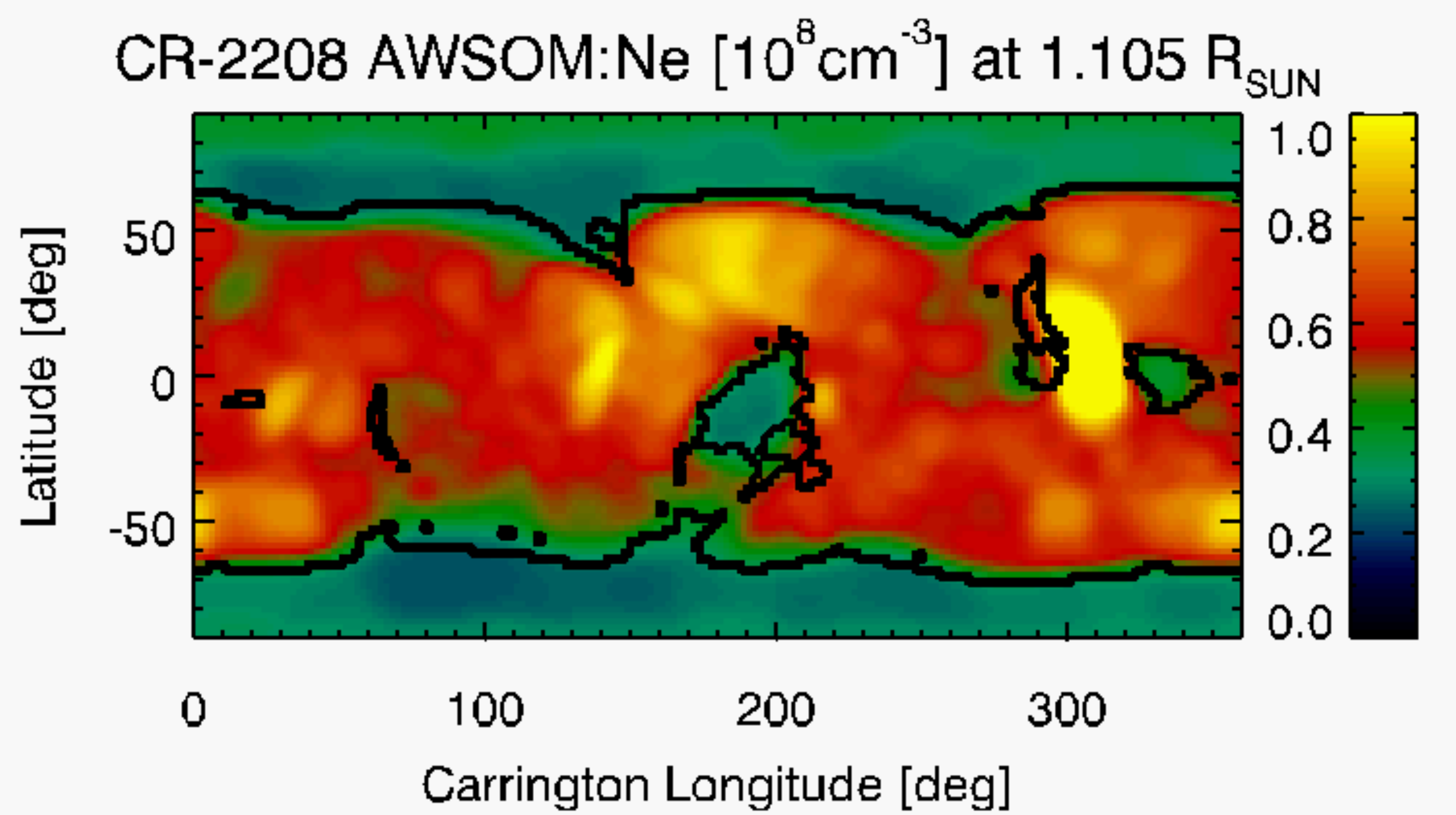}
\includegraphics[width=0.495\textwidth]{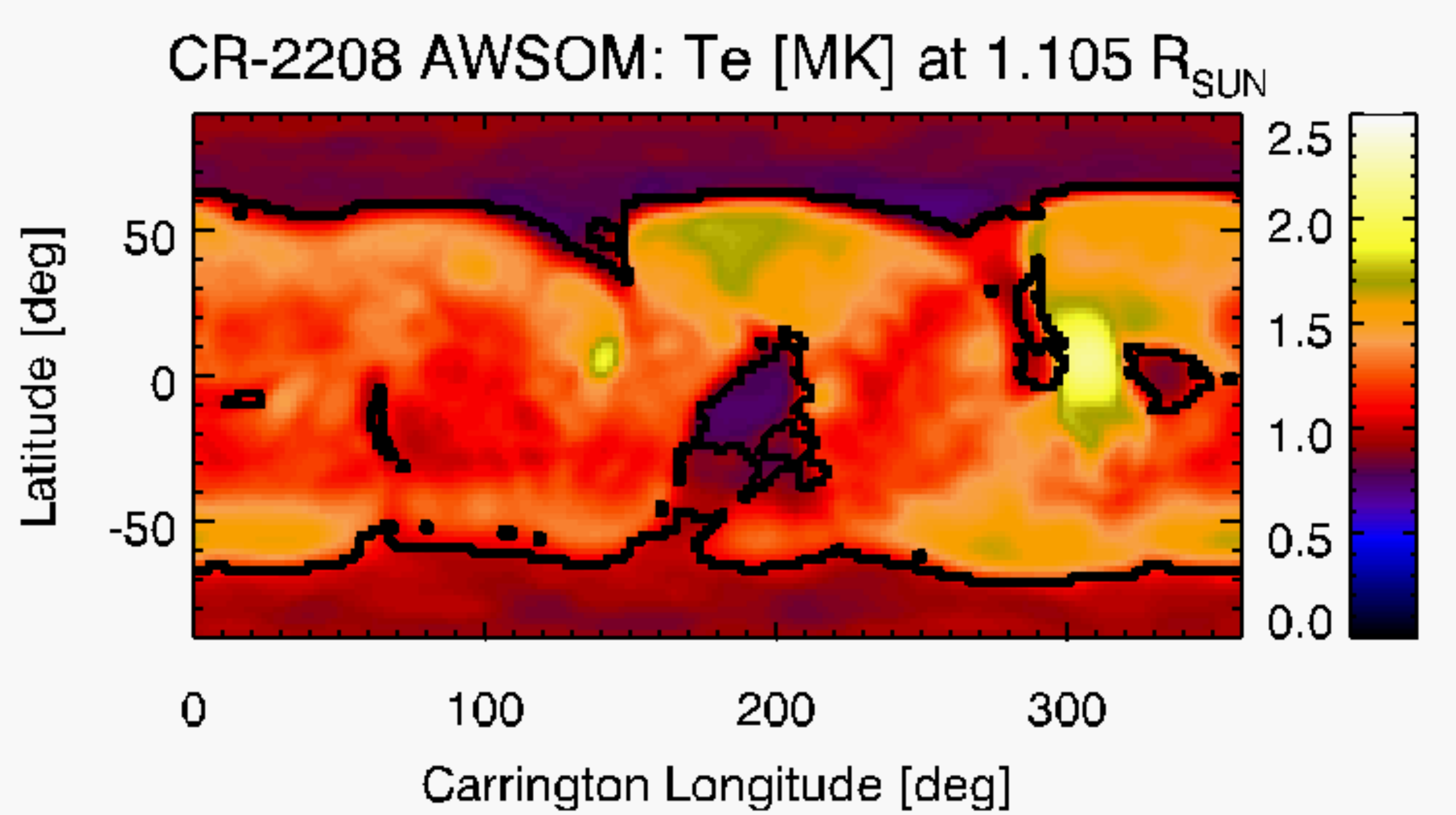}
\caption{Same as Figure \ref{carmaps_awsom_2082} for CR-2208.}
\label{carmaps_awsom_2208}
\end{center}
\end{figure}

{As described in Section \ref{awsom}, the AWSoM model includes an artificially thick TR, achieving coronal conditions above height $\approx 1.06\,\mrsun$. Results for the AWSoM model are shown here above that height. For both target rotations,} Figures \ref{carmaps_awsom_2082} and \ref{carmaps_awsom_2208} show latitude-longitude maps of the AWSoM electron density and temperature. {Maps are shown} at the two largest heights selected for visualization of the DEMT results in Figures \ref{carmaps_demt_2082} and \ref{carmaps_demt_2208}. {Thick-black curves} indicate the magnetic open/closed boundaries based on the magnetic field of the AWSoM model. Visual inspection of these maps shows that the AWSoM model for both rotations is highly {axisymetric}, as the tomographic {results}.

{When compared to DEMT results (Figures \ref{carmaps_demt_2082} and \ref{carmaps_demt_2208}), the latitude-longitude maps of the AWSoM model for heights $1.065$ and $1.105\,\mrsun$ capture well the denser and hotter equatorial streamer belt surrounded by the less dense and colder CHs. Furthermore, for both rotations, the temperature maps show the low latitudes of the equatorial streamer belt to be characterised by lower temperatures than its mid-latitudes, as also seen in the DEMT results. The latitude-longitude maps of the AWSoM and DEMT results are shown in the same units and scales, so that a {visual comparison reveals} similar values of electron density and temperature in both models.}

\begin{figure}[h!]
\begin{center}
\includegraphics[width=0.495\textwidth]{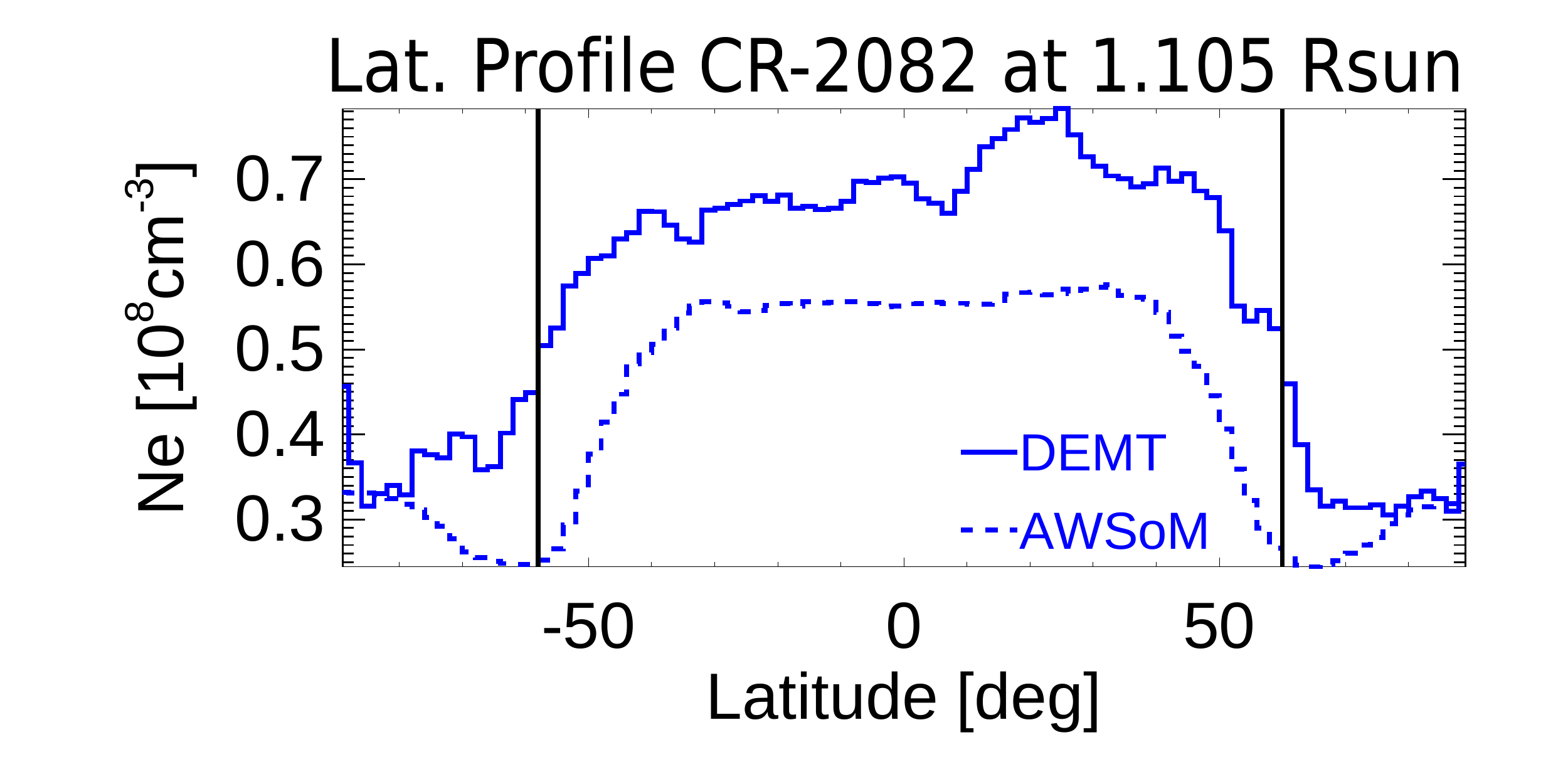}
\includegraphics[width=0.495\textwidth]{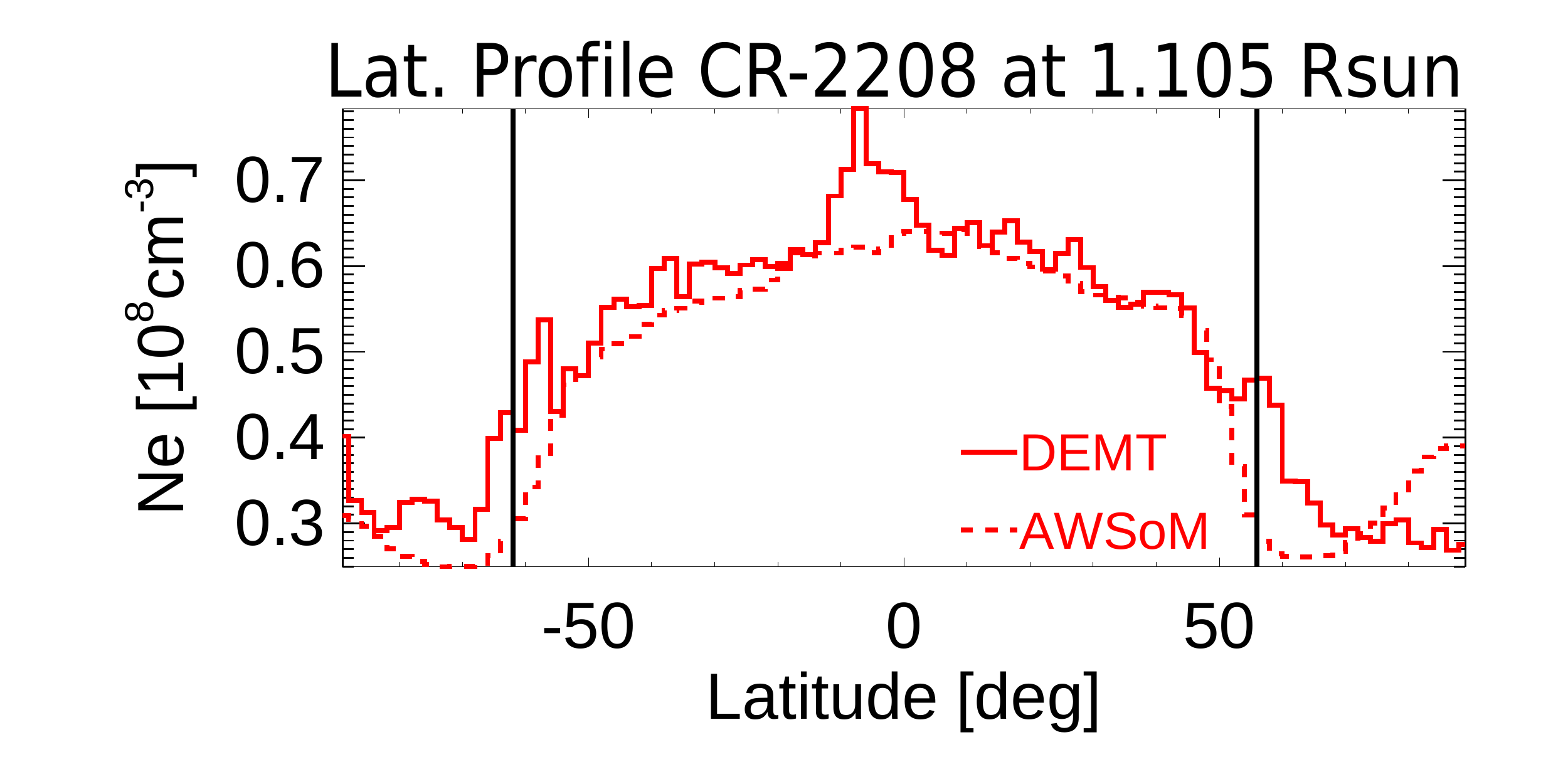}
\includegraphics[width=0.495\textwidth]{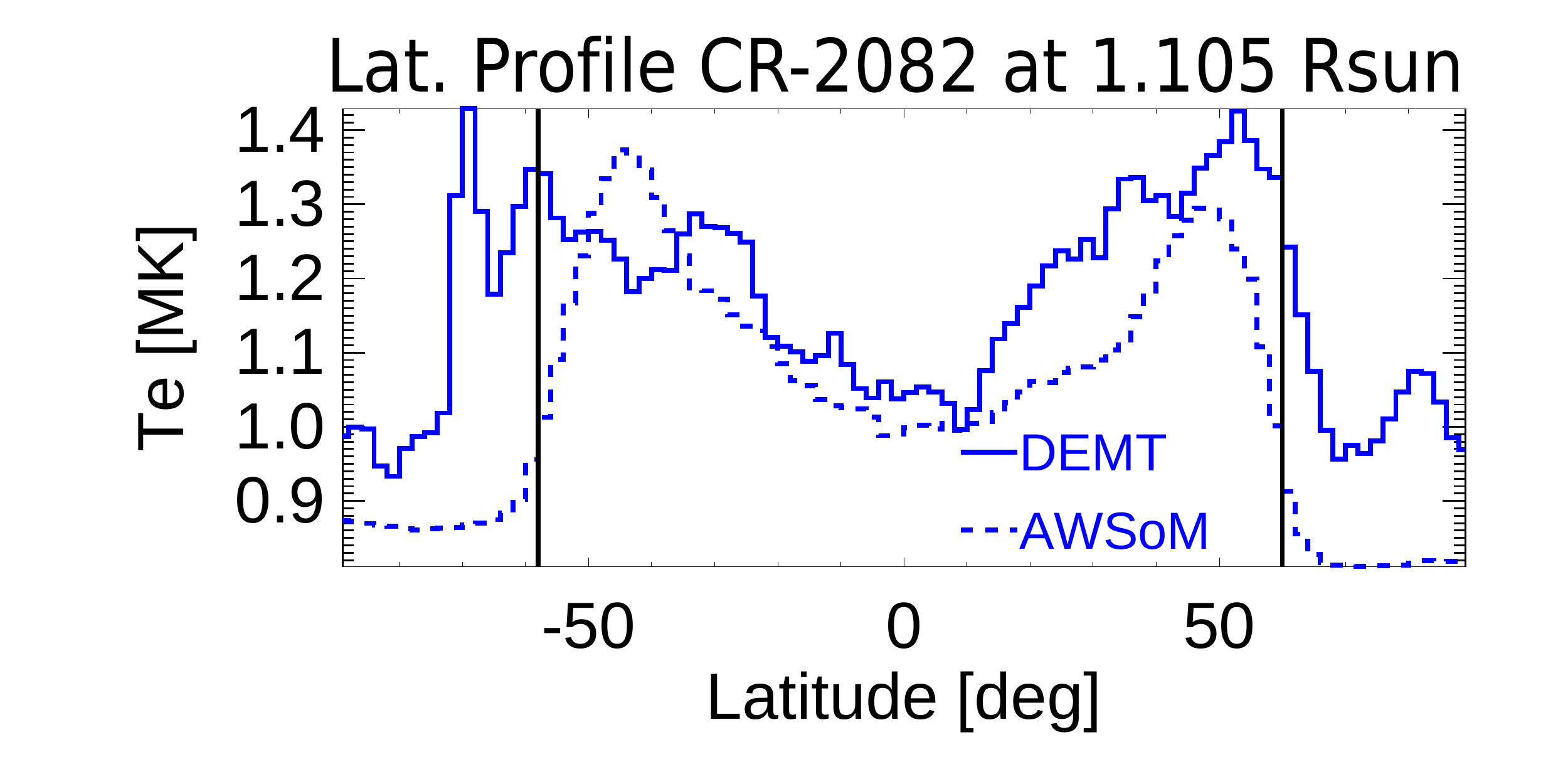}
\includegraphics[width=0.495\textwidth]{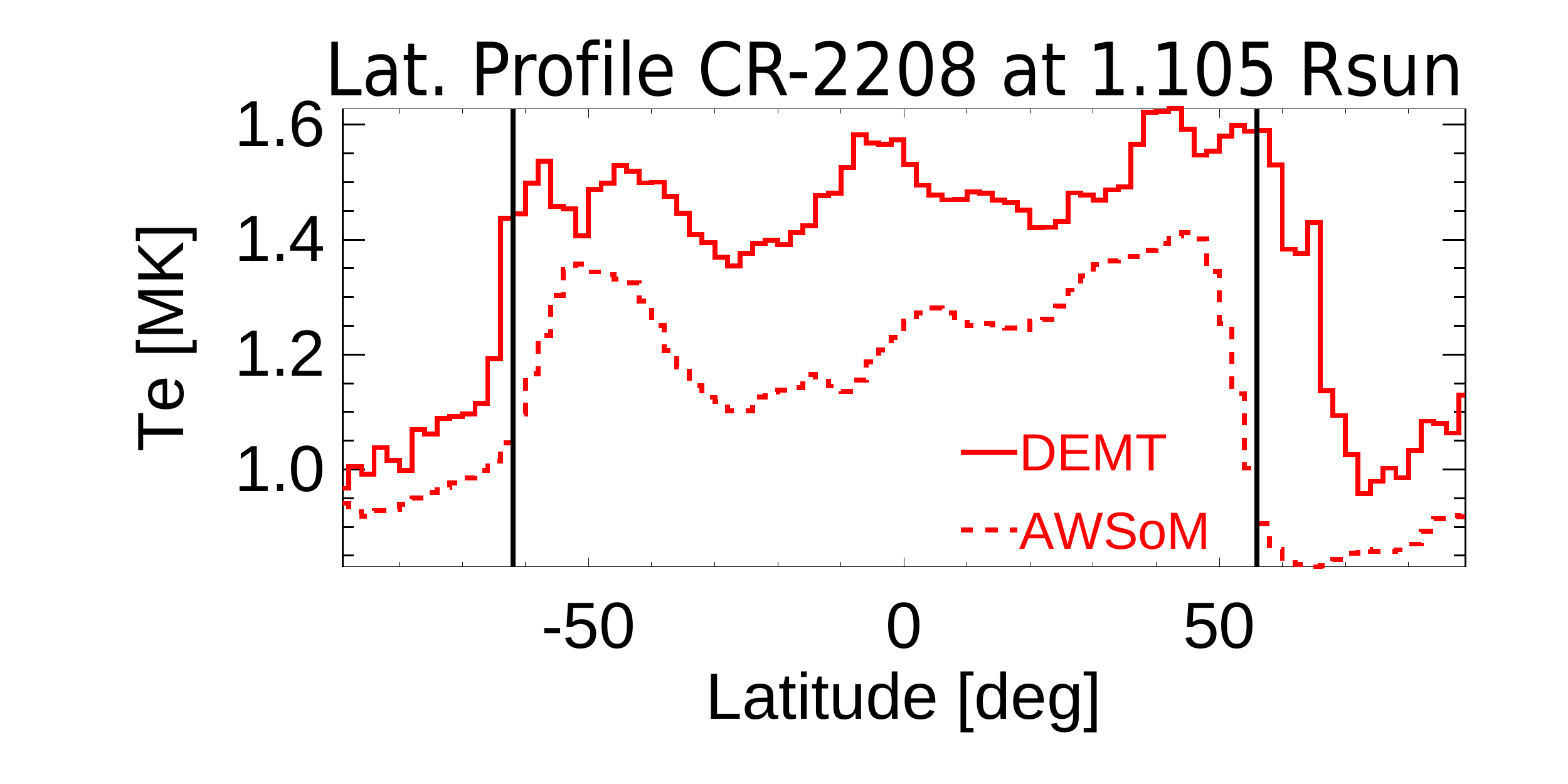}
\includegraphics[width=0.495\textwidth]{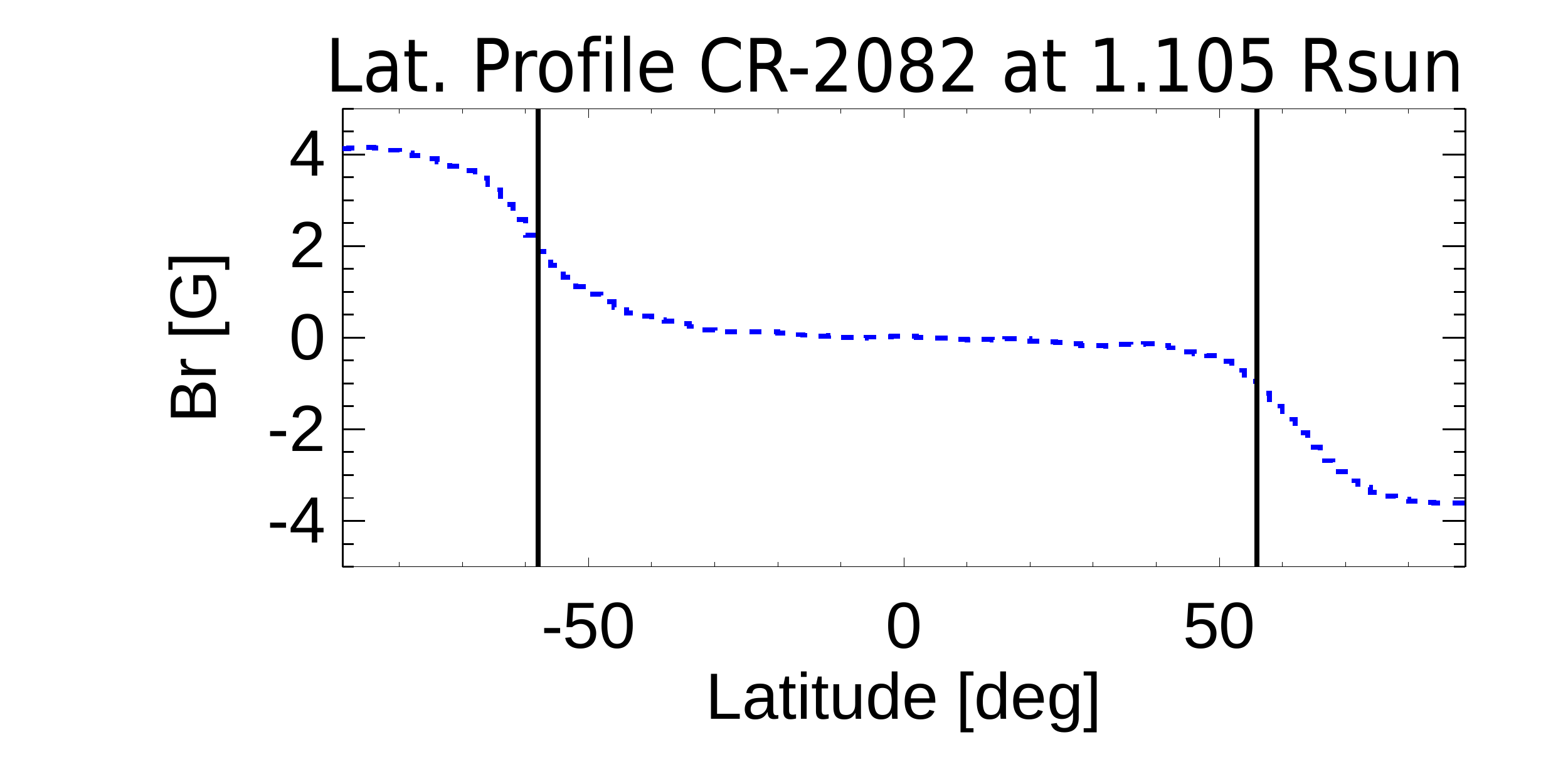}
\includegraphics[width=0.495\textwidth]{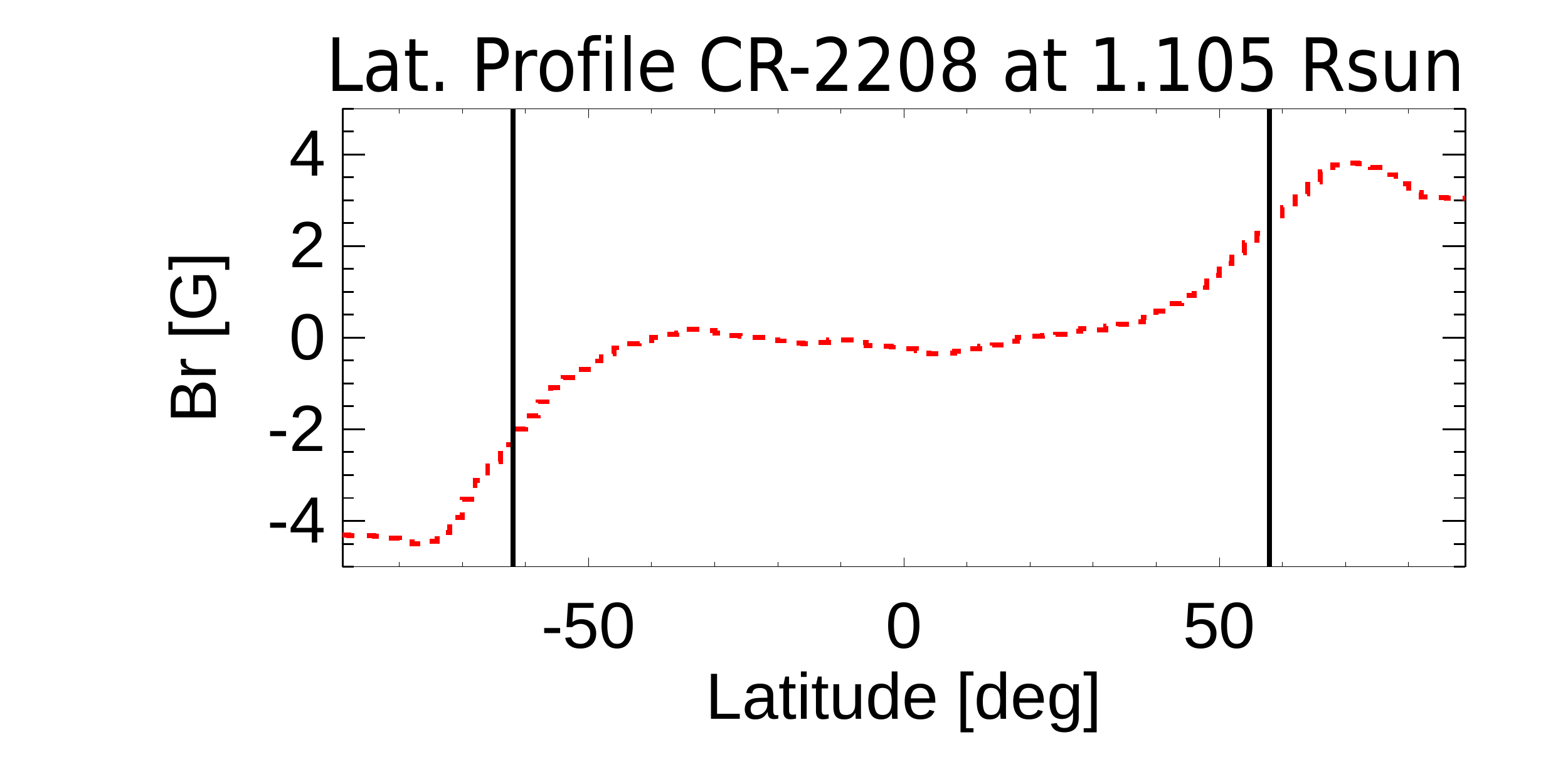}
\caption{Longitude-averaged latitudinal variation of the electron density ({top panels}), {electron temperature ({middle panels})} and {radial magnetic field $B_r$ (bottom panels), for rotations} CR-2082 ({blue color, left panels}) and CR-2208 ({red color, right panels) at} $1.105\,\mrsun$. Dashed and solid lines indicate AWSoM and DEMT results, respectively. Vertical black lines indicate the longitude-averaged latitude of the open/closed magnetic boundary in both hemispheres.}
\label{perf_lat}
\end{center}
\end{figure}

{Being highly axisymmtric rotations, the longitude-averaged latitudinal {profile} of the results of both models is an informative way to compare their large-scale structure. {Such a comparison} is shown in Figure \ref{perf_lat} at height $1.105\,\mrsun$, {with the top} panels comparing electron density and mid panels electron temperature. {The longitude-averaged latitudinal profile of $B_r$ is shown in the bottom panels.} In these longitude-averaged profiles, longitudes containing ARs or low latitude CHs were excluded. In each panel the averaged latitudinal variation for the DEMT model is shown in solid-line style, while the result for the AWSoM model is shown in dashed-linesyle. Left panels show the comparison for CR-2082 (in blue) and right panels for CR-2208 (in red). In each panel the vertical black lines denote the corresponding longitude-averaged latitude of the open-closed boundary in both hemispheres.}

{Several details from Figure \ref{perf_lat} are {worth highlighted}. Firstly, at most latitudes the overall agreement of the electron density of both models is within $\approx 20\%$ for CR-2082, and $\approx 5\%$ for CR-2208. The noticeable exception is to be found near the open/closed boundaries of both target rotations, where the disagreement between both models can be up to twice as much. In the case of the electron temperature, for both rotations the models agree within $\approx 15\%$ at all latitudes. Secondly, for both rotations, and for both models, these plots clearly show the relatively lower temperatures characterizing the low-latitudes of the equatorial streamer belt compared to its mid-latitudes. Thirdly, for both rotations, the latitude of the open/closed magnetic boundary in both hemispheres matches the location of the strongest latitudinal gradient of the DEMT electron density. Note this is not the case for the AWSoM model, that shows a minimum density at the open/closed boundary. {Lastly, the DEMT} electron density decreases from the open/closed boundary towards the poles (in both hemispheres of the two rotations), while the AWSoM model shows the opposite trend. {For comparison, {$B_r$ in the CHs increases} from the open/closed boundary towards the poles for CR-2082, while showing local maxima around latitudes $-75\mdeg$ and $+70\mdeg$ in the case of CR-2208.}}

\begin{figure}[h!]
\begin{center}
\includegraphics[width=0.495\textwidth,clip=]{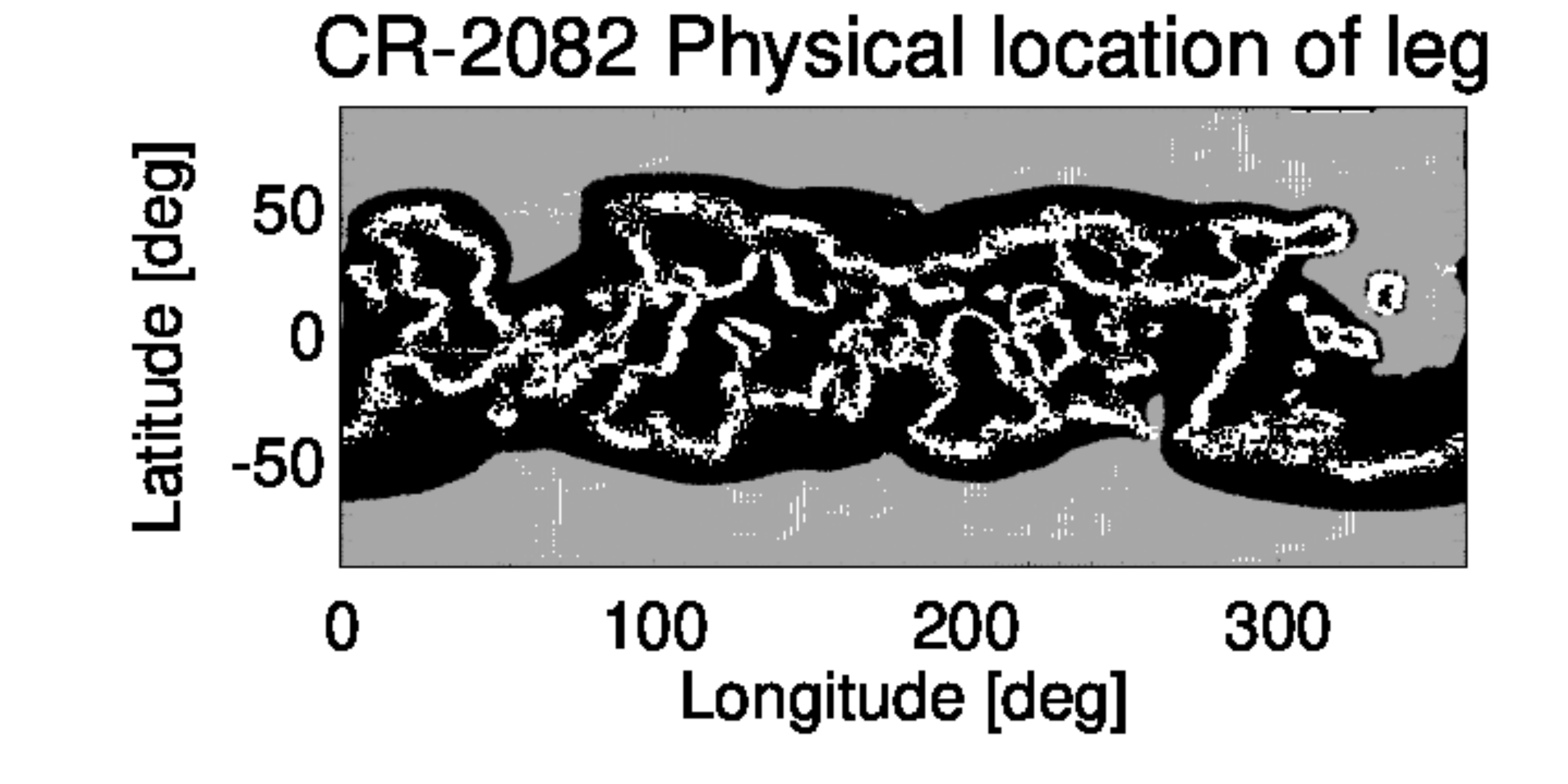}
\includegraphics[width=0.495\textwidth,clip=]{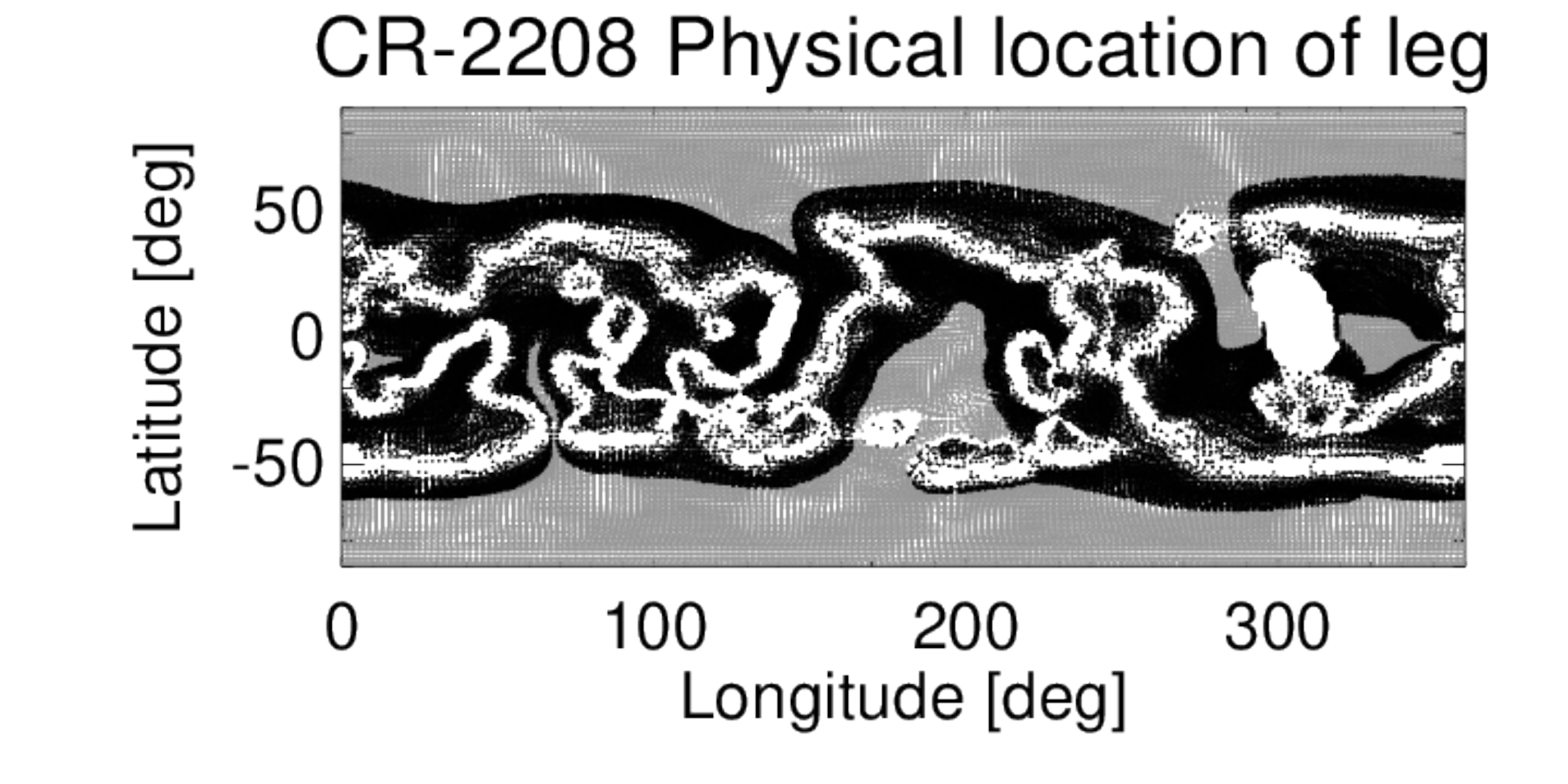}
\includegraphics[width=0.495\textwidth,clip=]{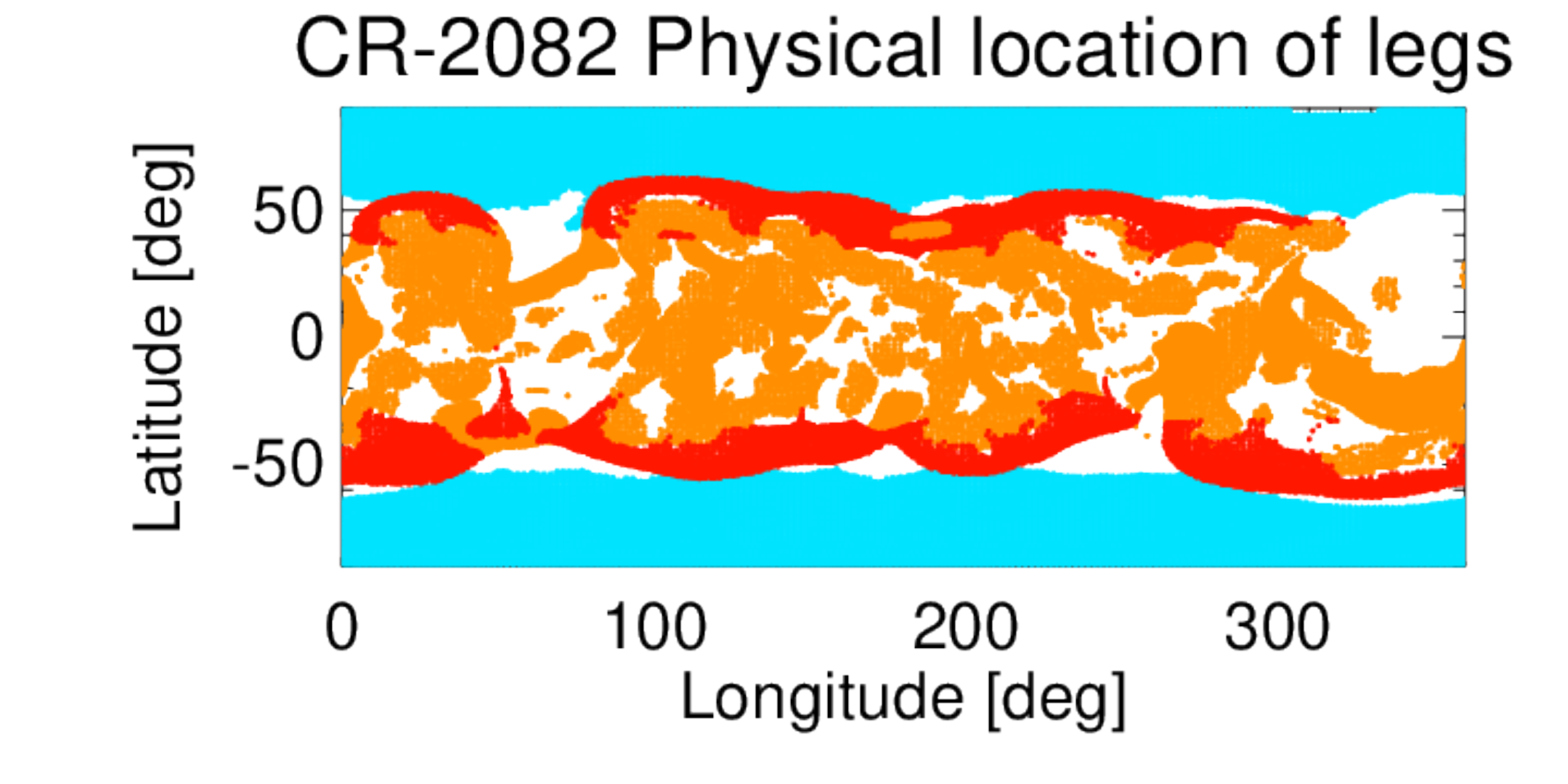}
\includegraphics[width=0.495\textwidth,clip=]{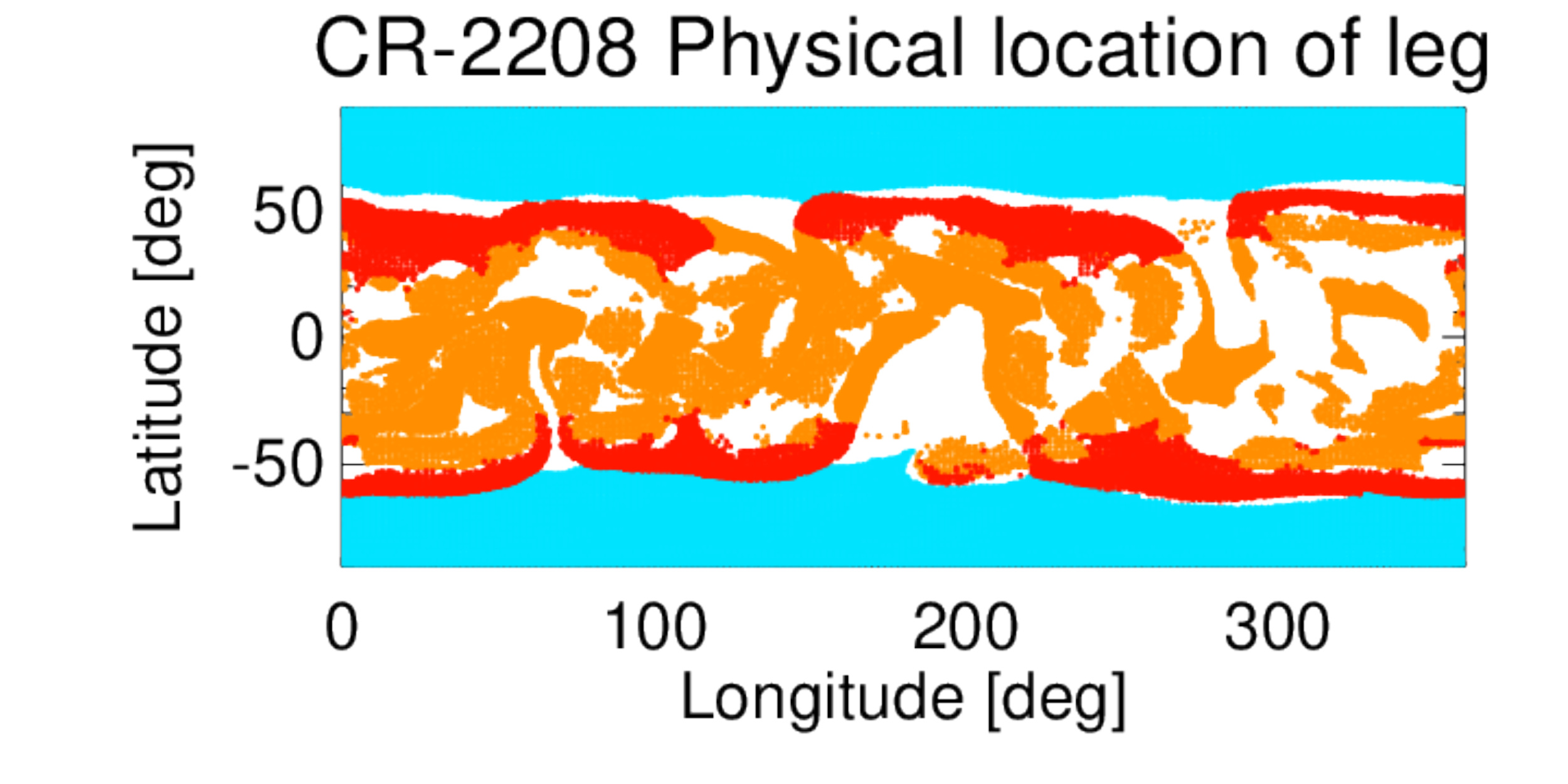}
\caption{Same as Figure \ref{rpoint_demt}, but using the density and temperature of the AWSoM model to classify {its legs in types I, II and III. The model does not exhibit legs of type 0.}}
\label{rpoint_awsom}
\end{center}
\end{figure} 

To characterize the results of the AWSoM model in distinct magnetic structures, its results for electron density and temperature were traced along its magnetic field lines. For each field line leg, the results were then fit to Equations \ref{Nfit} and \ref{Tfit}{, considering only data points above heliocentric height $1.055\,\mrsun$. We then classified the traced legs into types I, II and III, according to the criteria described in Section \ref{trace}.  Legs of type 0 are not included for AWSoM as it currently can not simulate down loops, as discussed in Section \ref{discu}.}

\begin{figure}[h!]
\begin{center}
\includegraphics[width=0.31\textwidth,clip=]{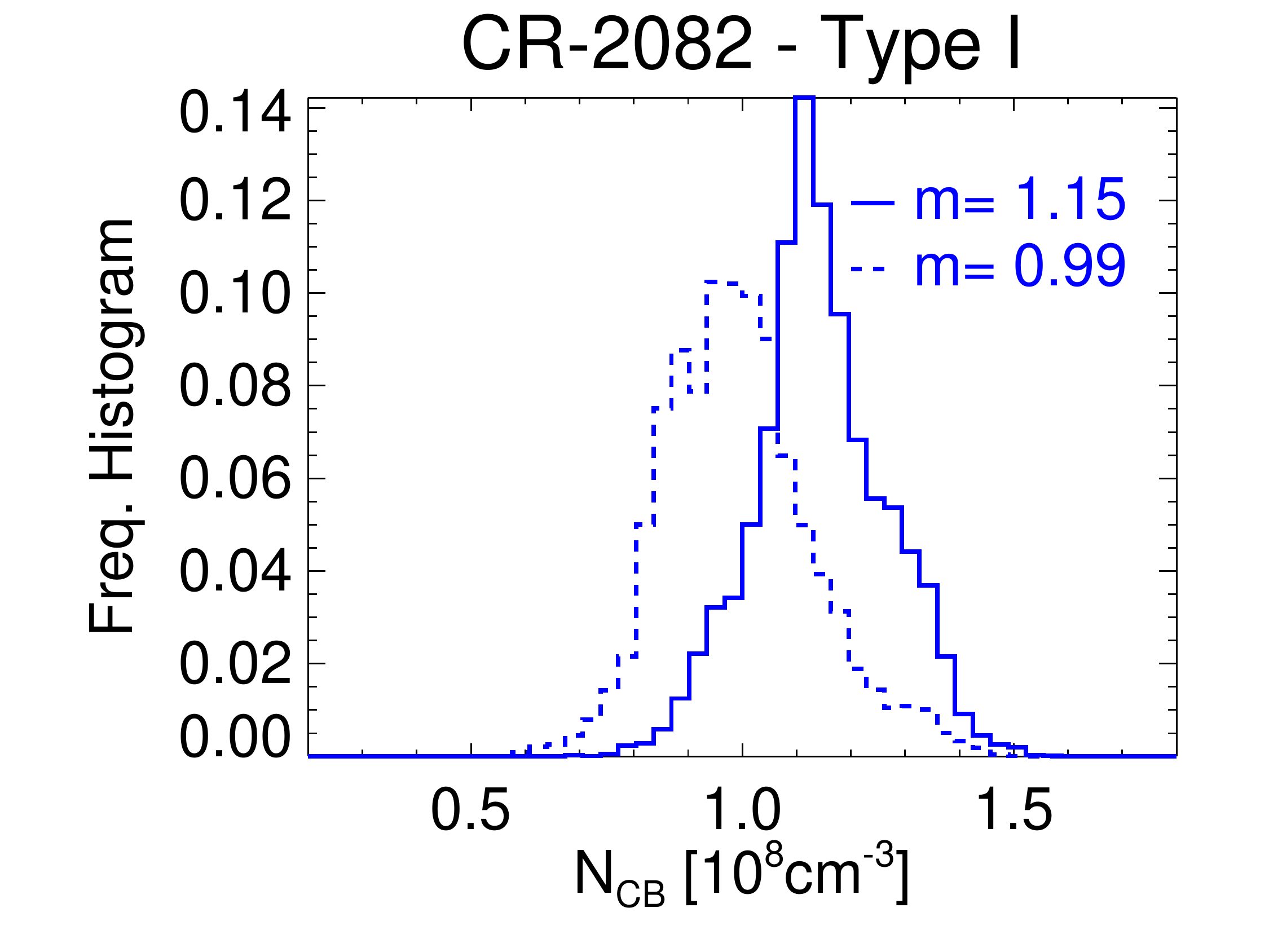}
\includegraphics[width=0.31\textwidth,clip=]{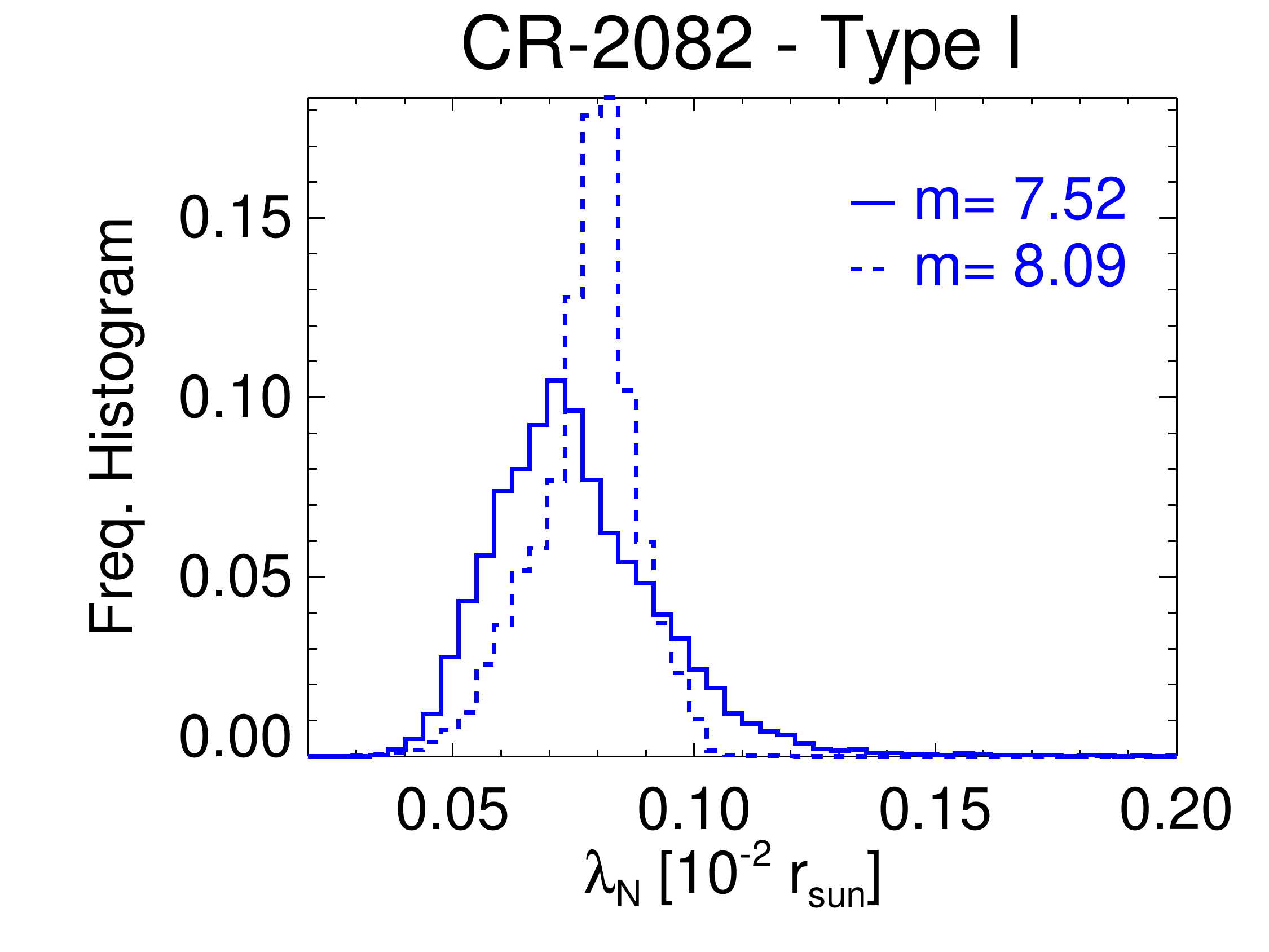}
\includegraphics[width=0.31\textwidth,clip=]{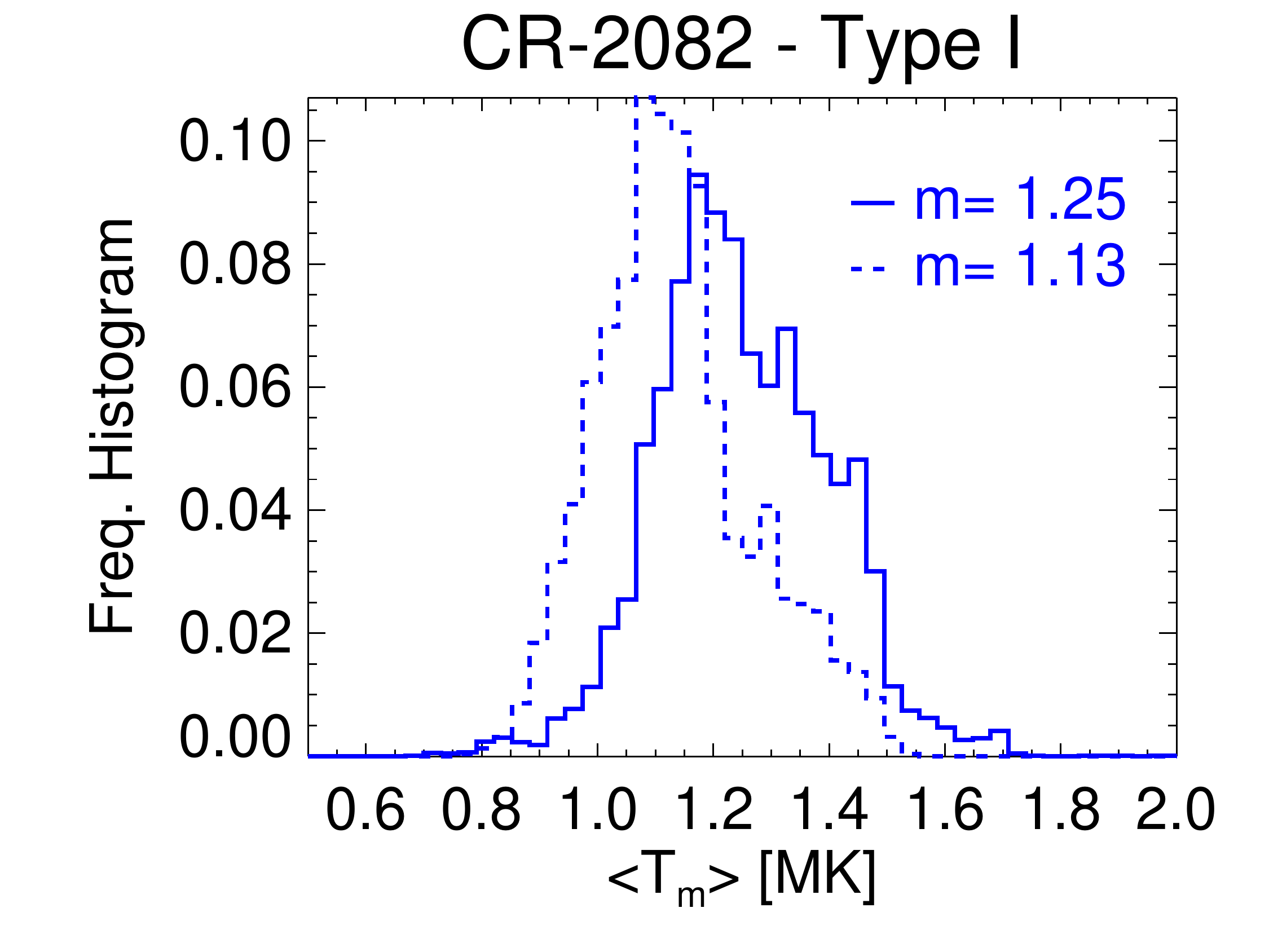}
\includegraphics[width=0.31\textwidth,clip=]{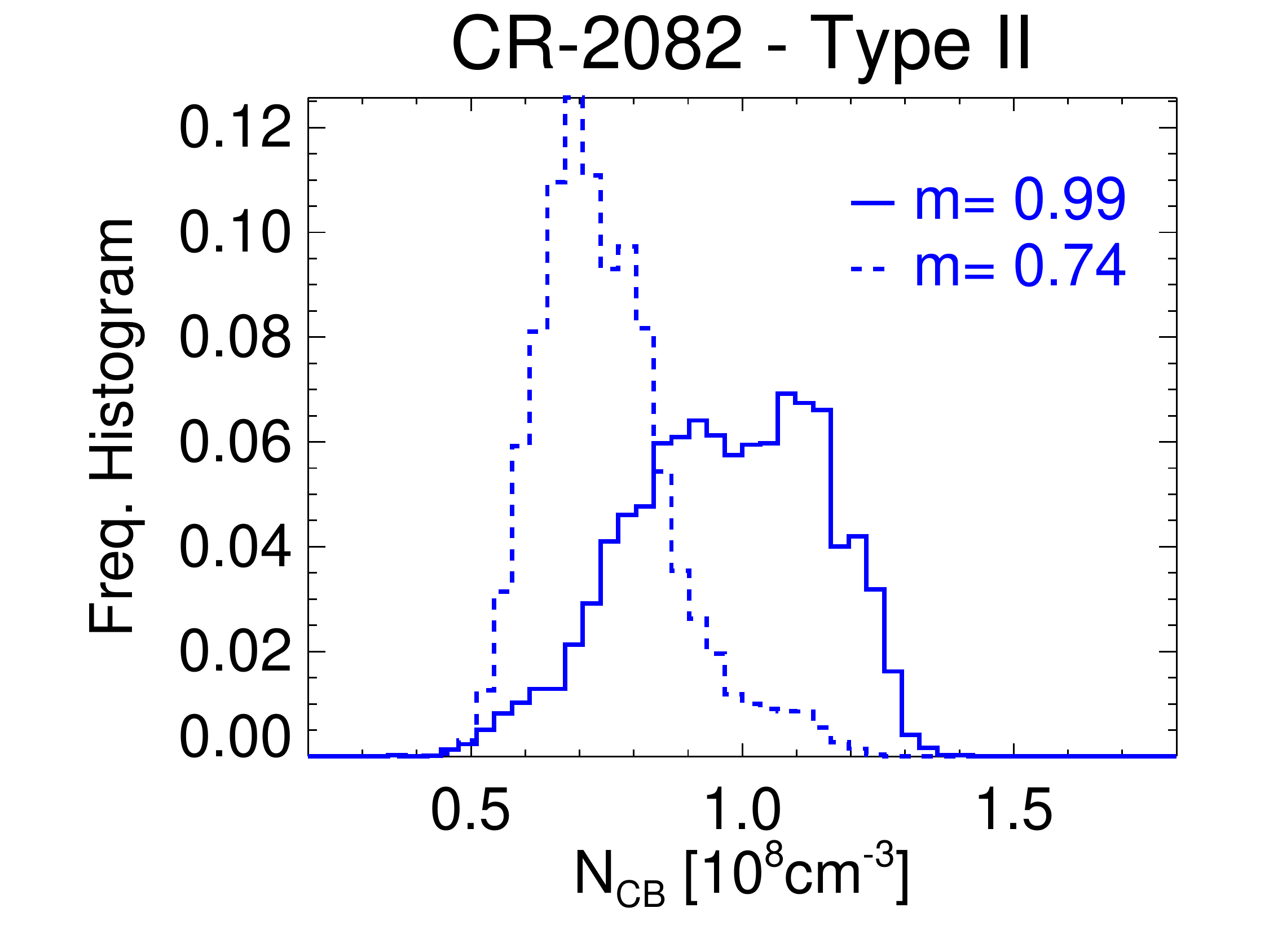}
\includegraphics[width=0.31\textwidth,clip=]{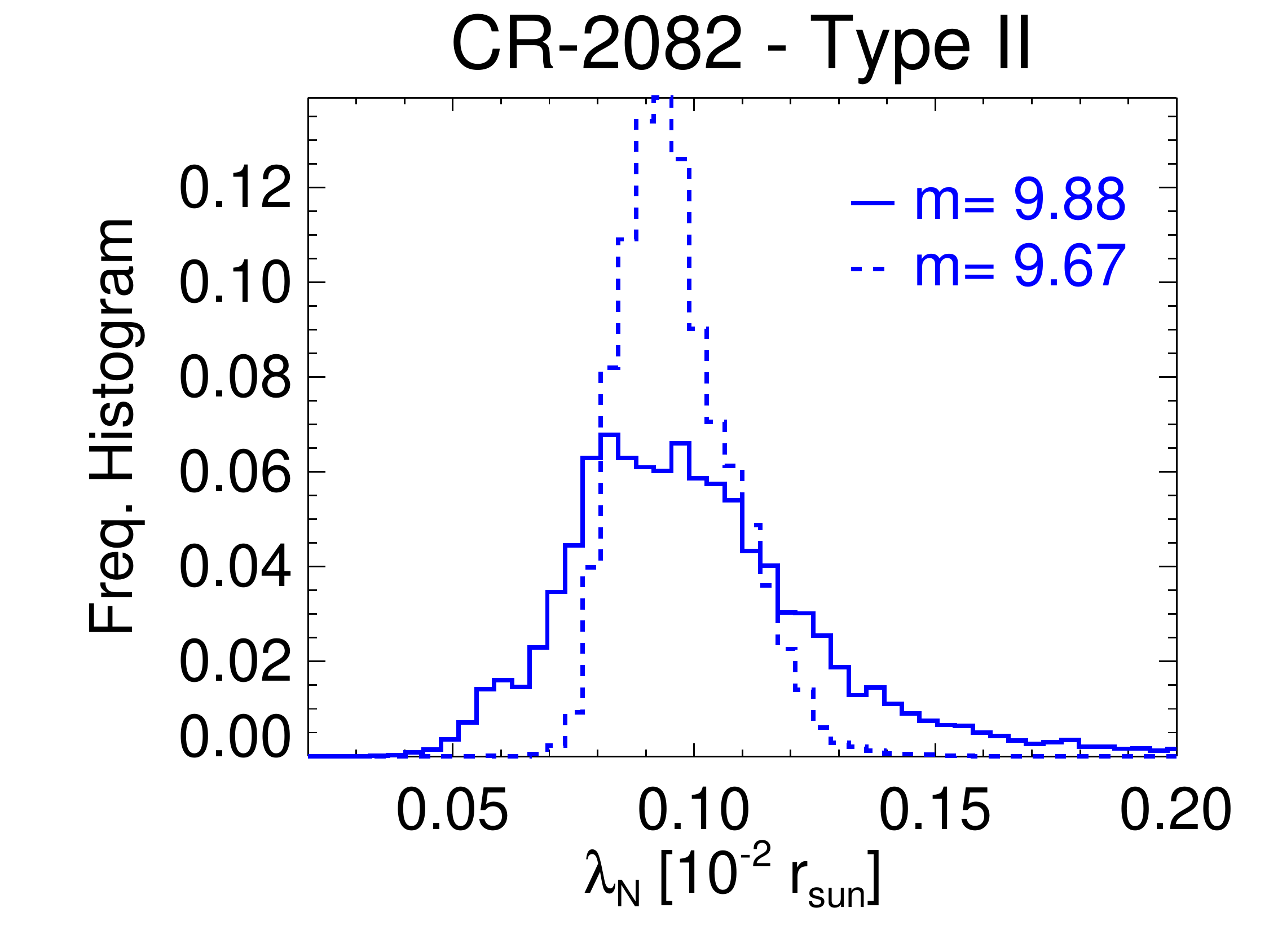}
\includegraphics[width=0.31\textwidth,clip=]{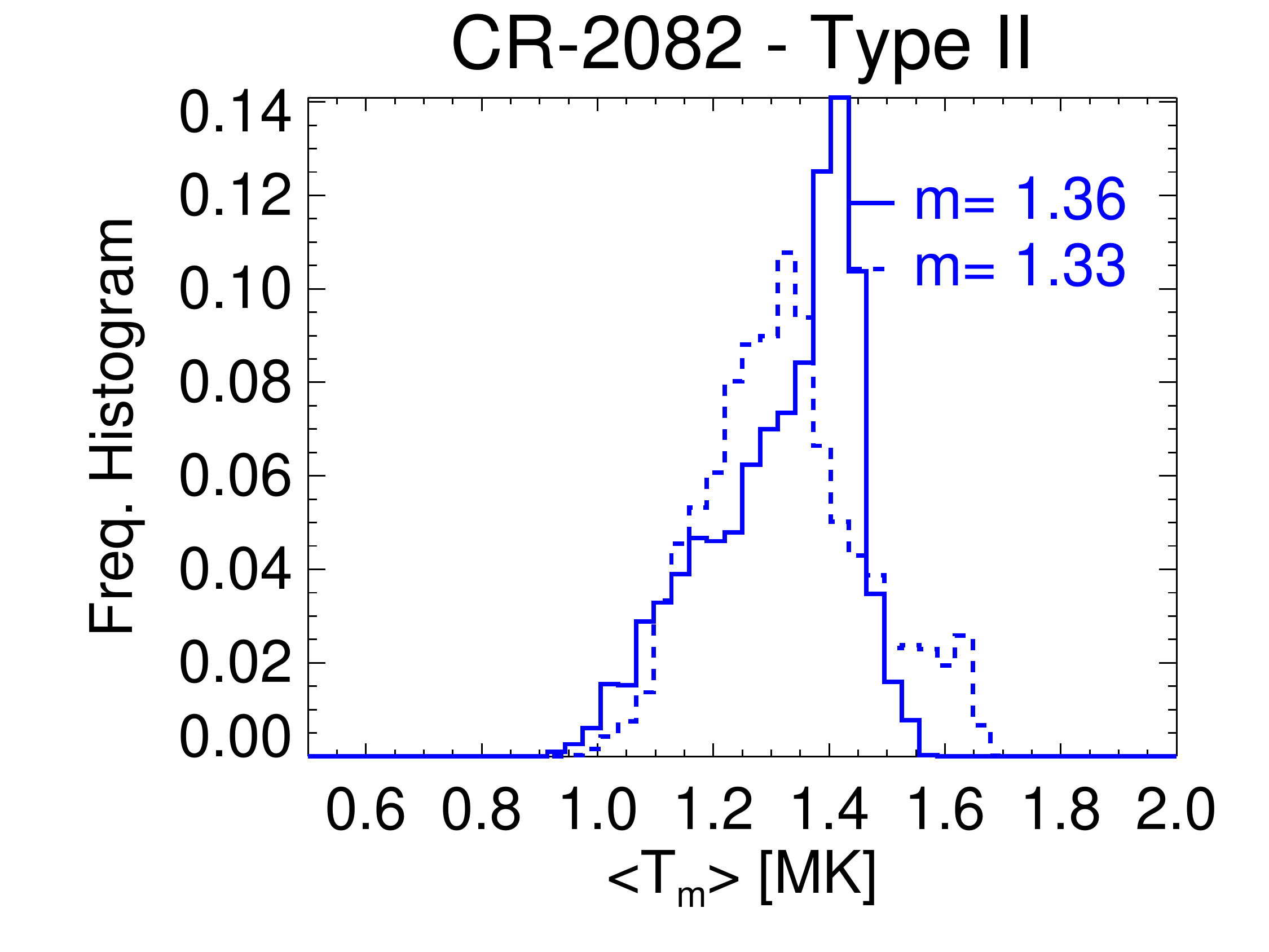}
\includegraphics[width=0.31\textwidth,clip=]{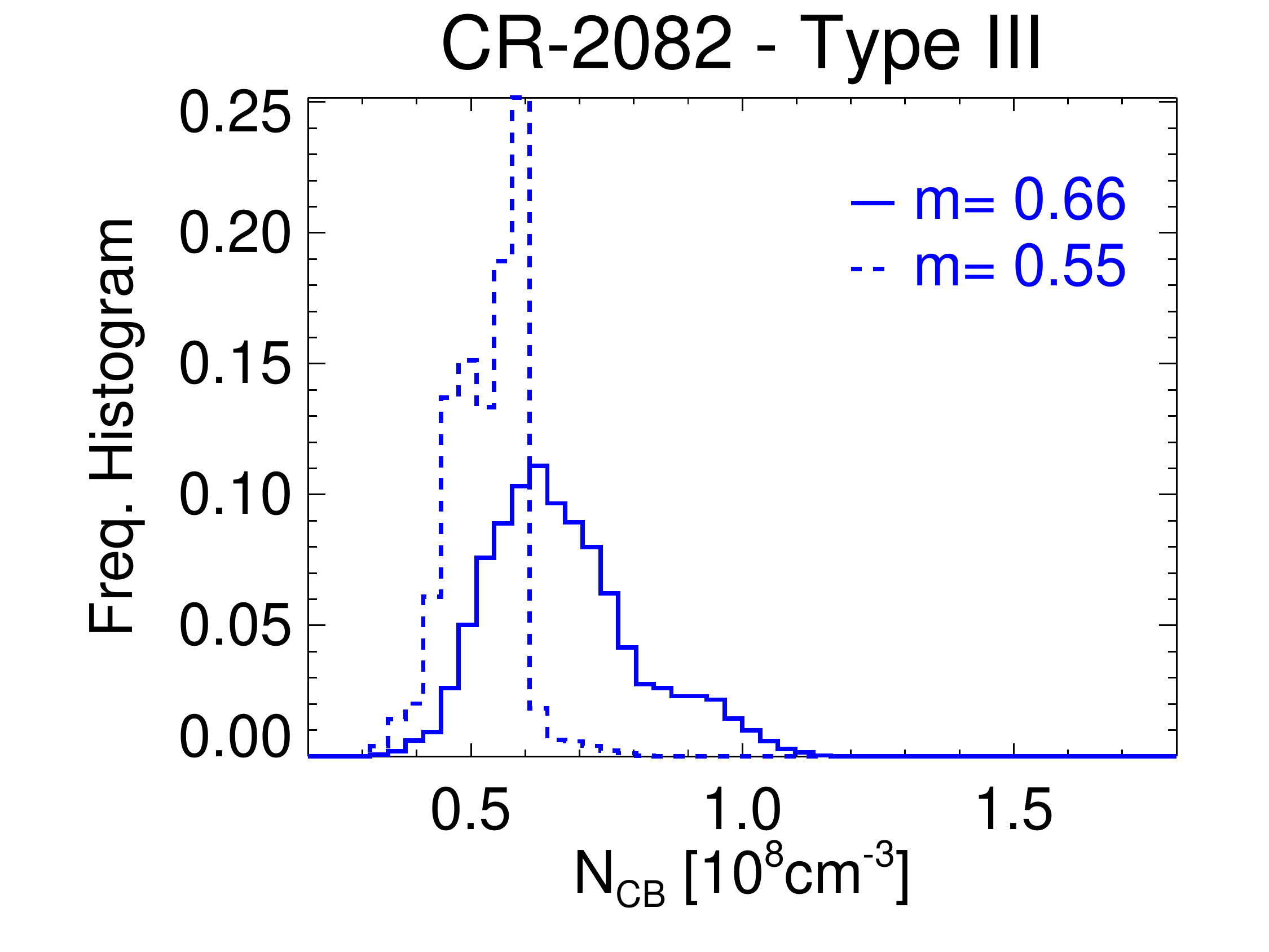}
\includegraphics[width=0.31\textwidth,clip=]{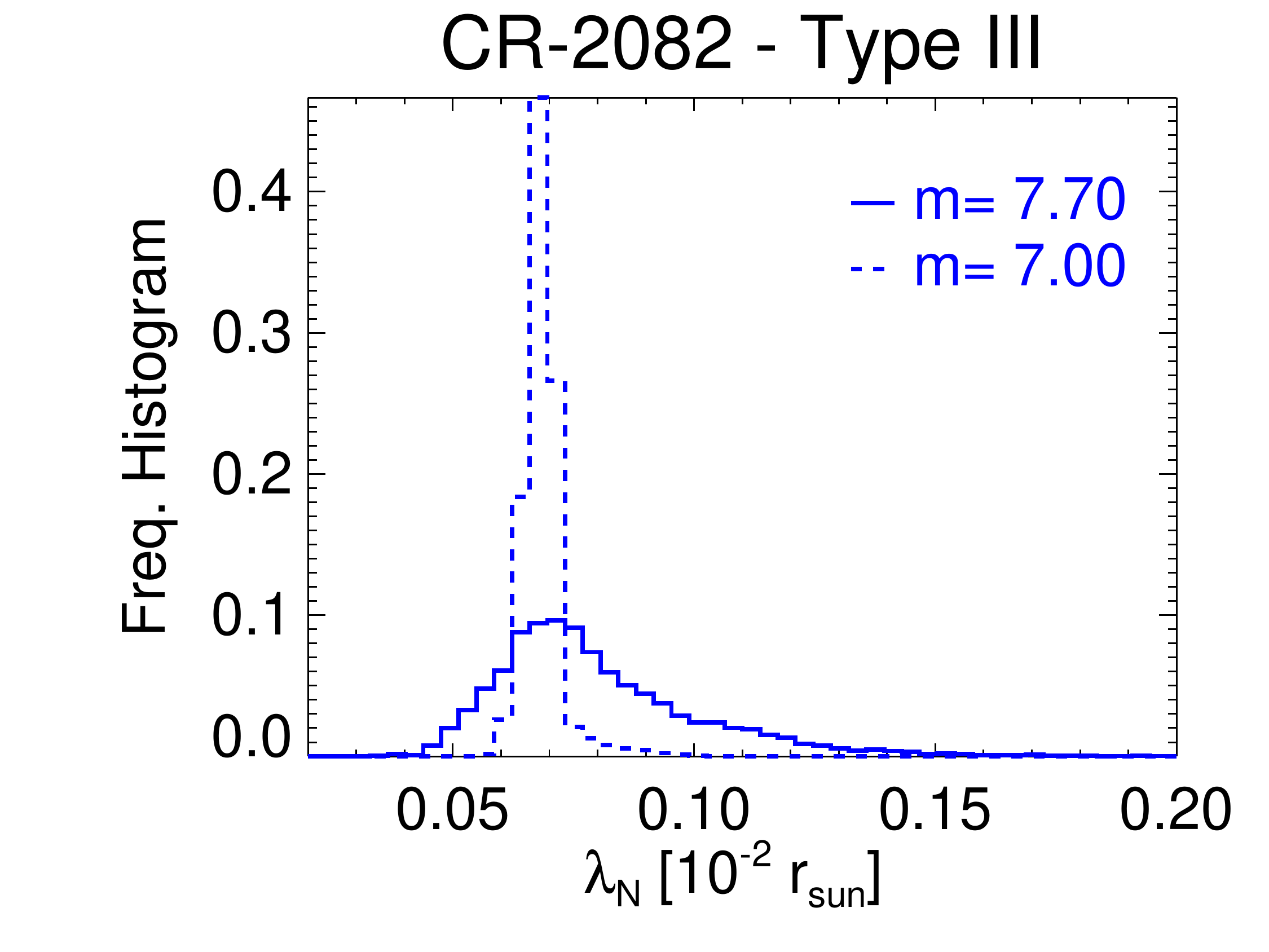}
\includegraphics[width=0.31\textwidth,clip=]{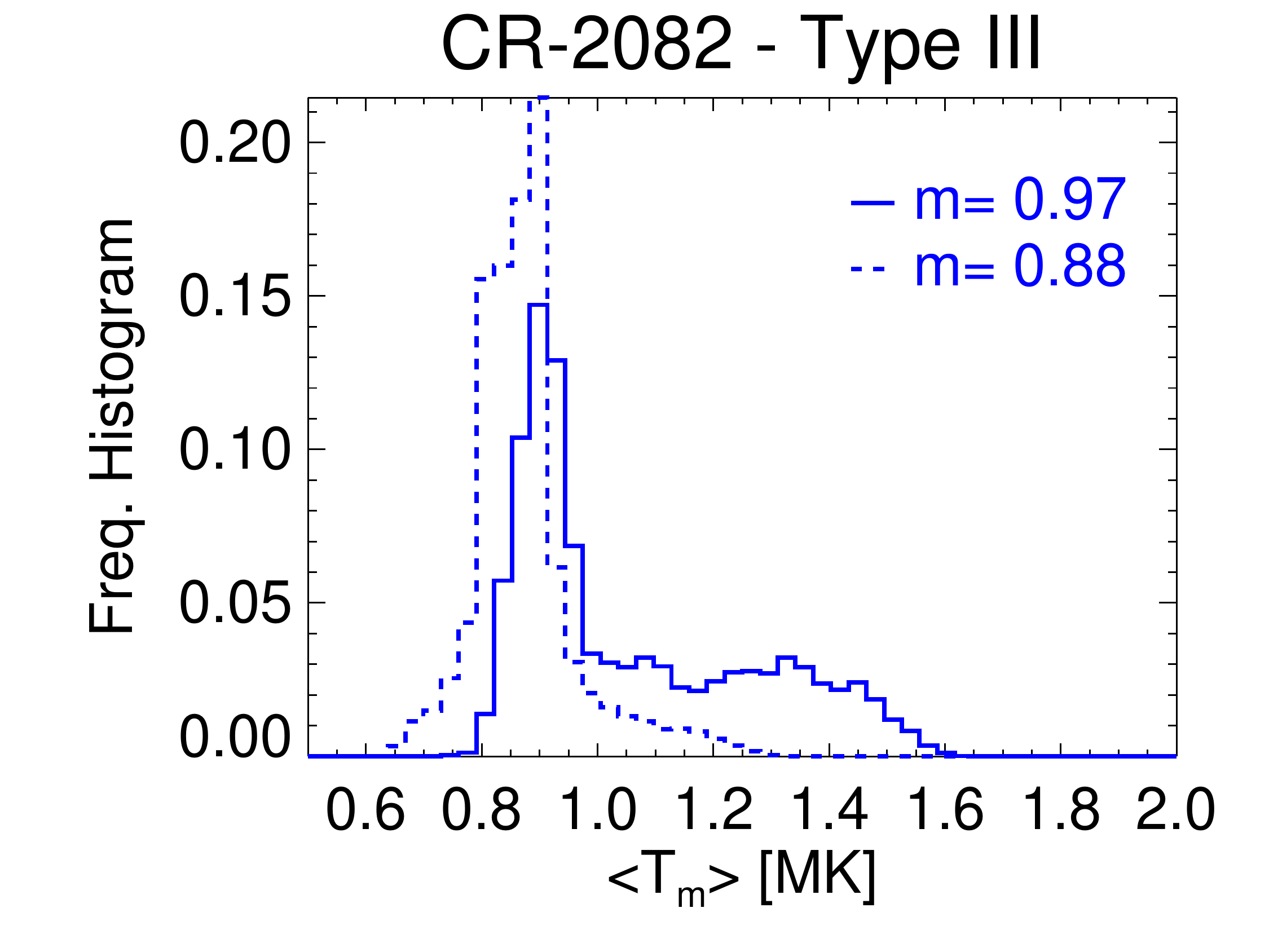}
\caption{{Statistical distribution {of the results of the DEMT (solid line-style) and AWSoM (dashed-line style) models} traced along legs of type I, II and III (from top to bottom), as defined in {Section \ref{trace}.} From left to right: electron density at the lowest coronal height of the AWSoM model $\Ne(r=1.055\,\mrsun)$, electron density scale height $\lN$, and leg-averaged electron temperature $\aTm$. In each panel the median {values $m$ are} indicated.}}
\label{histos_2082}
\end{center}
\end{figure} 

{For both target rotations, the top panels of Figure \ref{rpoint_awsom} show the latitude-longitude location (at heliocentric height $1.105\,\mrsun$) of all traced field line legs for which criterion (i) of Section \ref{trace} is met. That criterion is adapted here, requiring that at least five voxels of the tomographic grid are threaded by the leg. Open legs are indicated in gray color and closed ones in black color. For each leg, the fits to tomographic temperature and density were applied, as given by Equations \ref{Nfit} and \ref{Tfit}. Considering the AWSoM data points and the resulting fits along each leg, the bottom panels of Figure \ref{rpoint_awsom} show the latitude-longitude location of the subset for which also both criteria (ii) and (iii) of Section \ref{trace} are met. Using a three-color code, type I, II and III legs are shown in red, magenta and cyan color, respectively. This figure {can be compared} with the corresponding Figure \ref{rpoint_demt} for DEMT results. {The AWSoM maps are more densely populated} than those of DEMT. This is due to the 3D MHD model having {having spatially smoother distributions of electron density and temperature than those of DEMT}.}

\begin{figure}[h!]
\begin{center}
\includegraphics[width=0.31\textwidth,clip=]{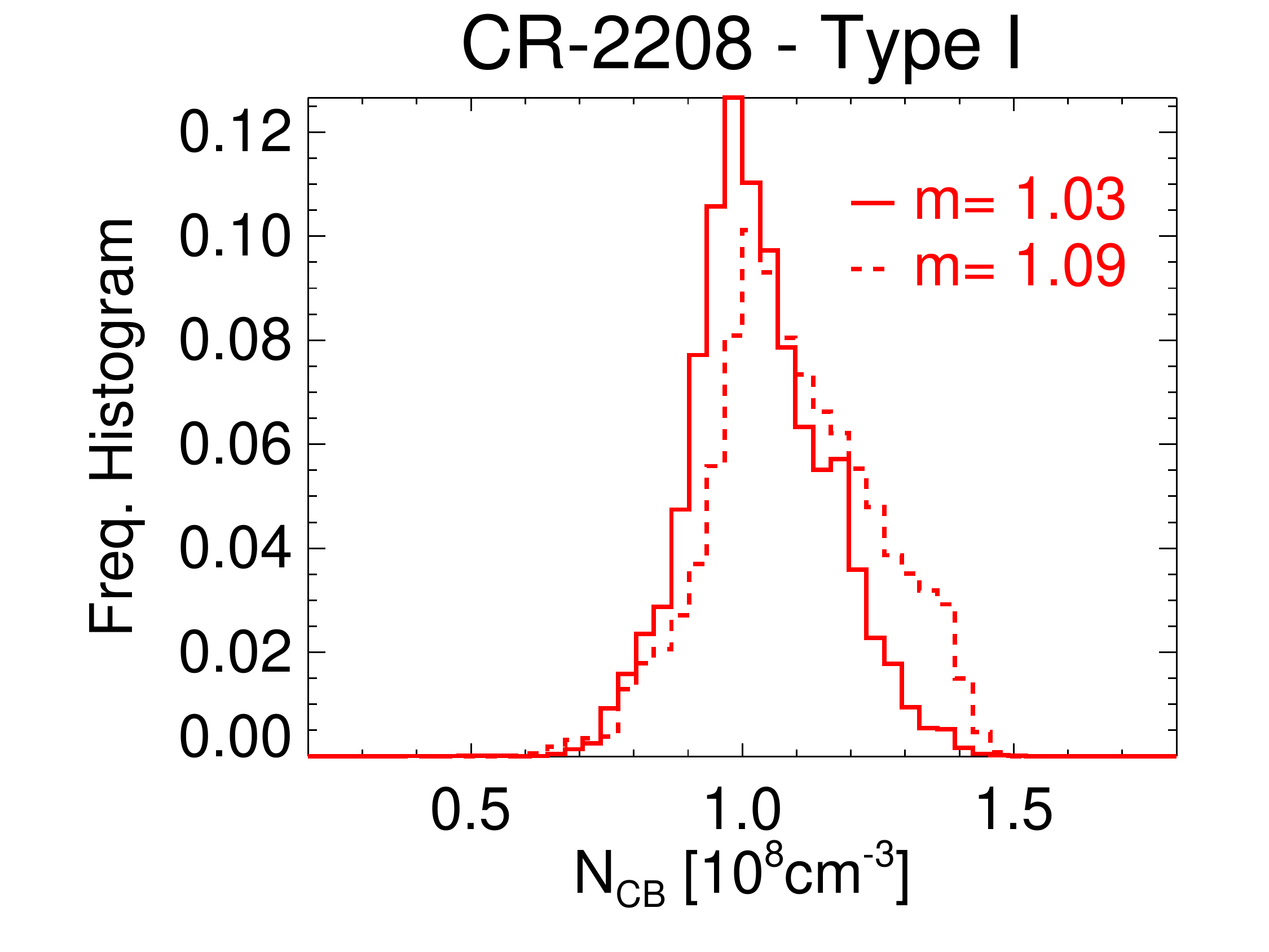}
\includegraphics[width=0.31\textwidth,clip=]{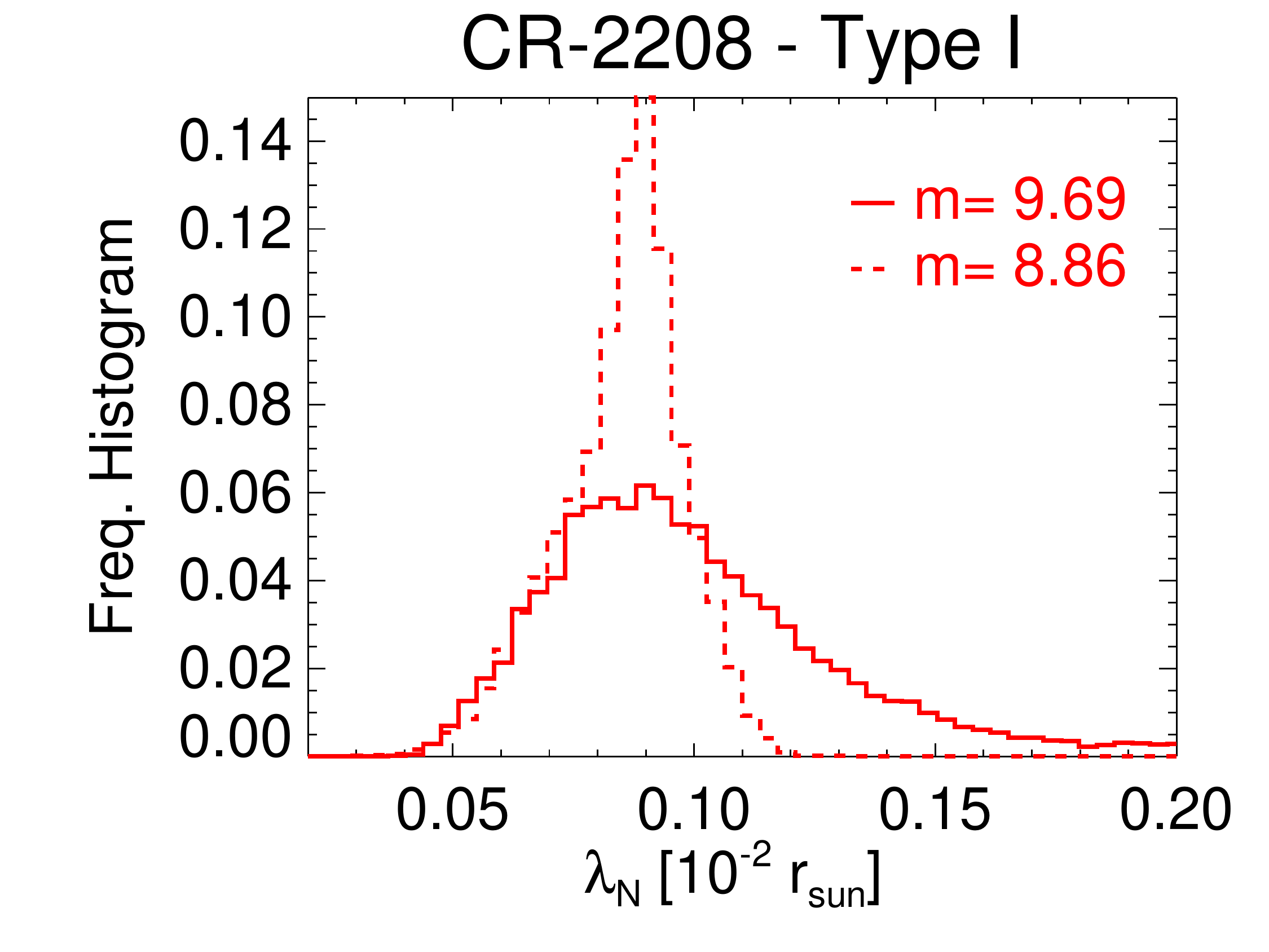}
\includegraphics[width=0.31\textwidth,clip=]{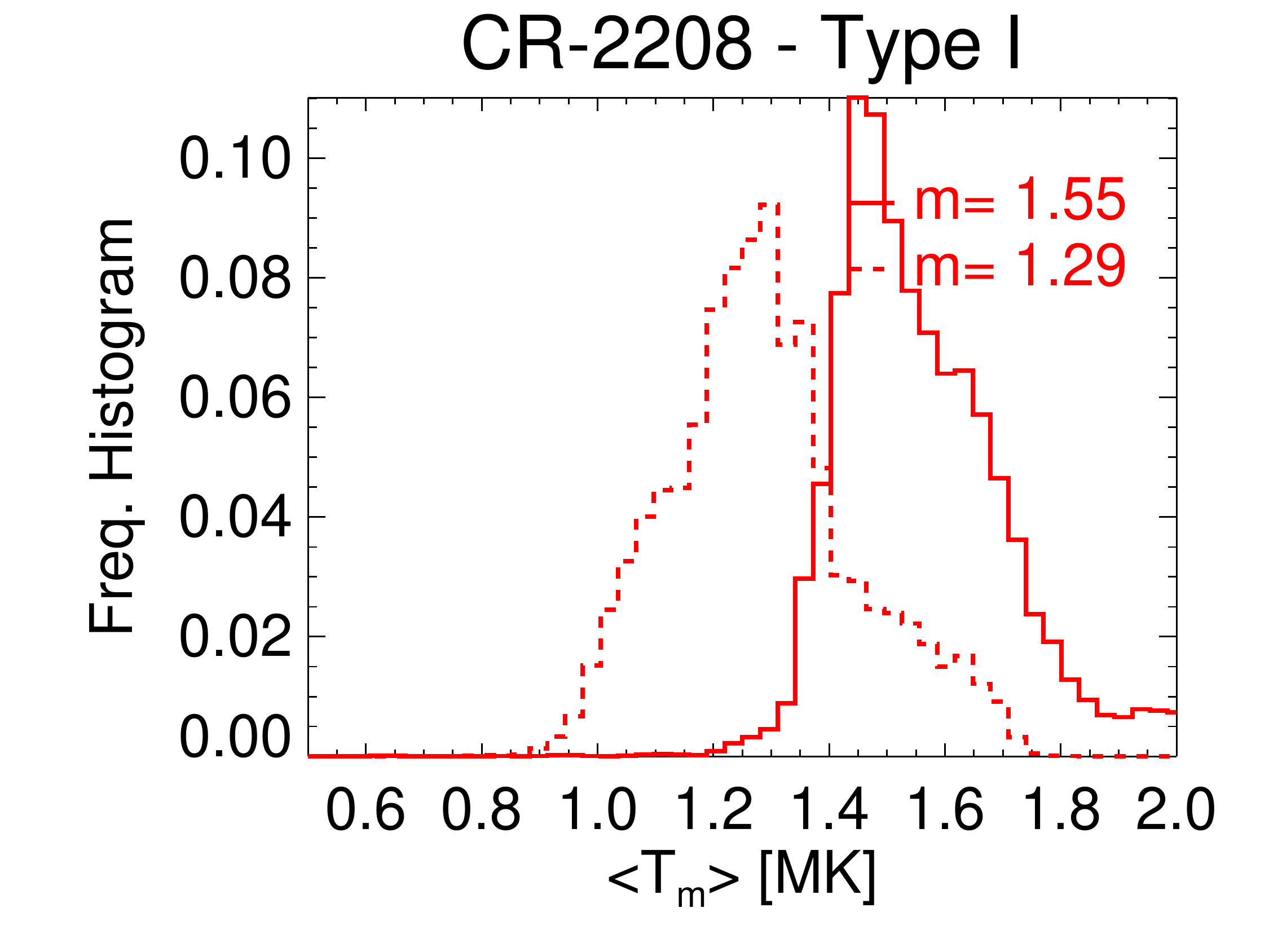}\\
\includegraphics[width=0.31\textwidth,clip=]{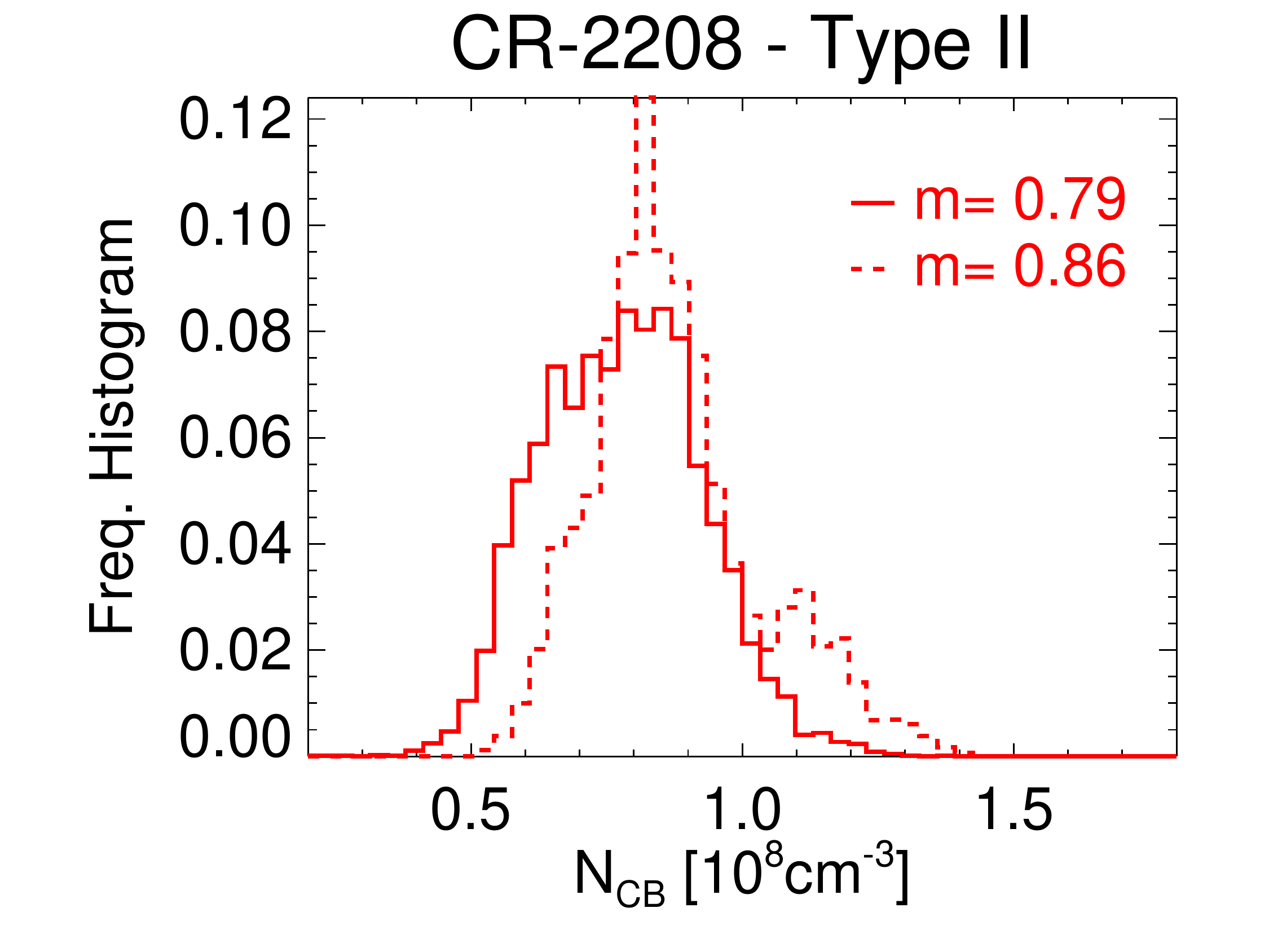}
\includegraphics[width=0.31\textwidth,clip=]{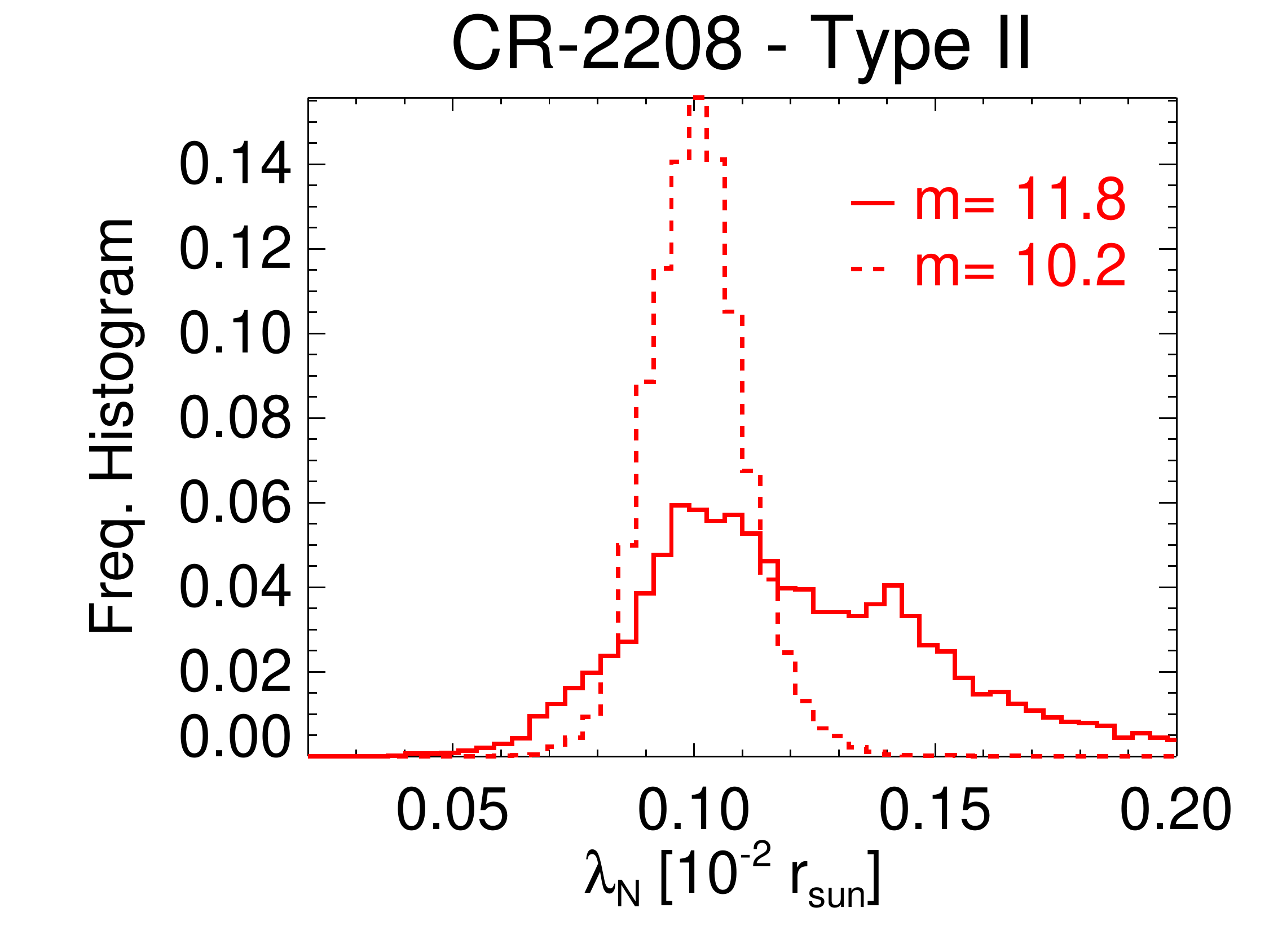}
\includegraphics[width=0.31\textwidth,clip=]{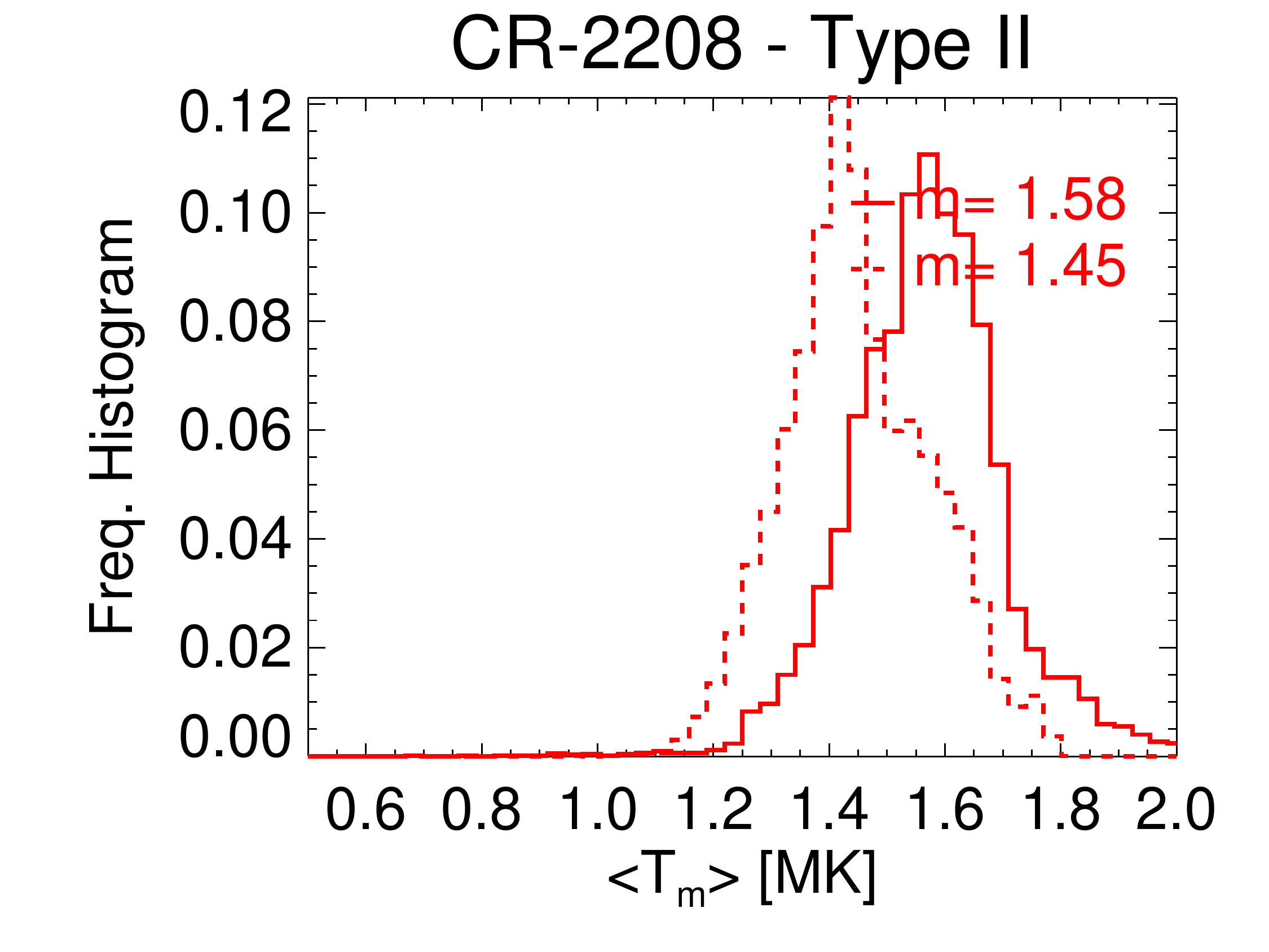}\\
\includegraphics[width=0.31\textwidth,clip=]{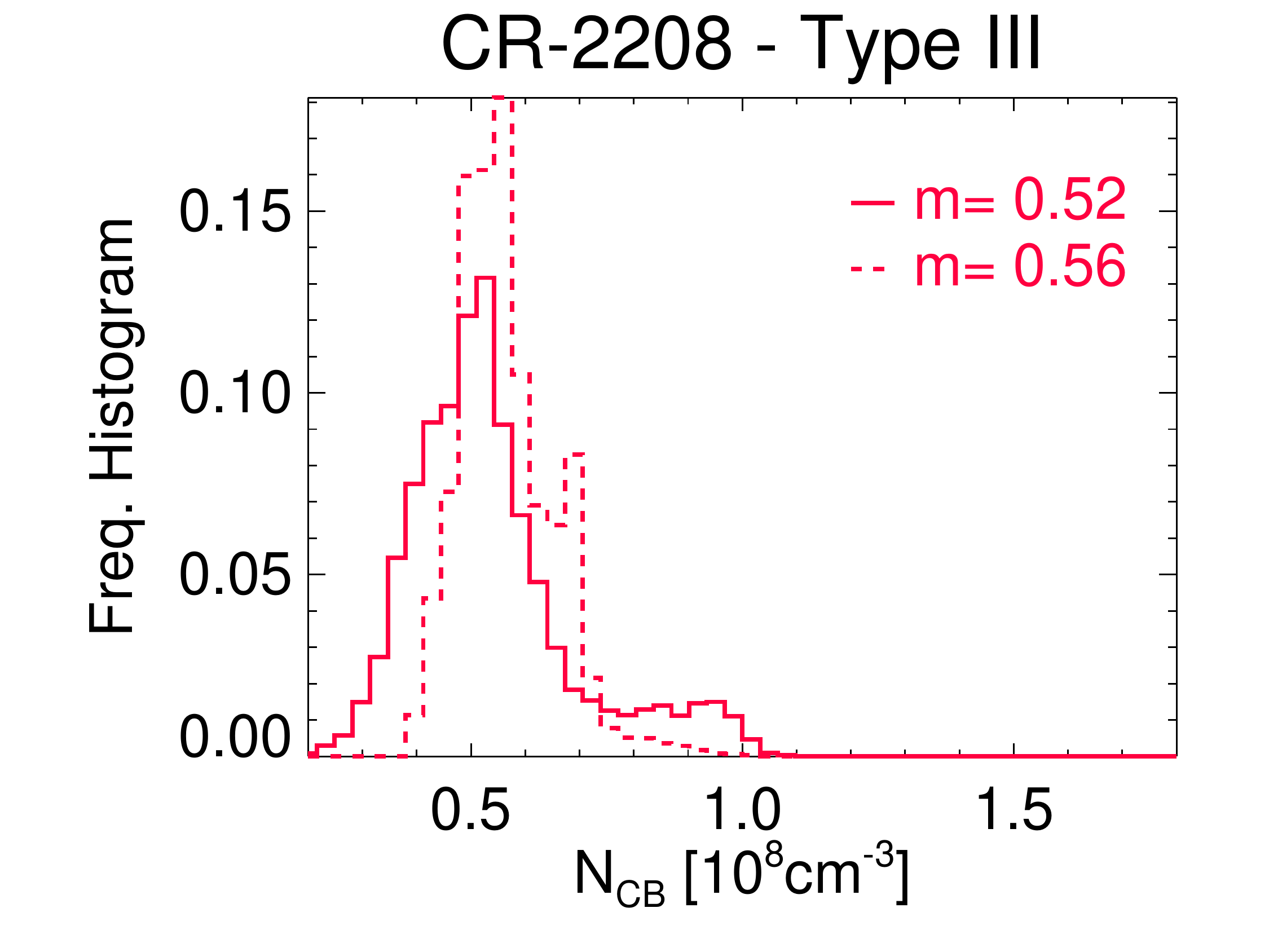}
\includegraphics[width=0.31\textwidth,clip=]{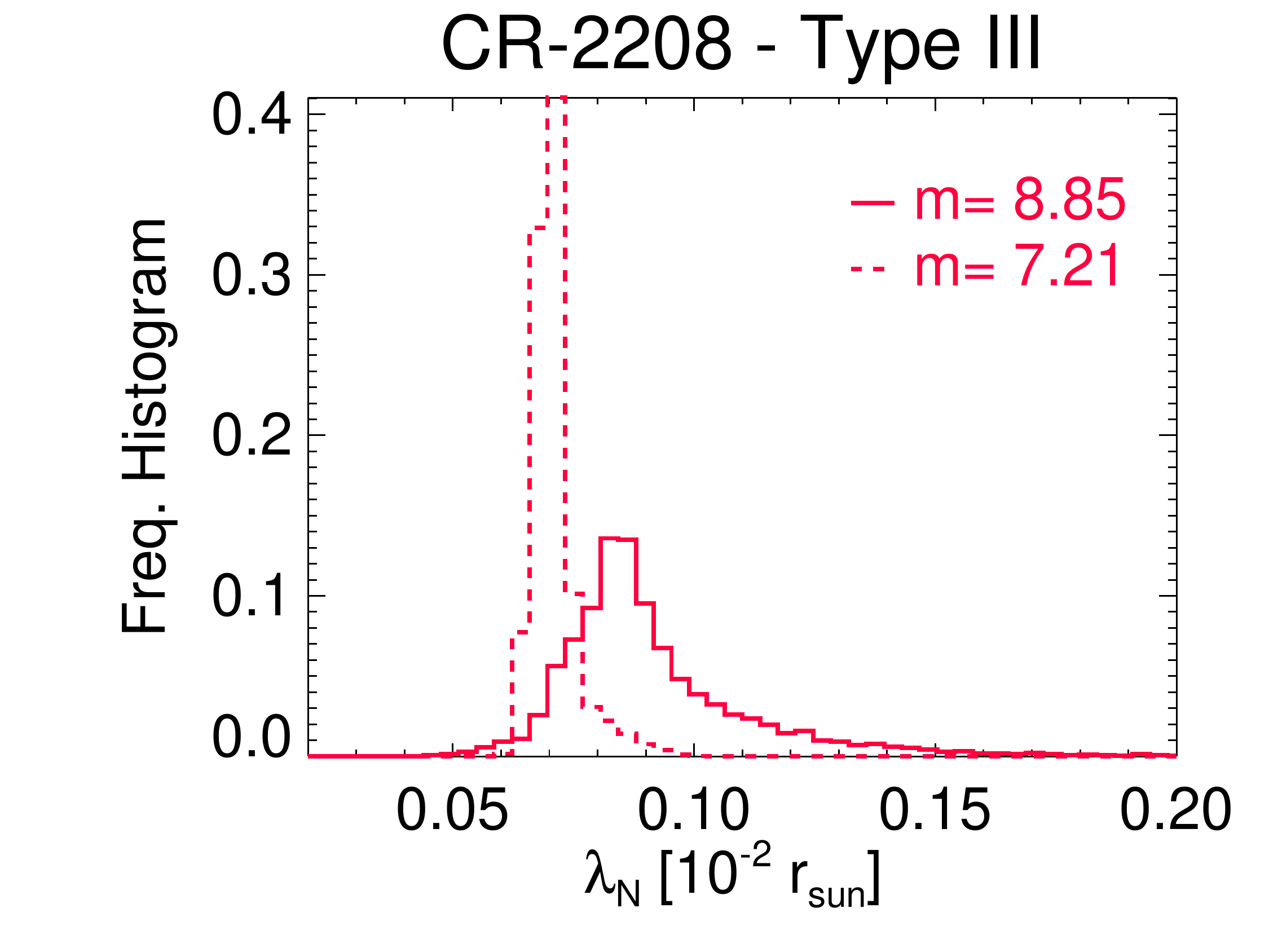}
\includegraphics[width=0.31\textwidth,clip=]{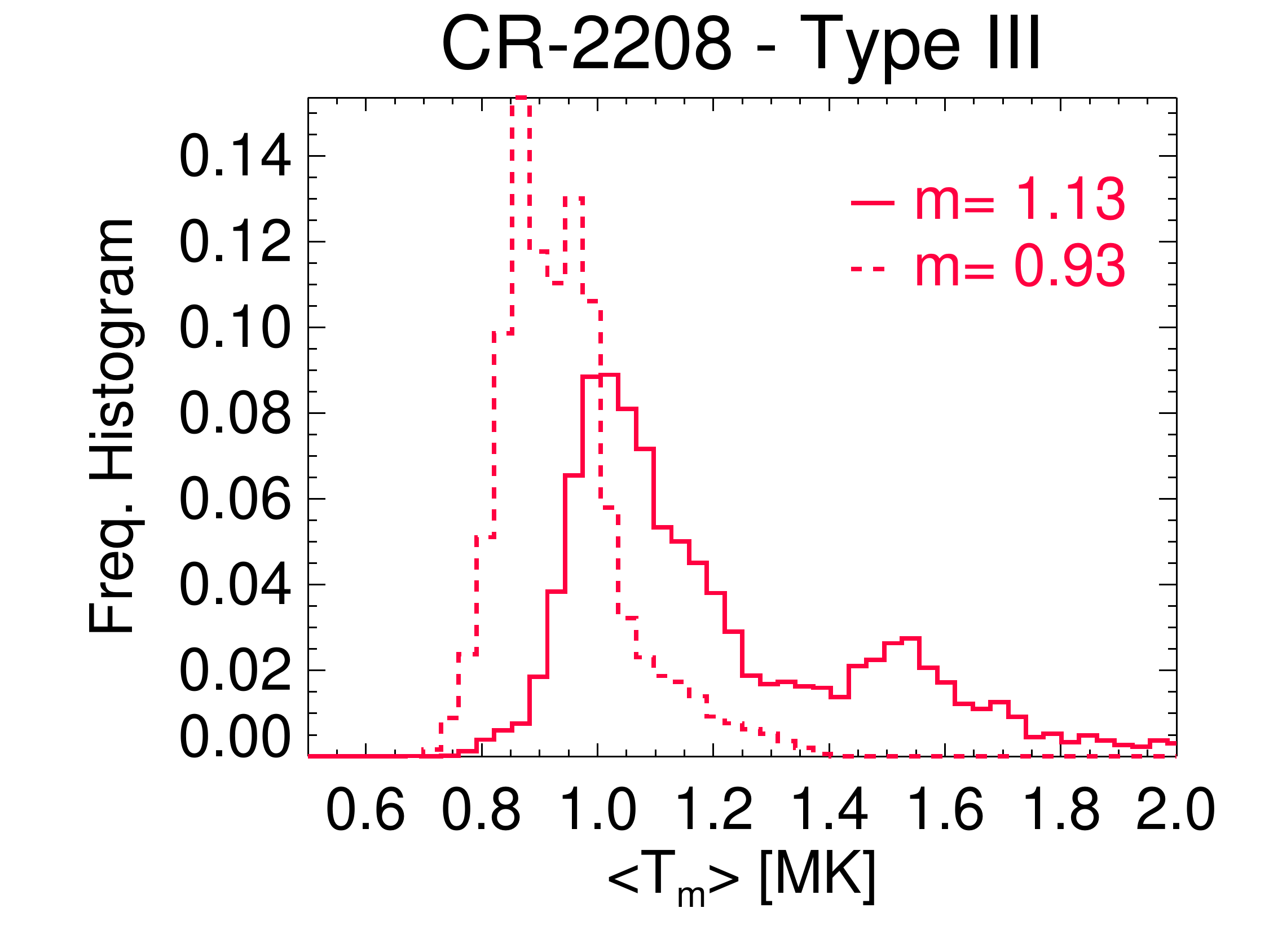}
\caption{Same as Figure \ref{histos_2082} for CR-2208.}
\label{histos_2208}
\end{center}
\end{figure}

{For rotation CR-2082, Figure \ref{histos_2082} shows the statistical distribution of the results of the DEMT (solid line-style) and AWSoM (dashed line-style) models traced along legs of type I, II and III (from top to bottom), as defined in Section \ref{trace}. From left to right: electron density at the lowest coronal height of the AWSoM model $\NCB\equiv\Ne(r=1.055\,\mrsun)$, electron density scale height $\lN$, and leg-averaged electron temperature $\aTm$. In each panel the median values $m$ are indicated. Figure \ref{histos_2208} shows the same analysis for rotation CR-2208.}

{For the two target rotations, Table \ref{tabla_comp} summarizes a quantitative comparative analysis between the results of the DEMT and AWSoM models {based on the results shown in Figures \ref{histos_2082} and \ref{histos_2208}}. The DEMT results are expressed as absolute values, while the ASWSoM results are informed as a percentual variation relative to the corresponding result for DEMT.}

\begin{table}
\begin{tabular}{l r@{.}l@{\hskip 0.05in} r@{\hskip 0.01in} r  r@{.}l@{\hskip 0.05in} r@{\hskip 0.01in} r r@{.}l@{\hskip 0.05in} r@{\hskip 0.01in} r }
\hline
Type    & \multicolumn{4}{c}{$\med(\NCB)$}             & \multicolumn{4}{c}{$\med(\lN)$} & \multicolumn{4}{c}{$\med(\avgTe)$} \\
        & \multicolumn{4}{c}{$[10^8\,{\rm cm}^{-3}]$}  & \multicolumn{4}{c}{$[{\rm 10}^{-2}\,\mrsun]$} & \multicolumn{4}{c}{$[\MK]$} \\
\hline
CR-2082\\
I    & 1&15 &(\Mi&14\%)  &   7&5 &(\Pl&~8\%) &   1&25 &(\Mi&10\%) \\
II   & 0&99 &(\Mi&25\%)  &   9&9 &(\Mi&~2\%) &   1&36 &(\Mi&~2\%) \\
III  & 0&66 &(\Mi&17\%)  &   7&0 &(\Mi&~9\%) &   0&97 &(\Mi&~9\%) \\
\hline          
CR-2208\\
I    & 1&03 &(\Pl&~6\%)  &   9&7 &(\Mi&~8\%) &   1&55 &(\Mi&17\%) \\
II   & 0&79 &(\Pl&~9\%)  &  11&8 &(\Mi&14\%) &   1&58 &(\Mi&~8\%) \\
III  & 0&{52} &({\Pl}&~{8}\%)  &   8&9 &(\Mi&18\%) &   1&{13} &(\Mi&18\%) \\
\hline   
\end{tabular}
\caption{Median value (indicated as ``Md'') of the statistical distribution of $\NCB$, $\lN$, and $\aTm$ for each coronal type of leg defined in Section \ref{trace}. DEMT values are expressed in absolute terms, while AWSoM results are expressed relative to the corresponding DEMT value.}
\label{tabla_comp}
\end{table}

{For rotation CR-2082, the median value of the electron density $\NCB$ of both models agree within $\approx 10-25\%$, {depending of the type of leg, with the largest discrepancy foud for legs of type II (near the open/closed boundary). The} median value of the scale height $\lN$ agree {within $\approx 10\%$ in all regions}. The leg-averaged electron temperature $\aTm$ of both models also agree {within $10\%$ in all regions}. For rotation CR-2208 the agreement of the median value of $\NCB$ and $\lN$ of both models {is within $10\%$, while} median values of $\aTm$ agree within $15\%$. {These detailed results, being consistent with the large-scale comparison provided in Figure \ref{perf_lat}, show in detail how the AWSoM model performs compared to DEMT in different magnetic structures.}}

\begin{figure}[h!]
\begin{center}
\includegraphics[width=0.495\textwidth,clip=]{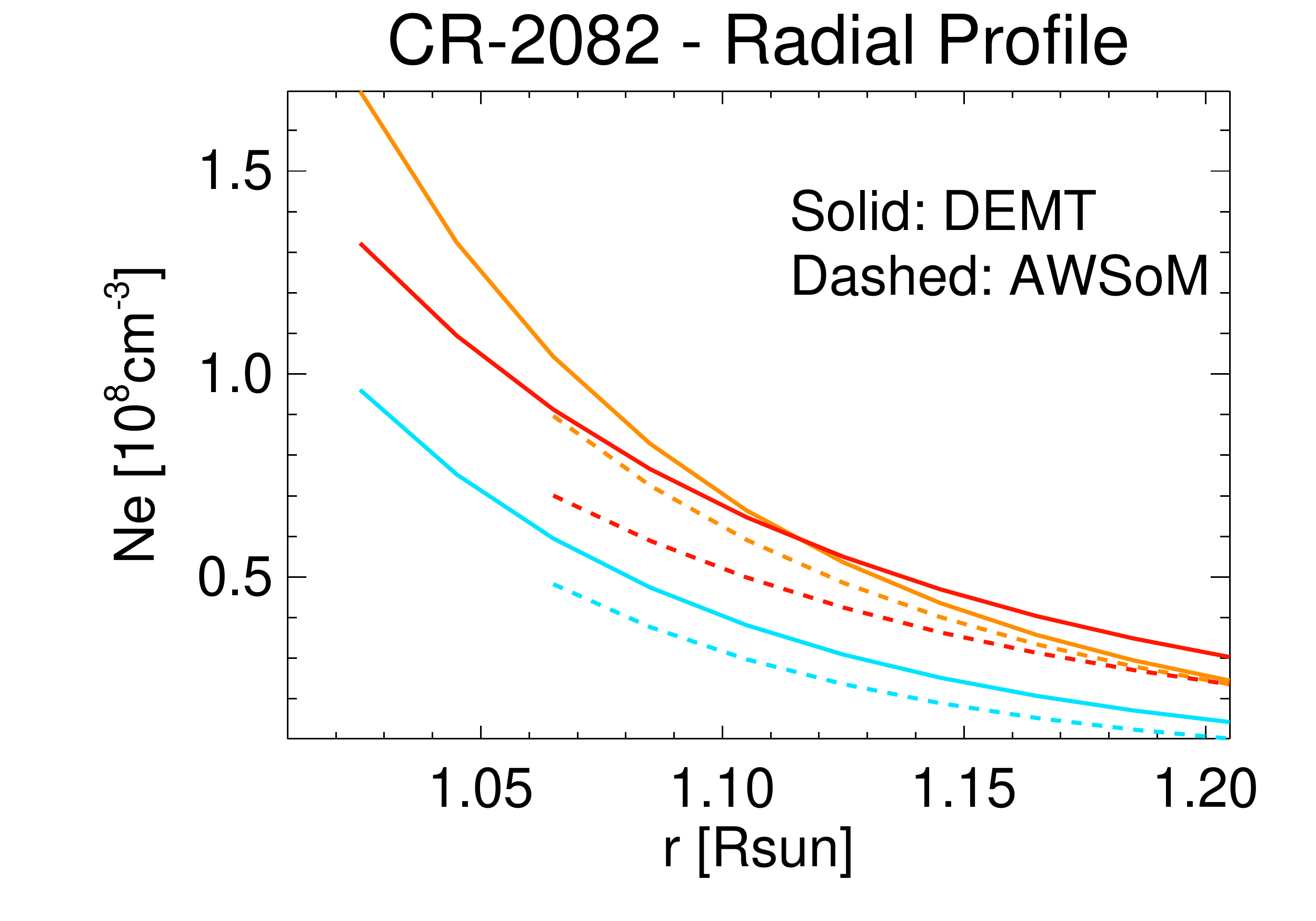}
\includegraphics[width=0.495\textwidth,clip=]{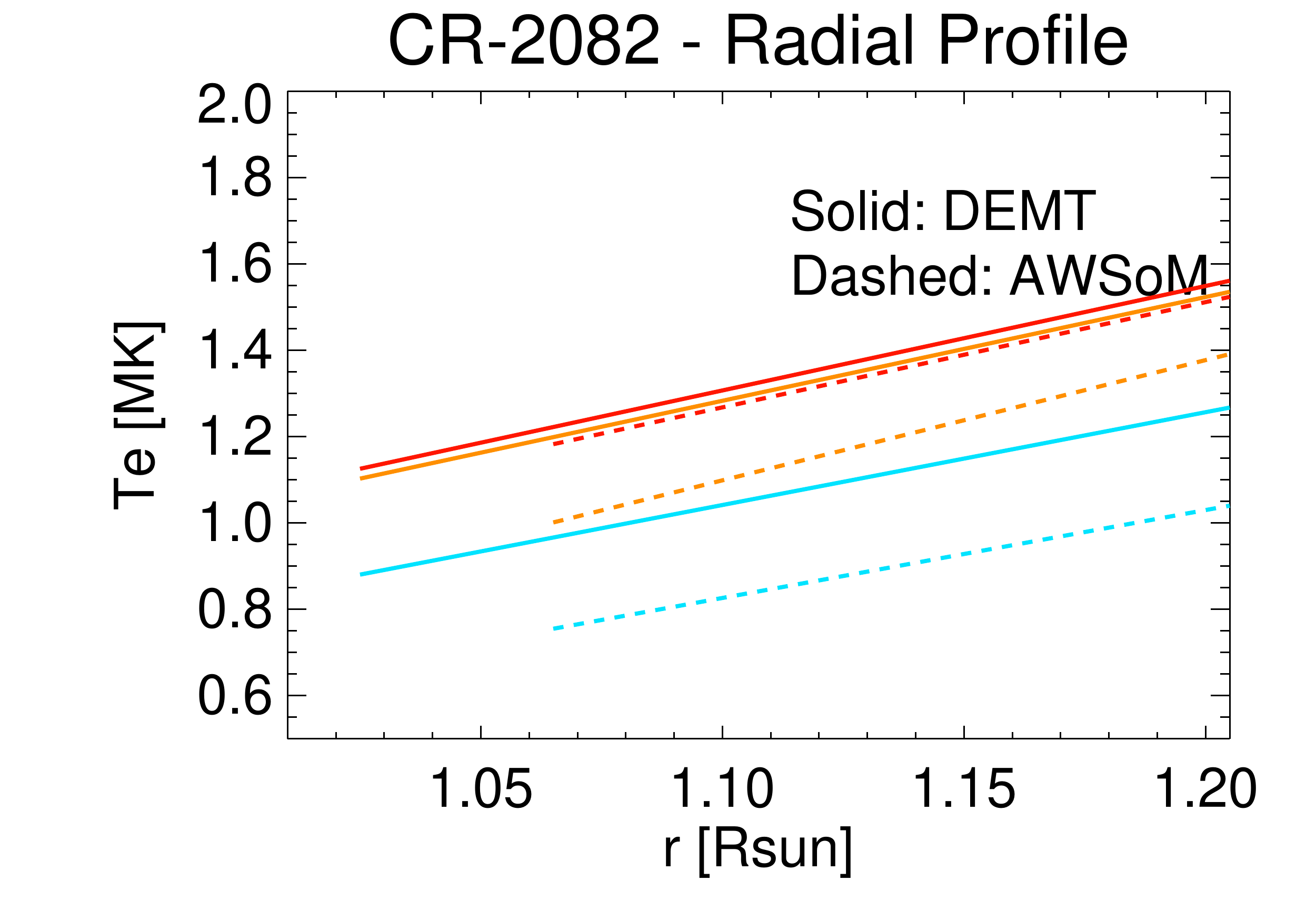}
\includegraphics[width=0.495\textwidth,clip=]{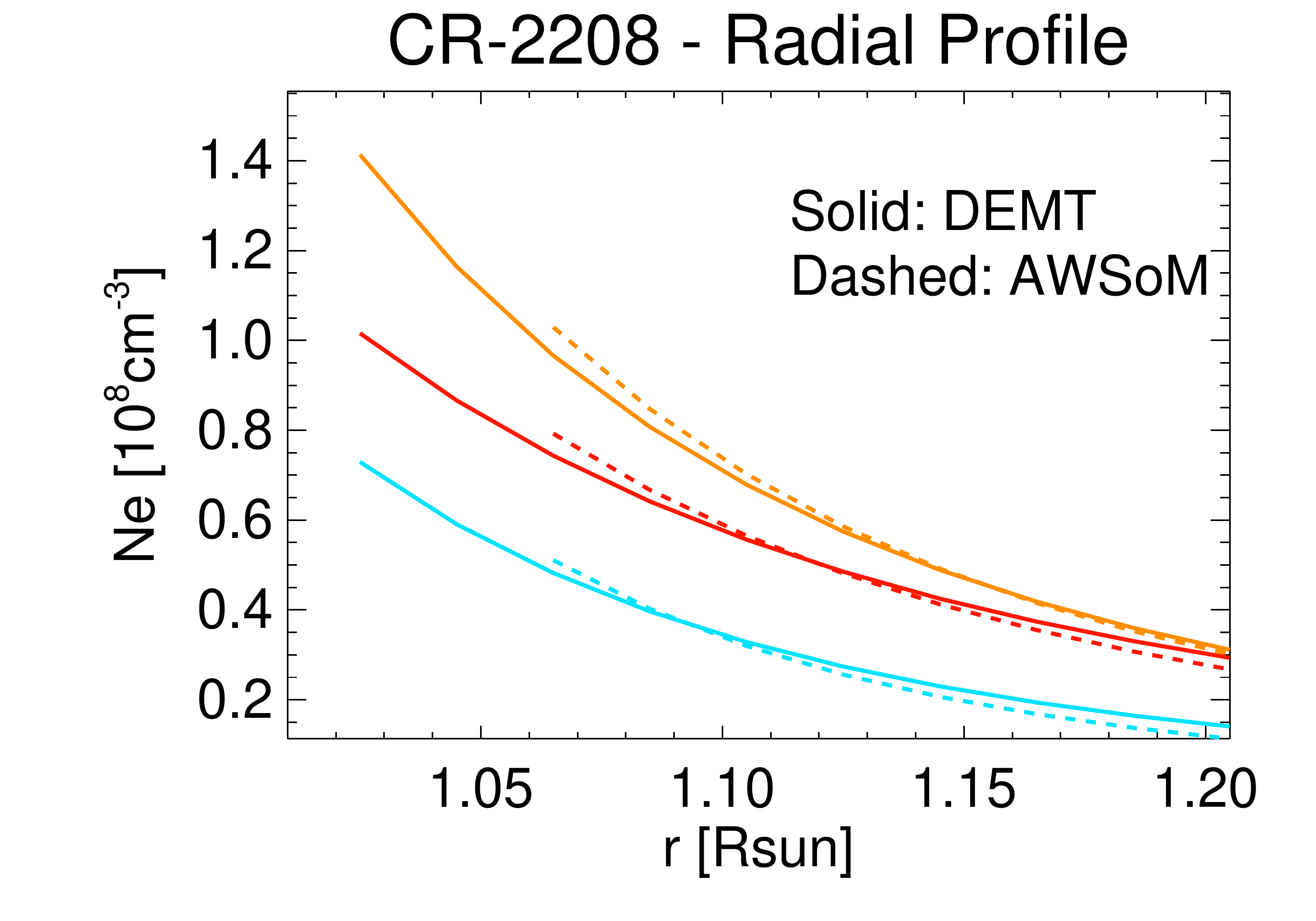}
\includegraphics[width=0.495\textwidth,clip=]{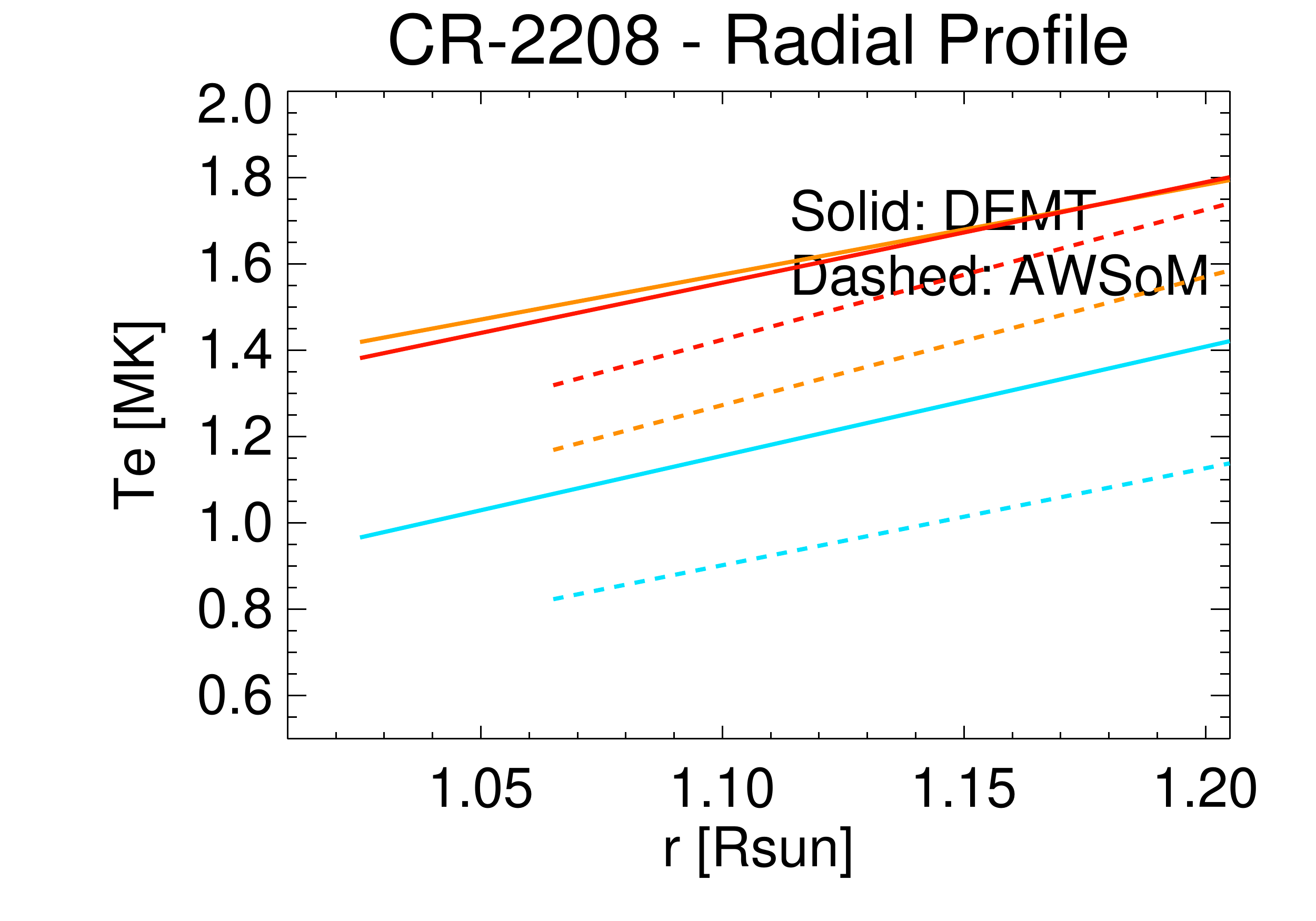}
\caption{{Average fits to $\Ne(r)$ (left panels) and $T_e(r)$ (right panels) for legs of {type I ({orange}), II ({red}), and III (cyan)}, for CR-2082 (top panels) and CR-2208 (bottom panels). Solid lines correspond to DEMT results while dashed lines correspond to AWSoM results.}}
\label{perfiles_promedio}
\end{center}
\end{figure}

{Finally, to provide a graphical comparison of both models across the full range of heliocentric heights covered by the DEMT results, Figure \ref{perfiles_promedio} shows the average fits of $\Ne(r)$ and $T_e(r)$ for legs of type I (red), II (magenta), and III (cyan) for both target rotations. In each panel the DEMT and AWSoM results are plotted in solid and dashed line styles, respectively.}

{As discussed above, Figure \ref{perf_lat} shows that the longitude-averaged latitudinal profile of the DEMT electron density in the CHs decreases towards the poles. Figure \ref{perf_lon_vr} below shows the longitude-averaged AWSoM radial wind speed $V_r$ at $6\,\mrsun$, where all field lines are open. The heliocentric current sheet (HCS) location is indicated by the minimum of the speed curve. For each rotation, all velocity data points to the south of the HCS position map down to the southern CH in Figures \ref{perf_lat}. Similarly, all velocity data points to the north of the HCS position map down to the northern CH in Figures \ref{perf_lat}. This clearly shows an anti-correlation between the DEMT electron density at low heights and the AWSoM wind speed at larger heights.}

\begin{figure}[h!]
\begin{center}
\includegraphics[width=0.495\textwidth]{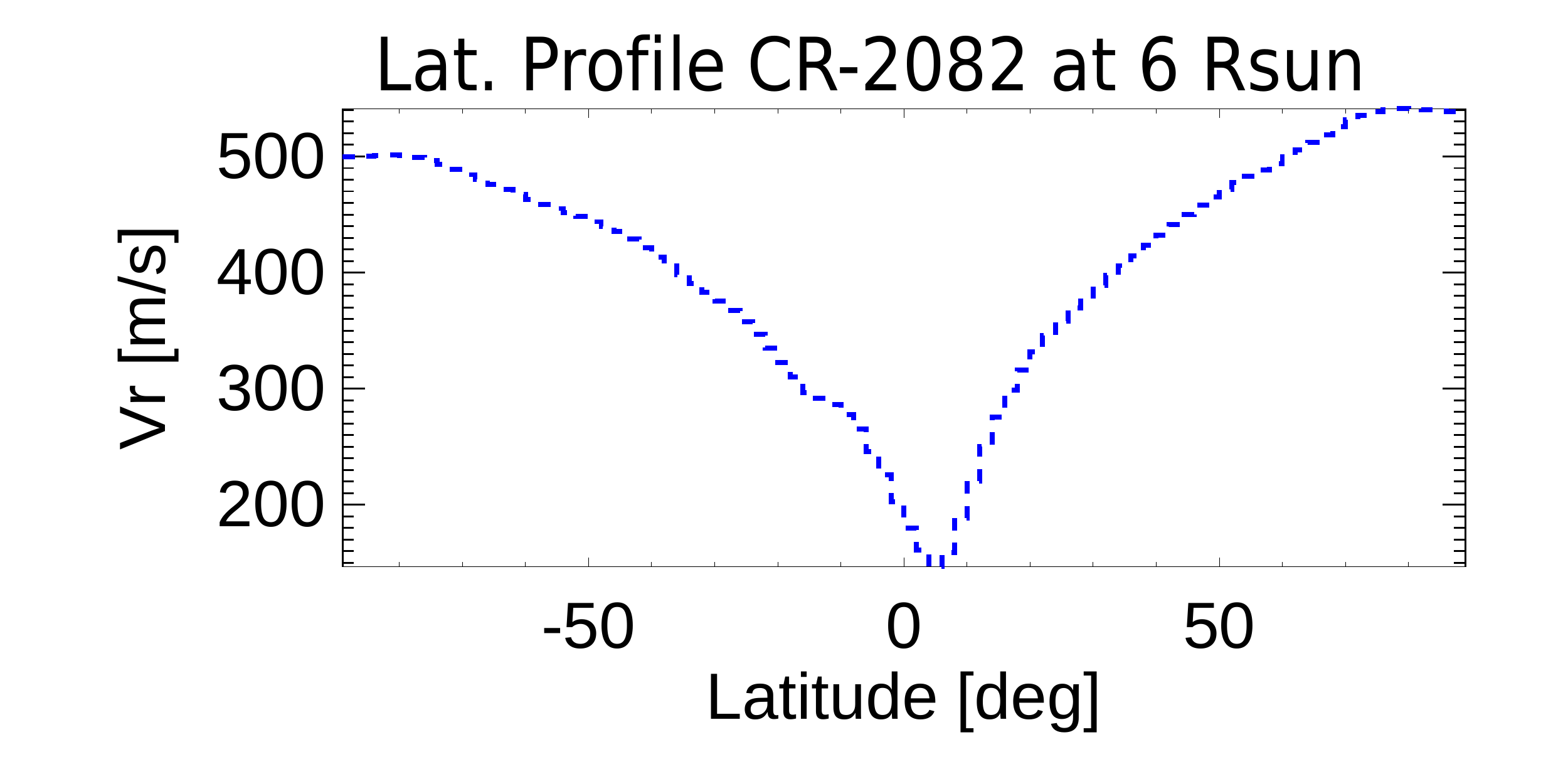}
\includegraphics[width=0.495\textwidth]{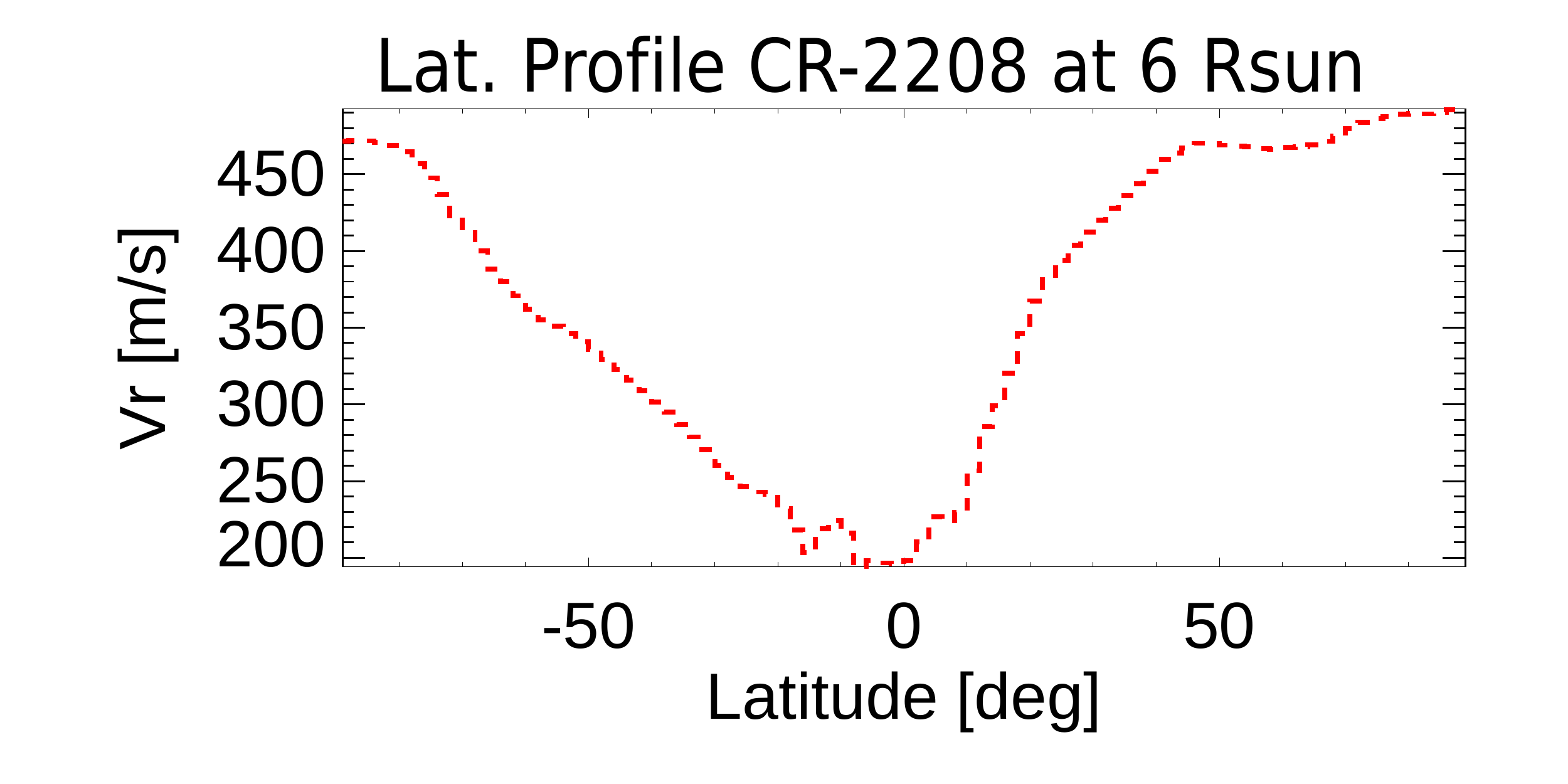}
\caption{{Longitude-averaged latitudinal dependence of the AWSoM model {wind speed $V_r$} at $6.0\,\mrsun$ for CR-2082 (left panel) and CR-2208 (right panel).}}
\label{perf_lon_vr}
\end{center}
\end{figure}

\section{{Discussion and Conclusions}}\label{discu} 

{Magnetic field lines of type 0, I and II were selected to be associated with increasingly outer layers of the equatorial streamer belt (Figure \ref{rpoint_demt}). These magnetic structures progressively exhibit decreasing coronal base density, increasing density scale height, and increasing electron temperature, as informed in 3D quantitative detail in Figure \ref{histos_fulldemt} and Table \ref{tabla_demt}.} For both rotations we find that down legs populate the low latitudes of the streamer belt, while up legs dominate its mid-latitudes. Also, in the case of CR-2082 the fraction of down legs is significantly larger than for CR-2208. These findings are consistent with previous studies by \citet{huang_2012} and \citet{nuevo_2013}. In the case of the latter, they include in their analysis target CR-2081, which is a rotation almost identical to our target CR-2082. Our results for target CR-2082 compare very well with those of \citet{nuevo_2015} and \citet{lloveras_2017} for target CR-2081. As our study uses the improved version of the DEMT technique, such comparison provided a consistency {check.} {For both rotations, type III field lines in the CHs are characterised by sub-MK temperatures, and electron density values of order $\approx 1/2$ of those observed for the type 0 and type I lines in the core of the equatorial streamer.}

{The energy input flux $\phi_h$ at the coronal base, required to maintain stable coronal loops, is in the range} $\phi_h\approx 0.5-1.5 \times 10^5\,\erg\,\cminvs\,\s^{-1}$, depending on the rotation and the type of {loop, matching the values reported by} \citet{maccormack_2017}. Based on spectroscopic data of the EIS instrument in quiet-Sun regions \citet{hahn_2014} showed that, if the observed non-thermal broadening are assigned to Alfvén waves, their energy flux at the coronal base is estimated to be in the range $\approx 1.5-2.5\times 10^5\,\erg\,\cminvs\,\s^{-1}$. A large fraction of the coronal base energy input flux $\phi_h$ estimated in this work, or even its totality, could then be accounted for by Alfvén waves (see the discussion in \citealt{maccormack_2017}).

{The comparison of the results of the AWSoM model to the DEMT reconstructions can be summarized as follows.} For CR-2082, the electron density of both models agree within $\approx 20\%$ {in most regions}, while for CR-2208 the agreement is within $\approx 5\%$. {The noticeable exception is to be found near the open/closed boundaries of both target rotations, where the disagreement between both models can be up to twice as much.} In the case of the electron temperature, both models agree within $\approx 10-15\%$. This level of agreement between both models {(within or slightly beyond the uncertainty level of DEMT results)} is considerably better than that reported in previous works. \citet{jin_2012} and \citet{oran_2015}{, who used previous versions of the AWSoM model,} reported electron density values of the AWSoM model {differing by $\approx 50\%$ compared to the DEMT reconstructions, both in the equatorial streamer and CH regions.} 

{The overall better match of the results of the current version of the AWSoM model compared to DEMT reconstructions is partly due to the improved energy partitioning scheme of the model, described in Section \ref{awsom}. {The} simulation of CR-2082 used GONG maps as boundary condition, while the simulation of CR-2208 used the improved ADAPT-GONG maps. This is likely the cause {of a more accurate match} to the DEMT reconstructions in the case of CR-2208.}

For both rotations, the AWSoM model reproduces the relatively lower temperatures found by DEMT to characterize the low-latitudes of the equatorial streamer belt compared to its mid-latitudes. On the other hand, while the latitude of the open/closed magnetic boundary in both hemispheres matches the location of the strongest latitudinal gradient of the DEMT electron density {(physically expected in transitioning from magnetically closed to open regions)}, this is not the case for the AWSoM model, that shows a minimum density at the open/closed boundary. Also, while the DEMT electron density decreases from the open/closed boundary towards the poles (in both hemispheres of the two rotations), {as expected in transitioning from the source region of the slow to the fast component of the solar wind}, the AWSoM model shows the opposite trend. This behavior is notoriously opposite to that reported in the AWSoM model version used by \citet{oran_2015}, in which the electron density decreases from the open/closed boundary towards the poles. These unphysical characteristics of the results of the AWSoM model in the {range} of low heights analysed here ($r\lesssim 1.2\,\mrsun$), may be attributed to the less reliable values of $B_r$ provided by both the GONG and ADAPT-GONG maps at subpolar latitudes. This will {be investigated} in a follow up article focusing on the current deep minimum epoch, during which the large-scale corona shows the simplest possible structure.

Down loops are to be expected if heating is enhanced at the footpoints of coronal structures. \citet{schiff_2016} numerically simulated stable down loops by means of a 1D steady-state model, requiring that the initial population of Alfvén waves is efficiently converted into compressive modes. Mode conversion is favored by the $\beta\gtrsim 1 $ condition found to characterize the regions where down loops are observed in DEMT analysis \citep{nuevo_2013}. In the AWSoM model, heating is controlled by two key parameters, namely, the Alfvén wave dissipation length, and its reflection coefficient. So far, attempts to reproduce observed down loops in AWSoM simulations by increasing the reflection coefficient have not been successful.

A detailed empirical description of the 3D thermodynamic structure of the inner corona at a global scale is currently only possible with tomographic techniques, such as DEMT. Using tomographic results for continuous validation of 3D MHD models is of high relevance for the continued improvement of models. In {follow up articles} we will {carry out the 3D DEMT reconstruction and MHD modeling of} new target rotations {selected from} the current solar minimum {epoch} between SCs 24 and 25, {as well as rotation CR-2219 corresponding to the July 2nd 2019 total solar eclipse.}

\begin{acks}
D.G.L. and C.M.C. {acknowledge} CONICET doctoral {fellowships} (Res. Nr. 4870) to IAFE that supported {their} participation in this research. {D.G.L, A.M.V., F.A.N. and C.M.C. acknowledge ANPCyT grant 2016/0221 to IAFE that partially supported their participation in this research. A.M.V. also acknowledges UBACyT grant 20020160100072BA to DCAO-UBA to FCEyN-UBA and IAFE that partially supported his participation in this research.} W.M. and B.v.H acknowledge NSF grant 1663800 that partially supported his participation in this research. W.M. also acknowledges NASA grants NNX16AL12G and 80NSSC17K0686.
\end{acks}

\begin{footnotesize}
Disclosure of Potential Conflicts of Interest: The authors declare that they have no conflicts of interest.
\end{footnotesize}

\bibliographystyle{spr-mp-sola}%{plainnat}%{spr-mp-sola}
\bibliography{lloveras_diego_2020_arxiv}  

\end{article} 
\end{document}